\documentclass[12pt]{report}
\pdfoutput=1
\usepackage{tamuconfig}
\usepackage{amssymb}
\usepackage{mathrsfs,amsmath} 

\usepackage{times}
\usepackage[T1]{fontenc}
\usepackage[hidelinks]{hyperref}
\usepackage{booktabs}
\usepackage[ruled,vlined]{algorithm2e}
\usepackage[noend]{algpseudocode}
\usepackage[caption=false]{subfig}
\usepackage[symbol]{footmisc}

\usepackage{graphicx}

\DeclareGraphicsExtensions{.png}

\graphicspath{ {./graphic/} }

\begin{document}

\renewcommand{\tamumanuscripttitle}{METHODS AND SYSTEMS FOR THE SPECTRAL CALIBRATION OF SWEPT SOURCE OPTICAL COHERENCE TOMOGRAPHY SYSTEMS}

\renewcommand{\tamupapertype}{Dissertation}

\renewcommand{\tamufullname}{Amir Tofighi Zavareh}

\renewcommand{\tamudegree}{Doctor of Philosophy}
\renewcommand{\tamuchairone}{Sebastian Hoyos}

\renewcommand{\tamumemberone}{Samuel Palermo}
\newcommand{\tamumembertwo}{Ulisses Braga-Neto}
\newcommand{\tamumemberthree}{Saurabh Biswas}
\renewcommand{\tamudepthead}{Miroslav M. Begovic}

\renewcommand{\tamugradmonth}{December}
\renewcommand{\tamugradyear}{2019}
\renewcommand{\tamudepartment}{Electrical Engineering}

%
%
%
%


\providecommand{\tabularnewline}{\\}

\begin{titlepage}
\begin{center}
\MakeUppercase{SYSTEMS AND METHODS FOR THE SPECTRAL CALIBRATION OF SWEPT SOURCE OPTICAL COHERENCE TOMOGRAPHY SYSTEMS}
\vspace{4em}

A \tamupapertype

by

\MakeUppercase{Amir Tofighi Zavareh}

\vspace{4em}

\begin{singlespace}

Submitted to the Office of Graduate and Professional Studies of \\
Texas A\&M University \\

in partial fulfillment of the requirements for the degree of \\
\end{singlespace}

\MakeUppercase{Doctor of Philosophy}
\par\end{center}
\vspace{2em}
\begin{singlespace}
\begin{tabular}{ll}
 & \tabularnewline
& \cr
Chair of Committee, & Sebastian Hoyos\tabularnewline
Committee Members, & Samuel Palermo\tabularnewline
 & Ulisses Braga-Neto\tabularnewline
 & Saurabh Biswas\tabularnewline
Head of Department, & Miroslav M. Begovic\tabularnewline

\end{tabular}
\end{singlespace}
\vspace{3em}

\begin{center}
December \hspace{2pt} 2019

\vspace{3em}

Major Subject: Electrical Engineering \par
\vspace{3em}
Copyright 2019 \hspace{.5em} Amir Tofighi Zavareh
\par\end{center}
\end{titlepage}
\pagebreak{}

%
%
%
%

\chapter*{ABSTRACT}
\addcontentsline{toc}{chapter}{ABSTRACT} 

\pagestyle{plain} 
\pagenumbering{roman} 
\setcounter{page}{2}

\indent This dissertation relates to the transition of the state of the art of swept source optical coherence tomography (SS-OCT) systems to a new realm in which the image acquisition speed is improved by an order of magnitude. With the aid of a better quality imaging technology, the speed-up factor will considerably shorten the eye-exam clinical visits which in turn improves the patient and doctor interaction experience. \\
\indent These improvements will directly lower associated medical costs for eye-clinics and patients worldwide. There are several other embodiments closely related to Optical Coherence Tomography (OCT) that could benefit from the ideas presented in this dissertation including: optical coherence microscopy (OCM), full-field OCT (FF-OCT), optical coherence elastography (OCE), optical coherence tomography angiography (OCT-A), anatomical OCT (aOCT), optical coherence photoacoustic microscopy (OC-PAM), micro optical coherence tomography ($\mu$ OCT), among others.\\
\indent In recent decades, OCT has established itself as the de-facto imaging process that most ophthalmologists refer to in their clinical practices. In a broader sense, optical coherence tomography is used in applications when low penetration and high resolution are desired. These applications include different fields of biomedical sciences including cardiology, dermatology, and pulmonary related sciences. Many other industrial applications including quality control and precise measurements have also been reported that are related to the OCT technology. \\
\indent Every new iteration of OCT technology has always come about with advanced signal processing and data acquisition algorithms using mixed-signal architectures, calibration and signal processing techniques. The existing industrial practices towards data acquisition, processing, and image creation relies on conventional signal processing design flows, which extensively employ continuous/discrete techniques that are both time-consuming and costly. The ideas presented in this dissertation can take the technology to a new dimension of quality of service.

\pagebreak{}

%
%
%
%

\chapter*{DEDICATION}
\addcontentsline{toc}{chapter}{DEDICATION}  

\begin{center}
\vspace*{\fill}
To my parents and my brother. Your selflessness is an example for me in the years to come.\\
To my beloved \textbf{you}. Without your help, I could not have succeeded. Thanks for keeping me organized, thanks for all your unconditioned support, thanks for bearing with me all these years. \textit{This is just the beginning}. 
\vspace*{\fill}
\end{center}

\pagebreak{}

%
%
%
%

\chapter*{ACKNOWLEDGMENTS}
\addcontentsline{toc}{chapter}{ACKNOWLEDGMENTS}  

\indent I would like to thank my committee chair, Dr. Sebastian Hoyos, for his guidance and support throughout the course of this research. His extraordinary full support was the motive force for carrying on this research forward. He provided me with opportunities that I could not have imagined I would accomplish when I first started my studies at Texas A\&M University.\\
\indent Thanks also go to my friends and colleagues at the Analog and Mixed Signal Center at the Electrical \& Computer Engineering Department at Texas A\&M University who without a doubt were always available for me through the ups and downs I faced during my years there.  \\
\indent I also want to thank the office of Entrepreneurship and Commercialization at the Texas A\&M Engineering Experiment Station who helped me and the research lab I worked in to try to transition the ideas presented in this dissertation to the commercial world. Special thanks go to Dr. Xiaomin Yang and Dr. Saurabh Biswas for their valuable pieces of advice and support. \\
\indent

\pagebreak{}
%
%
%
%

\chapter*{CONTRIBUTORS AND FUNDING SOURCES}
\addcontentsline{toc}{chapter}{CONTRIBUTORS AND FUNDING SOURCES}  

\subsection*{Contributors}
This work was supported by a  dissertation committee consisting of Professor Sebastian Hoyos [advisor], Professor Samuel Palermo, Professor Ulisses Braga-Neto of the Department of Electrical and Computer Engineering, and Professor Saurabh Biswas of the Department of Biomedical Engineering.

The data analyzed for this work was provided by Professor Javier Jo from the Department of Biomedical Engineering. \\
\indent All other work conducted for the dissertation was completed independently.
\subsection*{Funding Sources}
The author wants to acknowledge the awards that helped develop the ideas presented in this dissertation.  This dissertation has roots in multiple NSF funded projects during the past couple of years. This dissertation claims NSF lineage through the successful participation in the University of Texas at San Antonio Super Regional Program deployed in May-June 2017 by the NSF Southwest I-Corps Node (NSF Award - I-Corps Node: Southwest Alliance for Entrepreneurial Innovation Node (SAEIN), Award Number: 1444045). This project further claims NSF lineage through the successful participation in the national I-Corps program, where the project was awarded a \$50000 grant with the number 1760126 with the Division of Industrial Innovation \& Partnership.

\pagebreak{}
%
%
%
%


\chapter*{NOMENCLATURE}
\addcontentsline{toc}{chapter}{NOMENCLATURE}  


\hspace*{-1.25in}
\vspace{12pt}
\begin{spacing}{1.0}
	\begin{longtable}[htbp]{@{}p{0.35\textwidth} p{0.62\textwidth}@{}}
		OCT	&	Optical Coherence Tomography\\	[2ex]
		OLCR & Optical Low Coherence Reflectometry \\	[2ex]
        OFDI & Optical Frequency-domain Imaging\\	[2ex]
        OCM & Optical Coherence Microscopy \\	[2ex]
        FF-OCT& Full Field Optical Coherence Tomography \\	[2ex]
        OCE & Optical Coherence Elastography \\	[2ex]
        OCT-A & Optical Coherence Tomography Angiography \\	[2ex]
        aOCT & Anatomical Optical Coherence Tomography \\	[2ex]
        OC-PAM & Optical Coherence Photoacoustic Microscopy \\	[2ex]
        $\mu$OCT & Micro Optical Coherence Tomography \\	[2ex] 
		TD-OCT		&	Time Domain Optical Coherence Tomography\\	[2ex] 
		FD-OCT			&	Frequency Domain Optical Coherence Tomography\\	[2ex]
		SD-OCT & Spectrum Domain Optical Coherence Tomography\\ [2ex]
		SS-OCT & Swept Source Optical Coherence Tomography\\ [2ex]
		CW & Continuous-Wave\\ [2ex]
		NA & Numerical Aperture\\ [2ex]
		FOV & Field of View\\ [2ex]
		MZI & Mach-Zehnder Interferometer\\ [2ex]
		FPI & Fiber Fabry-Perot Interferometer \\ [2ex]
		FPF & Fiber Fabry-Perot Filter \\ [2ex]
		LCS & Level Crossing Sampler\\ [2ex]
		CTTE & Continuous Time Ternary Encoder\\ [2ex]
		CLK & Clock\\ [2ex]
		EKF & Extended Kalman Filter\\ [2ex]
		UKF & Unscented Kalman Filter\\ [2ex]
		FFT & Fast Fourier Transform\\ [2ex]
		DFT & Discrete Fourier Transform\\ [2ex]
		IpDFT & Interpolated Discrete Fourier Transformation\\ [2ex]
		BY-2 & Betrocco-Yoshida\\ [2ex]
		RVCI & Rife-Vincent Class I\\ [2ex]
		RTL & Register Transfer Level\\ [2ex]
		FPGA & Field Programmable Gate Array\\ [2ex]
		CORDIC & Coordinate Rotation Digital Computer\\ [2ex]
		ASIC & Application-Specific Integrated Circuit\\ [2ex]
		RAM & Random Access Memory\\ [2ex]
		CPU & Computing Processing Unit\\ [2ex]
		ADC & Analog to Digital Converter\\ [2ex]
		DAC & Digital to Analog Converter\\ [2ex]
		VCSEL & Vertical Cavity Surface Emitting Lasers\\ [2ex]
		FDML &  Fourier Domain Mode Locked Laser\\ [2ex]
		SLED & Superluminescent diode \\ [2ex]
		ODL & Optical Delay Line \\	[2ex]
		PC & Polarization Controller \\ [2ex]
		SNR &  Signal to Niose Ratio\\ [2ex]
		RIN & Relative Intensity Noise \\ [2ex]
		SOA & Semiconductor Optical Amplifier \\ [2ex]
		MEMS & Micro Electro Mechanical Systems \\ [2ex]
		FWHM & FWHM Full-Width Half Maximum \\ [2ex]
		USPTO & United States Patent and Trademark Office \\ [2ex]
		FDA & Food and Drug Administration \\ [2ex]
		
	\end{longtable}
\end{spacing}

\pagebreak{}

%
%
%
%

\phantomsection
\addcontentsline{toc}{chapter}{TABLE OF CONTENTS}  

\begin{singlespace}
\renewcommand\contentsname{\normalfont} {\centerline{TABLE OF CONTENTS}}

\setcounter{tocdepth}{4} 

\setlength{\cftaftertoctitleskip}{1em}
\renewcommand{\cftaftertoctitle}{%
\hfill{\normalfont {Page}\par}}

\tableofcontents

\end{singlespace}

\pagebreak{}


\phantomsection
\addcontentsline{toc}{chapter}{LIST OF FIGURES}  

\renewcommand{\cftloftitlefont}{\center\normalfont\MakeUppercase}

\setlength{\cftbeforeloftitleskip}{-12pt} 
\renewcommand{\cftafterloftitleskip}{12pt}

\renewcommand{\cftafterloftitle}{%
\\[4em]\mbox{}\hspace{1pt}FIGURE\hfill{\normalfont Page}\vskip\baselineskip}

\begingroup

\begin{center}
\begin{singlespace}
\setlength\cftbeforefigskip{\baselineskip}



\listoffigures

\end{singlespace}
\end{center}

\pagebreak{}

%
\phantomsection
\addcontentsline{toc}{chapter}{LIST OF TABLES}  

\renewcommand{\cftlottitlefont}{\center\normalfont\MakeUppercase}

\setlength{\cftbeforelottitleskip}{-12pt} 

\renewcommand{\cftafterlottitleskip}{1pt}

\renewcommand{\cftafterlottitle}{%
\\[4em]\mbox{}\hspace{4pt}TABLE\hfill{\normalfont Page}\vskip\baselineskip}

\begin{center}
\begin{singlespace}

\setlength{\cftbeforetabskip}{0.4cm}

\setlength\cftbeforefigskip{\baselineskip}

\listoftables 

\end{singlespace}
\end{center}
\endgroup
\pagebreak{}  

%
%
%
%


\pagestyle{plain} 
\pagenumbering{arabic} 
\setcounter{page}{1}

\chapter{\uppercase {Introduction and Literature Review}} \label{Chapter1}

\section{Motivation}

Optical Coherence Tomography (OCT) is an imaging technology that is mostly used to diagnose eye-related diseases. It serves millions of patients across the globe that visit ophthalmologists for initial and follow-up appointments with over 30 million OCT imaging procedures a year \cite{econ1}. Eye clinics rely on OCT as the only biomedical imaging modality with sufficient low penetration and high resolution for accurate diagnosis. The overall patient experience and the quality of the diagnosis depends on the speed and resolution of images acquired by this technology, which have made incremental improvements in the past 25 years since the time that the concept was initially introduced \cite{OCTOriginal}.  The ideas put forward in this dissertation promise to deliver a significant, one order of magnitude improvement to the scan speed of this imaging technology while maintaining the maximum possible image resolution that is set by the light source wavelength bandwidth. To accomplish such a breakthrough speedup factor, this dissertation proposes an innovative combination of electronic systems and circuits that increase the imaging rate and reduce the image artifacts via a real-time spectral calibration of swept source optical coherence tomography (SS-OCT). These techniques are designed to be implemented as a modular electronic device which can be integrated into an OCT instrument. In addition to providing real-time, three-dimensional, high-resolution images, a faster swept source OCT scan rate allows ophthalmologists to look at other eye tissues of medical importance, all in one scan. From a commercial standpoint, not only will this technology provide the manufactures of SS-OCT systems with the ability to perform spectral calibration using a simple plug-and-play device, but it also will embed important computational resources for other operations related to the image reconstruction, reducing the overall cost of the system by replacing the external computational device altogether.\\
\indent More advanced iterations of OCT technology have always come about with better but limited signal processing and data acquisition speed that use conventional mixed-signal architectures, calibration and signal processing techniques. The ideas presented in this dissertation can take the technology to a new dimension of quality of service. In the course of this research, it is discovered that in order to acquire high-quality images, the patient's eyes need to remain still for prolonged periods of time. Fast eye movements deteriorate image quality and diagnosis ability. While the existing industrial practices towards data acquisition, processing, and image creation rely on conventional signal processing design flows, which extensively employ continuous/discrete techniques that are both time-consuming and costly, ideas proposed here make the image scan an order of magnitude faster at the highest resolution setting. This will give medical practitioners the opportunity to diagnose more patients a day which translates to a lower cost per patient per machine.\\
\indent While current systems for generating OCT images function well for obtaining information on the basic anatomy of the retina, there are emerging technologies for OCT which the current machines are struggling to implement.\\
\indent As an example, the technique of OCT angiography (OCT-A), which employs serial scans of the retina to identify the location and patency of vascular structures, has been developed and promises another leap forward in the understanding of retinal pathology. However, the architecture of the current systems limits the implementation of this technology since the amount of data generated is massive. Current OCT systems measure multiple parameters for each scan point in a scan head and process the data in a separate attached computer. The limited transport capability of commercially viable data pipelines and the modest processing power of computer processing units available to the clinical community limit the application of this technology to very small scans of the retina in only the most cooperative patients; hence the need to provide ideas that enable the production of computational unites that can act as a central processing units (CPU) that are specifically designed for the purposes necessary for tomographic image creation via OCT devices.  \\
\indent The innovations proposed herein, stem from  a collaborative effort that results in presenting analog and mixed signal acquisition and processing systems and circuits. This innovation enables designers to provide SS-OCT device manufacturers with calibration engines that bring them to the next realm of image acquisition speed. The proposed, ultra-fast level crossing sampler (LCS), as well as the most efficient signal processing demodulation algorithms reported in the literature, can be combined to provide a viable solution for the fastest, all integrated, real-time, plug and play acquisition setup that can be used specifically in SS-OCT devices. Unlike the current OCT devices that require a certain processing time to provide a full cross-sectional image of the object under test, when the proposed engine is widely adopted in the market, next generations of SS-OCT devices will be able to acquire tomographic images as the material to be imaged undergoes any spontaneous change.\\
\indent Due to the possibility of integrating all the ideas presented in this dissertation in an onchip implementation, there exist the potential to produce a product that would be of a significant clinical advance and open new frontiers in the understanding of retinal disease.

\section{Commercial Opportunity}

\subsection{Broader Societal Need}

Optical Coherence Tomography (OCT) is a noninvasive imaging technique that provides cross-sectional images of the subject under test. Since it was first introduced in the early nineties \cite{OCTOriginal} as a modality for depth-resolved high-resolution imaging in transparent and turbid media, the device has been widely used in the clinical practice of ophthalmology where is used to acquire tomographic images of the retina of the human eye. The OCT device can be used for choroidal imaging of healthy eyes, or to diagnose age-related macular degeneration, central serous chorioretinopathy, diabetic retinopathy and other inherited retinal conditions \cite{OCT-CFA}. Such wide utilization of OCT devices makes them in fact the go-to device for ocular disease management in ophthalmology clinics. Besides the wide adoption of the technology in ophthalmic clinics, there exists an ever-expanding traction of using these devices in industrial applications for testing and evaluation purposes due to their price, speed, and accuracy. The OCT technology can be used for industrial inspections such as that of the silicon and plastic micro electro mechanical systems (MEMS). The technology can further be utilized to perform quality control of solar panels and devices made of composite materials such as turbine blades. Other applications of such devices include inspection of micro-engineered plastic prototype parts, polymer devices, and pearls \cite{OCTInsustrialapp}. Other than the aforementioned practices where the device can greatly benefit the public, OCT technology can be used in applications that involve dimensional metrology, contactless material characterization, and archaeological science and art diagnostics \cite{OCTInsustrialapp1}. It also has been reported in the literature that the OCT technology is being more increasingly accepted in areas of the life sciences realm other than ophthalmology such as pathology, oncology, and surgical guidance \cite{OCTEco}. In fact, \cite{OCTEco} shows that the OCT market is being expanded in terms of publications per year, number of startup companies founded, number of clinical procedures taken place to the point that the market size is reaching a \$1.8 billion value by the year 2024. \\
\indent Since its introduction in 1991 as a time domain interferometric technique (TD-OCT) \cite{OCTOriginal}, the OCT industry has undergone a rapid economic growth akin to ultrasound technology in previous decades \cite{UltrasoundMarket}. This market growth is mainly attributed to the new data acquisition techniques and signal processing algorithms that have evolved during the course of the past two decades. Starting with the introduction of spectral domain OCT (SD-OCT) in 2006, the industry experienced its first substantial growth where all major players in the field moved to produce the newly developed iteration of the technology.  At around this time, the number of companies active in the field increased by more than one hundred percent and continues to grow at similar rates to date. The United States Patent and Trademark Organization (USPTO) \cite{OCTPatent} data shows a significant boost in terms of the number of patents granted in the field at around the same timeframe. While the number of services provided by governmentally supported insurances such as Medicare rather plateaued for other imaging modalities in ophthalmology clinics, this number continues to grow for tests carried out with OCT devices \cite{OCTEco}.\\
\indent The main disadvantage of SD-OCT systems is its image acquisition speed and lack of sufficient resolution needed for an ophthalmologist to efficiently diagnose ocular diseases. This drawbacks pushed the OCT device manufacturers to design newer iterations of the OCT technology \cite{topcon}.  In recent years, two of the major players in the field, Carl Zeiss Meditec, and Topcon Corporation introduced their first ever SS-OCT products that were approved by the Food and Drug Administration (FDA). With the introduction of the third iteration of the technology, SS-OCT, in 2015, an even bigger growth in the field is expected to occur.
\subsection{OCT Market and Market Drivers}
\indent OCT market is estimated to have a 12\% compound annual growth rate through 2020 and reach \$1.32 billion due to an increase in the aging population. An increase in eye-related disorders is expected over the next 5 years at an annual growth rate of 7\% \cite{OCTMarket}. Besides the promising and continuous market size growth, research shows that the number of direct and indirect jobs in the field has unceasingly grown since the wide adoption of the OCT technology. OCT is now an academic and industrial trend in all major economic regions of the world such as North America, Europe, and Asia \cite{OCTEco}. \\
\begin{figure}[t]
	\centering
	\includegraphics[scale=0.7]{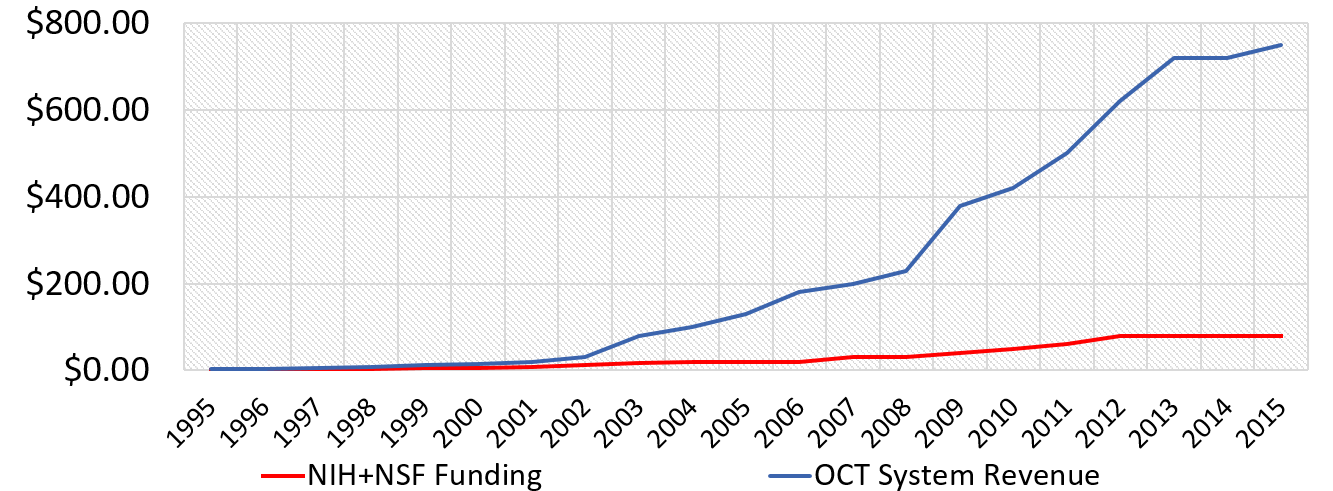}
	\caption{Estimated OCT annual system revenue (blue) in comparison to annual NIH and NSF OCT funding (red). The revenue includes the biometry market. Numbers are in million dollars. Reprinted from \cite{OCTEco}.}

	\label{MarketRevenue}
\end{figure}
\indent In the coming years, it is expected that the utilization of OCT devices in a variety of biosciences applications will be the main driver of the growth trend in the market. As an example, the application of such devices for dermatology is expected to rise in coming years. The main reason for such growth is attributed to the non-invasive, higher-quality, three-dimensional images that OCT devices can produce compared to the competitor applications such as ultrasound \cite{OCTMarket1}. Governmental support for an increased rate of medical diagnostic is another driver to the growth of the market. For instance, the food and drug administration (FDA) has launched multiple programs that aims to facilitate the routine screening of senior patients who constitute a great majority of the group that uses this medical test. Besides such facilitating programs, the government plays a key role in driving the growth by investing in this area. Data shows that the governmental investment in the field has been continuously increasing in recent years \cite{OCTEco}. The return on such investment has been outstanding from an economic standpoint, advancing science and clinical understanding, and from the perspective of improving patient care and quality of life. In fact, research shows (Figure  \ref{MarketRevenue}) that a worldwide governmental investment of around \$500M has resulted in a cumulative revenue of \$5.2B in the field of optical coherence tomography \cite{OCTGov}. Another driver of the market growth is the adoption of the technology in the Asia Pacific area. The presence of the device in countries such as Korea, China, India, Malaysia and other countries of South Asia is anticipated to be a vital force to grow the market to higher values. It can be expected that such technology adoptions in different regions of the world will be a competitive force that contributes to the growth of the market \cite{OCTMarket1}. 

\section{Innovation}
\indent The innovation related to the ideas proposed in this dissertation is based on significant improvements in accuracy and cost reduction of eye-related disease diagnosis by providing a high-quality, high-speed solution that can acquire tomographic images. Specifically, the findings of this dissertation could result in shorten clinical scan time, and as a consequence, increased image quality and resolution while reducing the number of required patient clinic visits. These improvements directly lower associated medical costs. In the course of our on the ground discovery, in the course of our participation in the national NSF I-Corps program, when we interviewed multiple ophthalmology clinics around the nation, these centers sometimes have to recall their patients for extra imaging due to the low quality of the original image acquired. The goal of the dissertation is to create a new-generation of OCT devices that are able to acquire images at much faster rates. Currently, the biggest challenge faced by OCT clinical applications is their relatively slow scan speed, which leads to low image resolutions and therefore potential unsatisfying clinical experience and also lack of confidence in the diagnosis process. A high-quality, high-speed processing tool is needed for reliable clinical decision-making and disease monitoring with OCT systems. This device should be able to perform all the necessary signal processing and data acquisition for OCT applications on the fly and as the tomographic data corresponding to real-time changes within the ocular medium is being produced. Besides the direct aim of this project to provide the ophthalmology community with such revolutionary product, the results presented in this dissertation can be used in other industrial applications to provide them with real-time solutions to acquire cross-sectional images in the application this product is intended to be used in.\\
\indent When OCT devices were first introduced as a concept \cite{OCTOriginal}, they were built such that the tomographic image information was embedded in the signal as a time encoded information. Different depths within the object to be imaged corresponded to a time span of the entire duration of an OCT output signal resulted from imaging that object. To acquire an image from different depths of an object, a reference optical component had to mechanically move. Each physical position of that reference component resulted in a tomographic image from a depth within the object. The mechanical movement of the reference components restricted the imaging speed that was possible to achieve with such devices. Furthermore, there was always the need to calibrate the position of that reference component so that the imaging could be possible as it had to correspond to the position of the object under test. The literature shows how limited the speed of TD-OCT devices is compared to later iterations of the OCT technology \cite{TDOCTBasic}. Such low imaging rates motivated the transition to more innovative versions of the technology. This new generation of the technology encodes all the tomographic information in the frequency domain; hence, removing the need for mechanical movements of TD-OCT devices. As the result of such innovation, the new generation of the technology (spectral domain OCT, SD-OCT) could acquire images at much higher rates \cite{SDOCTVsTDOCT}. Yet again, in the SD-OCT devices, there was the need to use slow spectrometers that acquired all information encoded in frequency all at once. The usage of such bulky elements limited the speed at which SD-OCTs could acquire images \cite{SDOCTVsTDOCT}. Swept Source Optical Coherence Tomography (SS-OCT) systems, as the latest iteration of the technology, does not suffer from the limitations that the previous generations of OCT devices have. In fact, with recent advancements in optical technology, there exists the chance to acquire images with these new devices at rates an order of magnitude higher than the currently widely adopted SD-OCT devices. The imaging acquisition speed in OCT devices has always been a force to move to more innovative OCT technologies. While the SS-OCT devices are the fastest yet reported, there can be more novel processing algorithms that enable real-time data acquisition and processing of SS-OCT tomographic images. The goal for OCT devices is to be able to capture images as changes are occurring within the object under test at any time.\\
\indent This dissertation reports a tested  group of signal processing algorithms and circuit topologies that would enable designers to acquire real-time, ultra-fast tomographic images using an all integrated specifically designed chip within SS-OCT devices. The chip enables full OCT manufacturers in the field of OCT design to transition to an order of magnitude higher image acquisition rate. While the current commercial image acquisition rate available in the market is restricted to a couple of hundreds of thousands of line scans per second, the techniques proposed in this dissertation will enable SS-OCT devices to acquire line scans in the MHz rate regime. This is a 10X improvement in imaging speed. The innovations reported in this dissertation could result in a component of SS-OCT devices that do the signal processing and acquisition process necessary for SS-OCT devices and is called an SS-OCT engine. 
\section{Background}
This section first will provide a general overview of different related tomography techniques that are used for material imaging and characterization. This section will then go over different categories of OCT that have already been introduced to this field of science and finally will thoroughly review an specific type of OCT devices named SS-OCT and examine the problems that these kinds of devices face.  
\subsection {Different Imaging Technologies}
OCT is an unfolding imaging technology that can be used in many clinical and academic applications. Similar to the ultrasonic imaging technology, OCT provides cross sectional images by performing transversal scanning of the tissue under test. At any particular time, the light source projects a beam at a specific point of the tissue under test. The back scattered light is then further processed to produce a cross-sectional tomographic image that maps the light reflectivity information of the tissue under test to a color code. This color code is what is called an A-Scan in the field. The light source is then scanned laterally and a cross sectional 2-Dimensional tomographic image is produced. Compared to ultrasound, OCT is capable of providing tomographic images with a resolution that is an order of magnitude higher.  The practical strength of this tomographic images, ultrasound and OCT, is that they can provide optical biopsy without actually cutting a tissue open \cite{Opticalbiopsy, Opticalbiopsy1}. There obviously exists numerous applications where invasive imaging applications are not preferred.\\      
\begin{figure}[!t]
	\centering
	\includegraphics[scale=0.5]{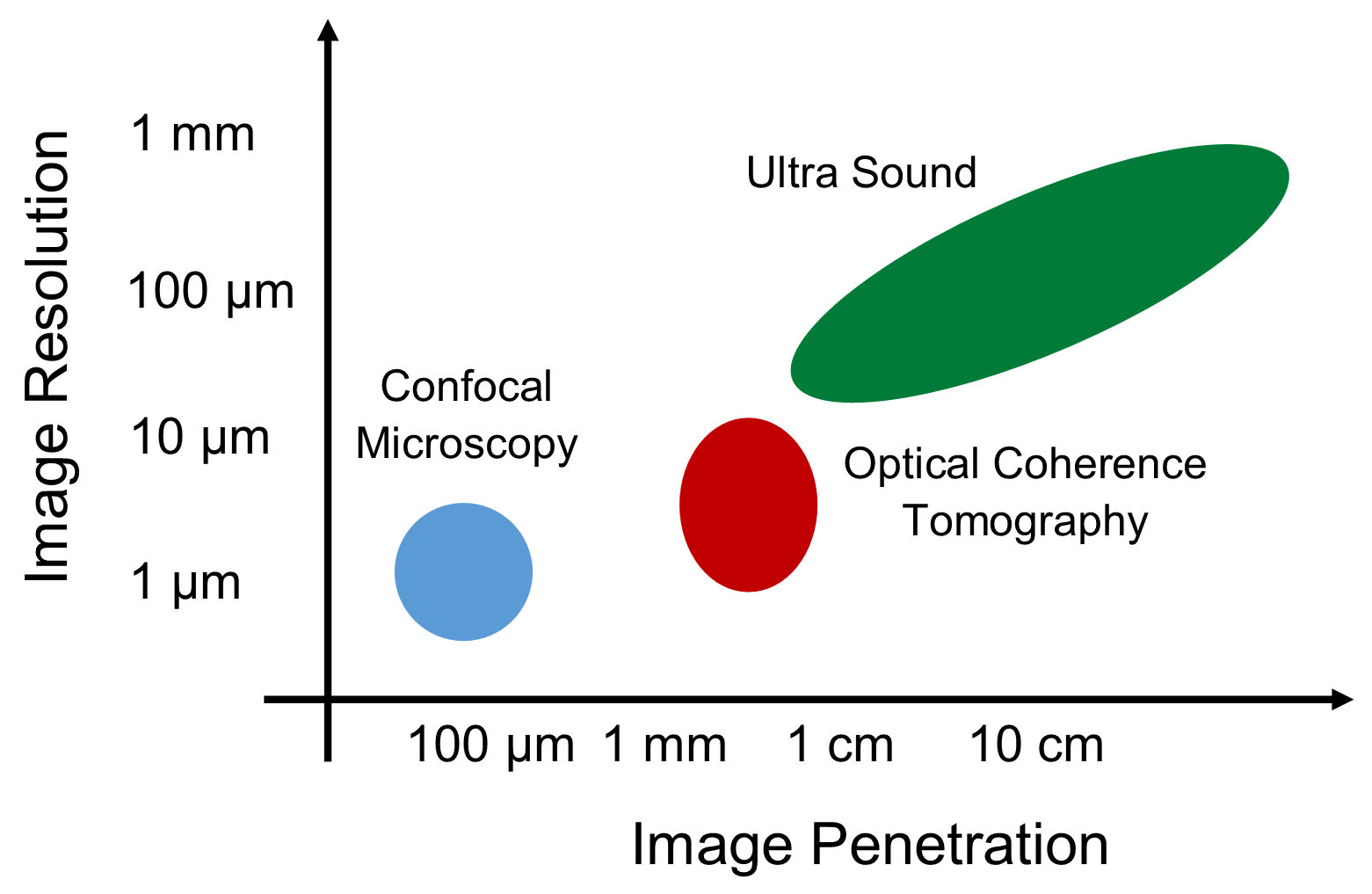}
	\caption{The Image Resolution Vs. the Imaging Penetration for three main tomographic technologies (confocal microscopy, optical coherence tomography and ultrasound). Higher frequencies of sound, yield finer resolutions in ultrasonic application yet limit the amount the sound can penetrate in tissue. Confocal Microscopy achieves sub micron (cell dimension) resolutions, nevertheless can acquire images from ultra thin layers of tissue. OCT act as an intermediate modality with imaging penetration and resolution between those of ultrasound and confocal microscopy. Reprinted from \cite{bookOCTIntroduction1}.}
	\label{OCTUltraMicro}
\end{figure}
\indent OCT has features that are analogous to other imaging technologies such as microscopy and ultrasound. Figure  \ref{OCTUltraMicro} shows a comparison of the imaging penetration and resolution between the three technologies. Depending on the frequency of sound used in an ultrasound device, the technology exhibits the widest range of imaging resolution among the three \cite{Ultrafrequency, Ultrafrequency1, Ultrafrequency2}. While higher sound frequencies yield better resolution in an ultrasonic imaging modality, it also results in a lower imaging depth as higher frequency sounds are more drastically absorbed by the tissue that is intended to be taken an image from.\\
\indent In the ultrasound technology, the echos of sound are recorded and the reflectivity of the tissue is calculated by measuring the amplitude and the delay of the sound that is being received by the device. These sonic signals are capable of penetrating deep in most kinds of tissue making it possible to take images from internal organs within the body.  Due to the potential of light being scattered when is directed on the tissue, unlike ultrasound, OCT penetration is limited to a couple of millimeters. However, the capability of using probes such as endoscopes, catheters, etc. makes it possible for OCT devices to acquire images from internal body organs as well.  \\
\indent Confocal Microscopy on the other hand can provide images of the tissue with resolutions in submicron levels. At such high resolution, it is possible to  observe cells. Although the specific type of light, with higher numerical aperture (NA), results in a higher resolution, it at the same time limits the depth that is possible to acquire images from. It also should be noted that in this imaging modality and unlike the prior two imaging modalities an image is acquired from a thin layer of tissue in an \textit{en face} manner.\\
\indent It is much easier in the ultrasound technology to measure the time delay of sound echoes reflected back from the tissue under test. Such phenomenon is due to the much lower propagation velocity of sound compared to that of light. While a typical ultrasound resolution (100 $\mu$m) results in an echo delay difference of around 100 ns, well within the detection resolution of electronic devices, it is impossible to measure the delay differences of the back scattered light from two particles within a tissue that are apart from one another by typical OCT resolutions. As such, directly measuring light echo delay differences is practically impossible and other techniques such as interferometry should be used to measure the properties of the tissue under test.
\subsection{Low Coherence Interferometry}
As stated above, measuring optical echos is practically impossible due to the ultra fast propagation velocity of light. Different devices such as the Kerr Shutter \cite{Kerrshutter} have been tried to be used to measure the light echo delays with an accuracy of femto seconds. One alternative method to measure these echos is to use non-linear optical processes such as harmonic generation or parametric conversion \cite {nonlinearoptical, nonlinearoptical1, nonlinearoptical2} where short pulses of light illuminate the tissue under test. Back scattered light then is mixed with a delayed version of the reference light projected on the tissue under test. Such technique enables designers to detect echo time delays with a femto second accuracy.\\
\indent Low coherence interferometry (white interferometry), first introduced by Isaac Newton, is another technique using which the echo time delay can be measured. Low coherence interferometry lays the frame work for what is currently recognized in this field of science as OCT and was first used in biological applications in the late 1980s \cite{firstuseofinterferometry, firstuseofinterferometry1}. \\
\indent Interferometry measures the auto correlation between a reference light that has traveled for a known length and the backscattered light that is reflected back from the tissue under test. The interferometer does so by coupling the two aforementioned sources of light via an optical coupler. Rather than the intensity of light, interferometers measure the field of light that is resulted from mixing the reference light and the one reflected back from the tissue under test.\\
\indent Assuming that the amplitude of the field of light reflected from the reference is denoted by $E_R$ and that of the reflected light from the tissue under test by $E_S$, the intensity of light ($I_D$) after the sample and reference lights are coupled can be formulated as below \cite{bookOCTIntroduction1},
\begin{equation} \label{Equ.1}
I_D \approx |E_R|^2 +|E_S|^2 +2E_RE_S cos(k\Delta L)
\end{equation}
where $k$ is the spectral frequency of the light projected on the tissue under test and $\Delta L$ is the length difference between the paths that the light travels from the reference and the sample to be merged on the coupler.  \\
\indent Coherence length of a light beam refers to the length that the light can travel and remain coherent by a specific amount. The more a light source is coherent, the more the coherence length becomes. Coherent electromagnetic waves are those in which the wave only includes a single spatial frequency and is all in phase. A low coherence light source is picked to specify the scan to a particular range determined by the coherence length of the light source. If coherent light sources with large coherence lengths are picked, the interference will be observed over a wide range of length difference between the reference and the sample path. It should be noted that the resolution of a low coherence tomography is determined by the coherence length of the light source used. The less the coherence length is, the finer the resolution becomes. The coherence length of a light source is the definitive function of the light source bandwidth source used. The more the bandwidth, the less the coherence length and the finer the interferometric resolution. Below, this dissertation first covers the basic theory of OCT devices and then will introduce the different realization of the OCT system, its development and different applications that it is used in. 
\subsection{Optical Coherence Tomography}
\begin{figure}[!t]
	\centering
	\includegraphics[scale=0.6]{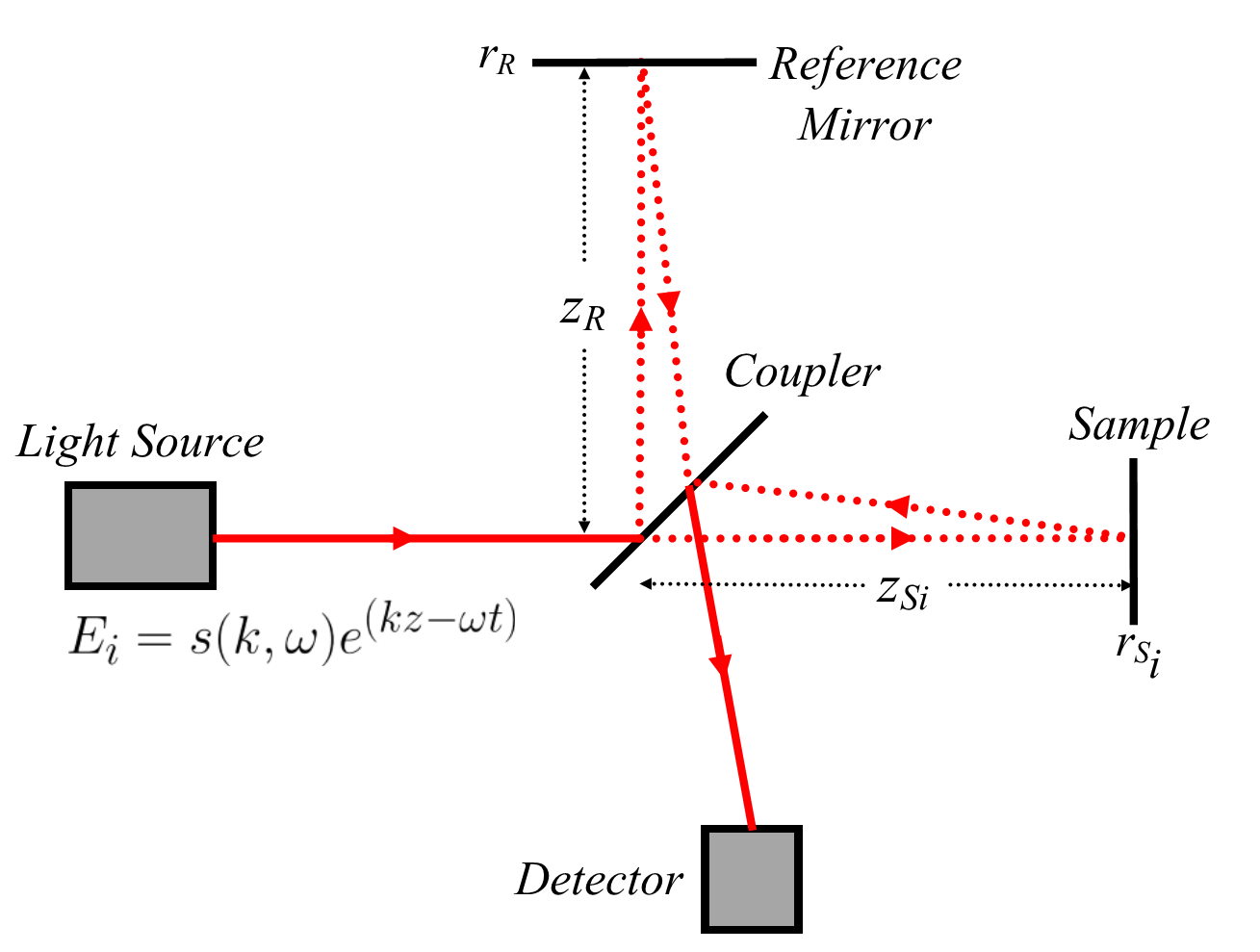}
	\caption{A simple Michelson interferometer. $E_i$ is the field of  the incident light, $E_s$ is the field of light reflected\textbackslash backscattered from the sample under test. $E_R$ is the field of light reflected back from a reference (typically a mirror).  $z_R$ is the distance that the reference mirror has from the coupler and $z_{Si}$ is the distance the sample have from the coupler. $r(z_{Si})$ is the reflectivity of light related to the sample located at position $z_{Si}$ and $r(z_{R})$ is the same of the reference mirror.}
	\label{MichelsonInter}
\end{figure}

A generic OCT system is consisted of a low coherent light source that is coupled into the interferometer (Figure \ref{MichelsonInter}). As Figure \ref{MichelsonInter} shows, OCT systems are almost universally realized by the use of a specific type of interferometer called Michelson. In a Michelson interferometer, the incident light coming our of the low coherent light source is projected on to a $2\times2$ coupler. The coupler is assumed to split the light into two beams, one traveling towards a reference, which typically is realized by a mirror, and one that travels toward the sample under test. The back reflected\textbackslash scattered light from the reference and the sample under test then will be merged on the coupler and are received by a photo detector. The intensity of light received by the detector will be further processed to produce a depth resolved light refelectivity map that is referred to as an A-scan. Computer controlled mechanisms focus the light source on a specific point of the sample under test. The beam is then laterally swept across the tissue under test to produce sequential A-scans from consequent points of the tissue under test. If those A-scan are stacked on top of one another, a cross sectional tomographic image is produced. The image produced by putting multiple A-scans beside each other is called a B-scan. To produce a volumetric tomographic image, multiple B-scans from different cross sections of the tissue under test are put besides each other. Multiple realizations of the technology exist, namely time domain OCT (TD-OCT), spectral domain OCT (SD-OCT), and swept source OCT (SS-OCT); they all share the same concept of interferometry as described above.
It also should be noted that the lateral resolution \footnote{Lateral resolution = $\Delta x$ = $0.37 \dfrac{\lambda_0}{NA}$, $\lambda_0$  is the center wavelength of the projected light source} \cite{bookOCTFormula} of the system is determined by the spot size of the beam projected on to the tissue under test and is a function of the numerical aperture (NA) of the sample arm used within the Michelson interferometer. It is possible to minimize the lateral resolution by increasing the numerical aperture. That however decreases the depth of focus (Field of view, FOV\footnote{$FOV=\dfrac{0.565\lambda}{sin^2\bigg[\dfrac{sin^{-1}(NA)}{2}\bigg]}$}) of the light beam projected on the sample under test. In these application where a very fine lateral resolution is required, the FOV also gets small to the point that it becomes comparable with the axial resolution of the systems which is unlike the FOV is determined by the physical properties of the light source. This mode of interferometry is what is referred to as optical coherence microscopy (OCM) \cite{OCM, OCM1, OCM2} where both confocal and coherent gating need to be done at the same time to more effectively reject the unwanted scattered light compared to when only confocal gating is used. OCM can achieve improved imaging depth and contrast compared with confocal microscopy that was introduced before. While spatial imaging constraints is the criteria that a confocal gate rejects light based on, a coherence gate rejects photons based on the path length they travel in tissue. $\mu$OCT is another OCT related technology that is used when very fine lateral resolution is desired. Such applications can include imaging tissues such as cochlea.
\subsubsection{Optical Coherence Tomography Formulation \cite{bookOCTFormula}}

\indent Consider the Michelson interferometer shown in 	Figure  \ref{MichelsonInter}. The electric field of the incident light being projected at the coupler can be formulated as, 
\begin{equation} \label{Equ.2}
E_i = s(k,\omega)e^{i(kz-\omega t)},
\end{equation}
where $s(k,\omega)$ is the electric field amplitude as a function of the wavenumber $k = 2\pi/\lambda$ and angular frequency $\omega = 2\pi\nu$, respectively. It is assumed that the power of the incident light is split in half by the coupler and is projected on the reference mirror and the sample under test. It can be assumed that the reference mirror has an electric reflectivity of $r(z_R)=r_R$ and a power reflectifity that equals $R_R = |r_R|^2$. The reference mirror is located at a position with the distance $z_R$ from the coupler. For the sample under test, it is assumed that it is consisted of discrete differential particles $1,2,...,n$ that each are located at positions $z_{S1}$, $z_{S2}$, ... , and $z_{Sn}$ with respect to the coupler. These discrete differential particles have electric field reflectivities equaling $r_{S1}$, $r_{S2}$, ... , $r_{Sn}$, respectively. With the differential particle assumption, the reflectivity of the sample under test can be written as $r_S(z_{S})=\sum_{n=1}^{\infty}r_{Sn}\delta(z_S-z_{Sn})$ where $\delta$ is the Dirac delta function.

Similar to the reference mirror, $R_{Sn} = |r_{Sn}|^2$. A tomographic image reconstruction then means that the power reflectivity function , $R_{Sn}$, should first be obtained and $\sqrt{R_{Sn}}$ then calculated. The electric field of reflected light from the the sample under test can be represented by,
\begin{equation} \label{Equ.3}
E_s = \frac{E_i}{\sqrt{2}}[r_S(z_{S})\circledast e^{i2kz_S}],
\end{equation}
where $\circledast$ represents convolution. The $\sqrt{2}$ stems from the fact that the coupler split the power of the incident light by a factor of two. With the discrete differential particle assumption, it can be obtained that, 
\begin{equation} \label{Equ.4}
E_s = \frac{E_i}{\sqrt{2}}\sum_{n=1}^{\infty}r_{Sn}e^{i2kz_{Sn}},
\end{equation}
and,
\begin{equation} \label{Equ.5}
E_R = \frac{E_i}{\sqrt{2}}r_{R}e^{i2kz_{R}}.
\end{equation}
These reflected lights are merged on the coupler and travel toward the photo detector that generates an electrical current that can be formulated as below.  
\begin{equation} \label{Equ.6}
I_D(k,\omega)= \big\langle \frac{\rho}{2}|E_R+E_S|^2 \big\rangle = \big\langle \frac{\rho}{2}(E_R+E_S)(E_R+E_S)^* \big\rangle,
\end{equation}
where $\rho$ is the responsivity of the detector and its dimension is $\big[\frac{Ampere}{Watt}\big]$. $\langle . \rangle $ denotes integration over the time response of the detector. At the coupler where $z=0$, Equation \ref{Equ.6} can be simplified as below,
\begin{equation} \label{Equ.7}
I_D(k,\omega)= \frac{\rho}{2}\bigg\langle\bigg|\frac{s(k,\omega)}{\sqrt{2}}r_{R}e^{i(2kz_{R}\omega t)}+\frac{s(k,\omega)}{\sqrt{2}}\sum_{n=1}^{\infty}r_{Sn}e^{i(2kz_{Sn}-\omega t)}\bigg|^2 \bigg\rangle,
\end{equation}
Expanding Equation \ref{Equ.7}, and taking the integral, it can be simplified as below. The assumption is that $\rho$ is so high that the result of integrating all terms that contain $e^{(i\omega t)}$ equals zero. 
\begin{equation} \label{Equ.8}
\begin{split}
I_D(k)& = \frac{\rho}{4}S(k)[R_R+R_{S1}+R_{S_2}+...+R_{Sn}] \\
&+ \frac{\rho}{4}S(k)\big[\sum_{n=1}^{\infty}\sqrt{R_{R}R_{Sn}}(e^{2ik(z_R-z_{Sn})}+e^{-2ik(z_R-z_{Sn})})\big] \\
&+ \frac{\rho}{4}S(k)\big[\sum_{n\ne m=1}^{\infty}\sqrt{R_{Sm}R_{Sn}}(e^{2ik(z_{Sm}-z_{Sn})}+e^{-2ik(z_{Sm}-z_{Sn})})\big],
\end{split}
\end{equation}
where $S(k) = \langle |s(k,\omega)|^2 \rangle$ which encodes the power spectral dependence of the light source. In most practical light sources, the power spectral dependence of light source can be modeled to have a Gaussian shape with a center spectral frequency of $k_0$ and a band width of $\Delta k$. The spectral bandwidth sets the value of the spatial frequency , $k_0 \pm \Delta k$, at which the light source power spectral dependence is $\dfrac{1}{e}$ times less compared to the same at $k_0$. For a Gaussian power spectral dependence $S(k)$ can be denoted as below,
\begin{equation} \label{Equ.9}
S(k) = \frac{1}{\Delta k\sqrt{\pi}}e^{-\big[\dfrac{k-k_0}{\Delta k}\big]^2}
\end{equation}
The inverse Fourier Transformation of Equation \ref{Equ.9} can be shown as below,
\begin{equation} \label{Equ.10}
\gamma(z) = \mathcal{F}^{-1}\{S(k)\}=  e^{-z^2\Delta k^2}.
\end{equation}
Later sections (\ref{FD_OCT Formulation}) of this dissertation show that the full width half maximum (FWHM) of the function $\gamma(z)$ is the coherence length of the light source and equals the axial resolution of an OCT system. Calculating the FWHM for the function $\gamma(z)$, the axial resolution , $l_c$, equals,  
\begin{equation} \label{Equ.11}
l_c = \frac{2\sqrt{ln 2}}{\Delta k} = \frac{2ln 2}{\pi}\frac{\lambda_0^2}{\Delta \lambda}.
\end{equation}

Equation \ref{Equ.8} can be simplified by using the Euler's equity,
\begin{equation} \label{Equ.12}
\begin{split}
I_D(k)& = \frac{\rho}{4}S(k)[R_R+R_{S1}+R_{S_2}+...+R_{Sn}] \\
&+ \frac{\rho}{2}S(k)\bigg[\sum_{n\ne m=1}^{\infty}\sqrt{R_{Sm}R_{Sn}}cos\big(k(z_{Sm}-z_{Sn})\big)\bigg]. \\
&+ \frac{\rho}{2}S(k)\bigg[\sum_{n=1}^{\infty}\sqrt{R_{R}R_{Sn}}cos\big(k(z_R-z_{Sn})\big)\bigg].
\end{split}
\end{equation}
It can be observed from the equation above, that the intensity of light detected at the photo detector is consisted of three different terms,
\begin{itemize}
    \item DC Term : The first line in Equation\ref{Equ.12} is independent of any path length in both the sample and reference arms. This term results in a DC offset in the current value at the output of the photo detector and can simply be rejected by electronic means available.
    \item Cross Correlation Term : The second line in Equation \ref{Equ.12} shows the result of the interference between the light that is reflected back from the reference mirror and all discrete differential particles withing the sample under test. As stated before, calculating the function $\sqrt{R_{Sn}}$ yields the tomographic image.
    \item Auto Correlation Term : The third line in Equation \ref{Equ.12} shows the result of the interference between the light that reflects back from different differential particles of the tissue under test. Compared to the second line, the term in the third line entails the term $R_{Sm}$ which is much less than the similar term $R_R$ in the cross correlation term, and hence can be ignored.
\end{itemize}

The discussion above yields that with a proper selection of power reflectivity for the reference mirror \cite{referencemirrorreflectivity, referencemirrorreflectivity1}, Equation \ref{Equ.12} can be simplified to the equation below,
\begin{equation} \label{Equ.13}
I_D(k) = \frac{\rho}{2}S(k)\bigg[\sum_{n=1}^{\infty}\sqrt{R_{R}R_{Sn}}cos\big(k(z_R-z_{Sn})\big)\bigg].
\end{equation}

Acquiring tomographic images then can be boiled down to finding methods that can demodulate the value $R_{Sn}$ out from Equation \ref{Equ.13}

\section{OCT When It Was First Introduced}

When OCT was first introduced \cite{OCTOriginal} it utilized a low coherence light source that was coupled to a Michelson interferometer. First clinical prototypes of the idea were developed at MIT Lincoln Laboratories and were later used in the New England Eye Center to perform the first tests in ophthalmology clinics \cite{OCTClinic,OCTClinic1,OCTClinic2,OCTClinic3}. OCT was and has been extensively used in ophthalmic applications to diagnose and monitor the prognosis of diseases such as glaucoma and macular degeneration where the anatomical structure of the retina need to be examined. Other signal processing algorithms enhance the diagnose ability of ophthalmologists by doing operations such as retinal thickness measurement or segmentation \cite{OCTsegmentation}.\\
\indent One other application of the OCT technology became possible by the utilization of catheters and endoscopes \cite{catheter,catheter2}. These probes enabled internal imaging of anatomical shapes of vessels as well as the anatomical shape of internal organs; hence, the name Anatomical OCT (A-OCT). Endoscopic probes became specially popular because of their utilization in gastrointestinal imaging \cite{gastrointens, gastrointens1} where cancer could be diagnosed. These images require relatively high quality as the cancer diagnosis needed to be done blindly without the use of histology. \\
\indent Another usage of endoscopic subsurface OCT, was introduced when it became relevant to observe subsurface morphological changes \cite{lungA-OCT,lungA-OCT1,lungA-OCT2,lungA-OCT3,lungA-OCT4,lungA-OCT5} or imaging the thickness of the airway wall \cite{lungA-OCT6}.
Extremely high resolution and high scanning ranges are required in this application. However, the characterizing airway lumen size and shape does require a significantly longer scanning range than is typical with current subsurface A-OCT. Future sections of this dissertation show how the ideas proposed in this dissertation help achieve such high imaging range specification in A-OCT devices \cite{A-OCTrelationtoSSOCT}. These early iterations of the OCT technology encoded the interferometric data in time, hence the name time domain OCT. The section below formulates the first iteration of OCT devices, Time domain OCT.

\section{Time Domain OCT, TD-OCT}
\begin{figure}[!t]
	\centering
	\includegraphics[scale=0.55]{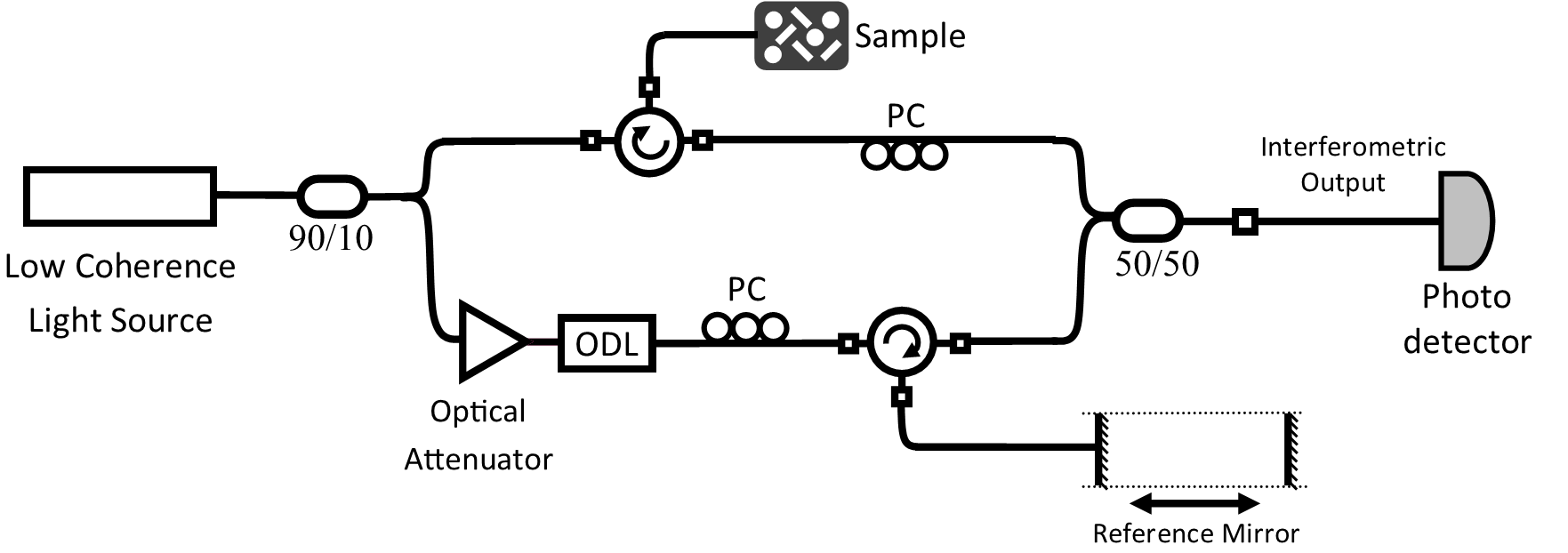}
	\caption{Early realization of OCT systems. A TD-OCT realization of an OCT system is shown. The low coherence light is split via a 90/10 coupler. Circulators are used to direct light to different parts of the OCT system. (PC) Polarization controller. (ODL) Optical delay line.}
	\label{TDOCT}
\end{figure}

Early iterations of OCT devices work by scanning the reference mirror in length and receiving the output of the Michelson interferometer as Figure  \ref{TDOCT} shows. At different time instances, the value of $z_R$ in Equation  \ref{Equ.12} changes as the reference mirror moves. In this iteration of the technology, the interferometric light intensity is encoded in time; hence the name TD-OCT. 
In this iteration of the technology, the interferometric output is received by a wavenumber dependant photo detector. The interferometric output then can be formulated as,
\begin{equation} \label{Equ.14}
\begin{split}
I_D(z_R)& = \frac{\rho}{4}S_0[R_R+R_{S1}+R_{S_2}+...+R_{Sn}] \\
&+ \frac{\rho}{2}S_0\bigg[\sum_{n=1}^{\infty}\sqrt{R_{R}R_{Sn}}e^{-(z_R-z_{Sn})^2(\Delta k)^2}cos\big(2k_0(z_R-z_{Sn})\big)\bigg],
\end{split},
\end{equation}
by integrating Equation \ref{Equ.12} over k. In the above equqtion, $S_0=\int_{0}^{\infty} S(k)dk$. \\
\indent The equation above is consisted of a DC term and a dynamic one that contains the light reflectivity information of the object under test. Assuming that the object under test is only consisted of two particles \{1,2\} that are located far from one another and are located at distances $\{z_{S1}, z_{S2}\}$, scanning $z_R$ will results in two amplitude modulated \textit{cosin} fringes of light intensity at the detector (modulated by $e^{-(z_R-z_{Sn})^2(\Delta k)^2}$) when $z_R$ approaches either $z_{S1}$ or $z_{S2}$. By measuring the amplitude of these fringes, the values $\sqrt{R_{R}R_{S1}}$, and  $\sqrt{R_{R}R_{S2}}$ and as a result $\sqrt{R_{S1}}$, and $\sqrt{R_{S2}}$, are determined respectively.\\
\indent The most important specifications in an OCT system can be listed as the axial resolution, lateral resolution, image acquisition speed, and the imaging range. All these parameters have been constantly improving since the technology was first introduced. Following sections of this chapter explain the underlying physical background that determine the value of these specifications and explain how these improvements resulted in the introduction of newer iterations of the technology. 
\section{OCT Image Axial Resolution, Lateral Resolution, and Depth of Field}

The resolution of a tomographic image acquired by an OCT system is a critical specification that should be taken into account when utilizing these devices. Unlike microscopy, OCT can achieve fine axial resolution independent of the light beam focusing and the spot size. This independence, gives another degree of freedom when designing OCT devices. The axial resolution of an OCT device equals the width of the autocorrelation of the electric field of the incident light \cite{OCTAxialRes}. This quantity being denoted by $l_c$ equals the coherence length of the light source and for one with Gaussian wave-number spectrum can be shown to be inversely proportional to the wavenumber bandwidth of the light source. The Wiener-Khinchin theorem \cite{wiener_1964} states the inverse relationship between the autocorrelation of a random process and its inverse Fourier transformation; hence the relationship in Equation \ref{Equ.11} for the axial resolution. As seen in this equation, the wider the spectral bandwidth of the light source, the finer the axial resolution becomes. \\
\begin{figure}[!t]
	\centering
	\includegraphics[scale=0.40]{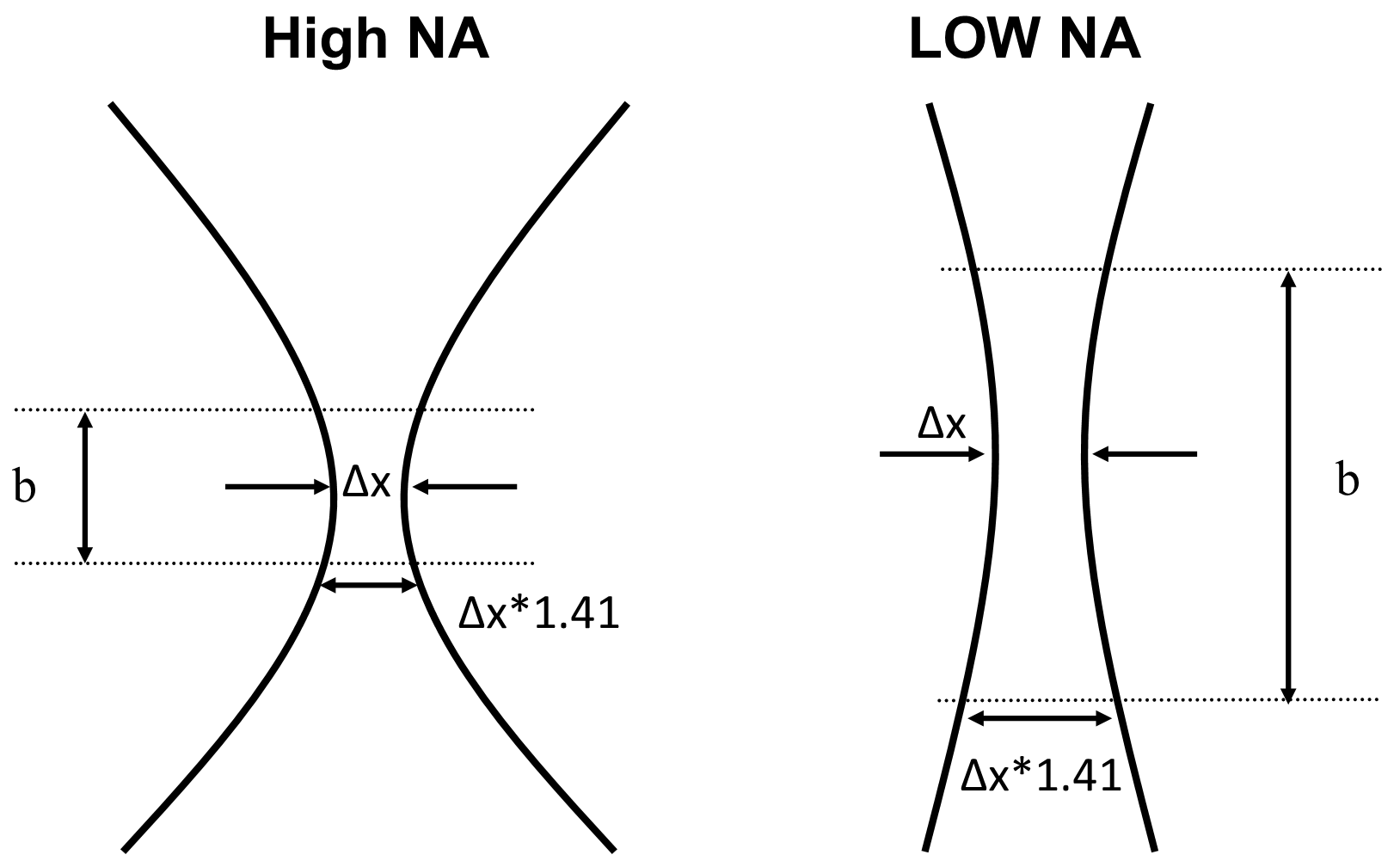}
	\caption{The comparison between two different light beams projected on the sample under test. (Left) A scenario where the sample arm has a relatively high numerical aperture (NA). (Right) A scenario where the sample arm has a relatively low numerical aperture (NA).}
	\label{NA}
\end{figure}
As stated above, the lateral resolution of an OCT device is similar to that of microscopy systems and is determined by the minimum spot size of the focused beam. It can be shown that the lateral resolution is inversely proportional to the numerical aperture of the beam. The lateral resolution is,
\begin{equation} \label{Equ.15}
\Delta x = 0.37 \frac{\lambda_0}{NA},
\end{equation}
where $\lambda_0$ is the center wavelength of the beam projected onto the sample under test. Finer lateral resolution can be achieved by increasing the numerical aperture as shown by Figure  \ref{NA}. As observed by Figure  \ref{NA}, increasing the NA decreases the depth of field at the same time. It can be shown that the depth of field can be formulated as, 
\begin{equation} \label{Equ.16}
FOV=\frac{0.565\lambda}{sin^2\bigg[\frac{sin^{-1}(NA)}{2}\bigg]}.
\end{equation}
Improving the lateral resolution is possible by increasing the numerical aperture. As stated before, for extremely high lateral resolutions where $b<l_c$, other techniques such as optical coherence microscopy of full field OCT are more desired to be used. In these techniques, it is more desired to acquire \textit{en face}\footnote{en face imaging acquires images from the surface of an object under test rather than the cross section of the object under test} images rather than cross sectional ones. The implications that improving the lateral resolution of the OCT system exhibits persuaded designers to improve the axial resolution of these systems at the early stages.
\section{Different Incident Light Sources and the Axial Resolution}
As stated in previous sections of this dissertation, the axial resolution of an OCT system is the direct function of the spectral bandwidth of the incident light that is projected on to the subject under test, hence the importance of using proper sources to achieve desired resolutions.\\
\indent Due to their compactness and inexpensiveness, superluminescent diodes are what typically used in OCT devices. Early OCT devices used GaAs light sources that projected infrared light with a central wavelength at around 800 nm and a spectral bandwidth of around 30 nm which yields an axial resolution of around 10 $\mu$m. Similar axial resolution can be achieved for light scattering biological tissues via more advanced SLEDs that are centered at 1300 nm and exhibit wider spectral bandwidth. In recent years, wider spectral bandwidths, upwards of 140 nm, have been reported in the literature \cite{SLEDWide,SLEDWide1} that yield axial resolutions in the 3-5 $\mu$m range. While SLEDs that project light centered at 800 nm can provide finer axial resolution due to their smaller central wavelength, SLEDs that are operating at center wavelengths around 1300 nm show better performances for light scattering tissues; SLEDs operating at around 1000 nm show the benefits of the two extremes where the penetration length as well as the axial resolution are both optimized.\\
\indent Femtosecond lasers are another type of light sources that are mainly used in research applications where ultra high resolutions are desired. These light sources can provide extremely wide spectral bandwidth at around infrared region central wavelength. These types of lasers can provide the wide bandwidth at center wavelengths at 800 nm, 1000 nm, 1300 nm. These types of lasers are capable of achieving axial resolutions less than 1 $\mu$m \cite{femtolaser}.
\section{Image Acquisition Speed Enhancement and Fourier Domain OCT}
For TDOCT systems to work, the position of the reference mirror is scanned to acquire interferometric information from different axial depths of the tissue under test. The mechanical movements limit the speed at which acquiring tomographic images is possible. Advances in the OCT technology led to the introduction of newer iterations of these devices, namely spectral domain OCT (SD-OCT) and swept source OCT (SS-OCT). These newly expressed iterations are also known as optical frequency domain interferometry (OFDI) \cite{OFDI,OFDI1,OFDI2}. Unlike TD-OCT, these frequency domain OCT (FD-OCT) techniques perform the image acquisition in the frequency domain and do not require any mechanical movement that is present in devices that encode the data in the time domain. In addition to FD-OCT systems being capable of acquiring images at much faster rates, they are shown to have much better sensitivity compared to their TD-OCT counterparts \cite{OFDI6,Sensitivity}.  \\
In SD-OCT systems, a broadband low coherence light sources is used where the interferometric patterns are measured via line scan cameras and a spectrometers \cite{OFDI,OFDI3,OFDI4}. In SS-OCT systems, however, a narrowband light source is used where its wavenumber is swept in a wide spectrum of frequencies \cite{OFDI1,OFDI5,OFDI6}. 
\subsection{Spectral Domain OCT, SD-OCT}
\begin{figure}[!t]
	\centering
	\includegraphics[scale=0.52]{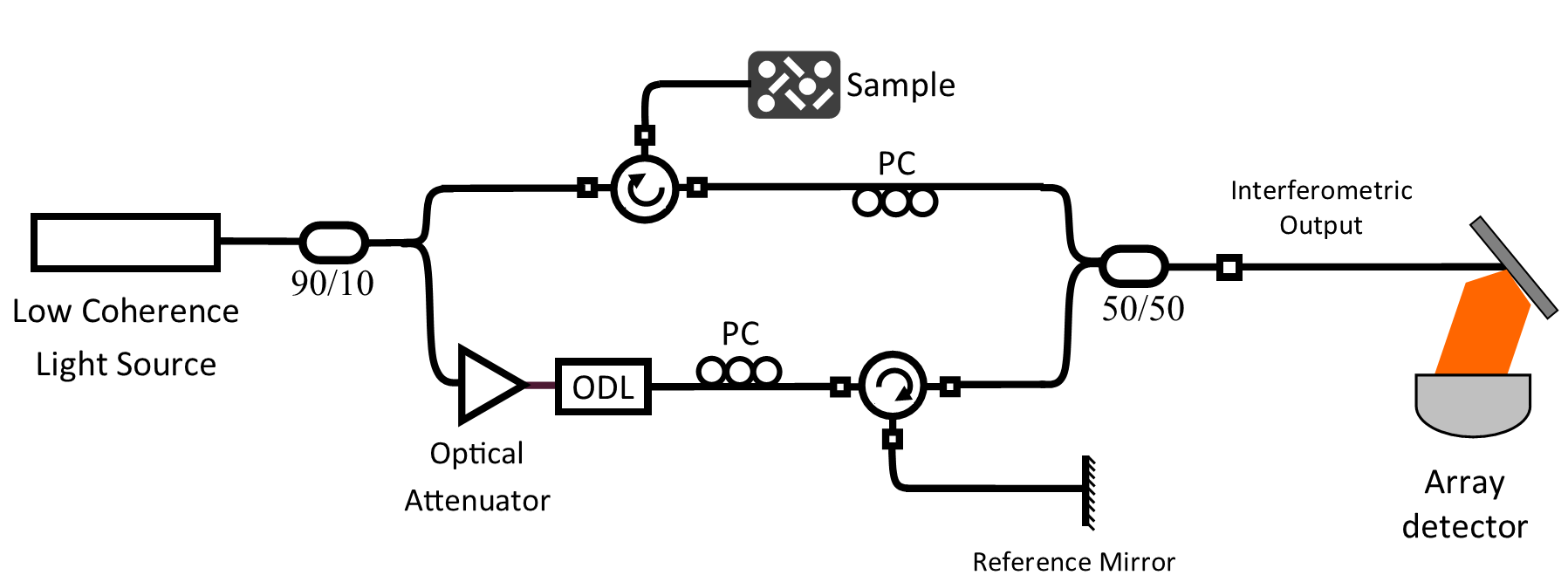}
	\caption{A generic realization of SD-OCT systems. The low coherence light is split via a 90/10 coupler. Circulators are used to direct light to different parts of the OCT system. An array detector (spectrometer) is used to detect the interferometric pattern. (PC) Polarization controller. (ODL) Optical delay line.}
	\label{SDOCT}
\end{figure}

In SD-OCT systems, a broadband light source is used to detect the interferometric patterns of the backscattered light in the frequency domain. As Figure \ref{SDOCT} shows, a Michelson Interferometer is used similar to the TDOCT version where the incident light is split in two paths. One is directed towards a fixed position reference mirror and the second is directed to the tissue under test. The backscattered light from the tissue under test is then re-coupled with the light reflected back from the reference mirror to form an interferometric pattern. The interferometric output then is received by an array detector (spectrometer). Later sections of this dissertation show how the necessary signal processing algorithms are carried out on the received signal to produce a tomographic image.\\
\indent Early realization of the SD-OCT technology exhibited axial resolution in the range of less than 10 $\mu$m axial resolution and an image acquisition of a couple of tens of thousands of A-scan per second \cite{OFDI1}. Ever since the first realization \cite{OFDI1}, the image acquisition speed has continuously improved to hundreds of thousands of A-scans per second. Such fast imaging speed, enables the acquisition of 3D volumetric images by stacking up two dimensional cross-sectional images. Acquiring volumetric images, enables the possibility of rendering \textit{en face} images as opposed to the usual cross sectional images, opening up a new realm of diagnosis especially in the most widely used application of the technology, namely ophthalmology. It should be noted that increasing the capacity of an OCT device to acquire more A-scans per second, enables designers to reduce the lateral resolution of the system and cover the same lateral length at the same time with an increased quality. Section \ref{FD_OCT Formulation} discusses how the back scattered light from the tissue under test entails the interferometric information in the frequency domain and how the information from different depths of the sample are coded to different frequencies of the interferometric output signal.
\subsection{Swept Source OCT, SS-OCT}
\begin{figure}[!t]
	\centering
	\includegraphics[scale=0.52]{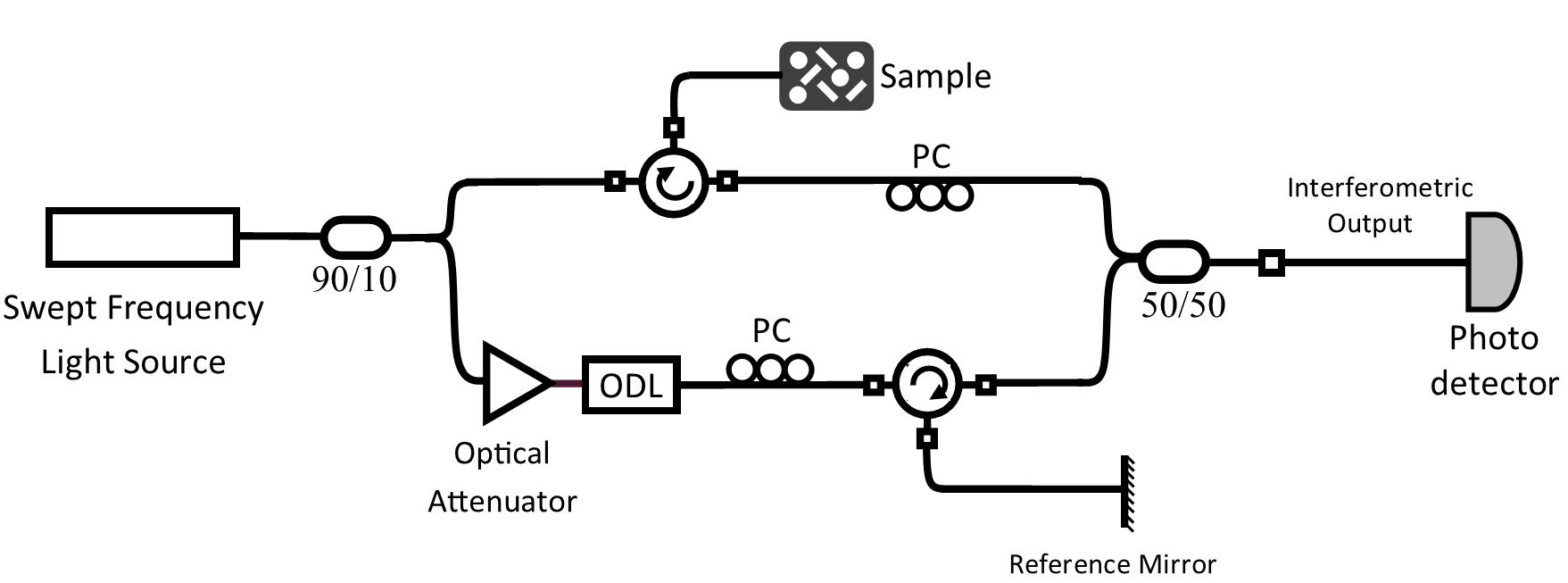}
	\caption{A generic realization of SS-OCT systems. The low coherence light is split via a 90/10 coupler. Circulators are used to direct light to different parts of the OCT system. The reference mirror is used and is not moving. A Swept source light source is used instead. (PC) Polarization controller. (ODL) Optical delay line.}
	\label{SSOCT}
\end{figure}
Unlike SD-OCT systems, SS-OCT systems rely on a narrowband light source that sweep the spectral frequency of the incident light over a wide range of values. Similar to SD-OCT systems, the incident light is projected to a Michelson interferometer (Figure  \ref{SSOCT}) in a way that the back scattered light from the subject under test is coupled with the light from the reference mirror and produce the interferometric pattern at the output. A photo detector is then used to record the output signal. These detectors usually are materialized via a specific type, namely InGaAS, and convert the intensity of the interferometric output to electrical current. Unlike the line scan cameras that only work at lower wavelength and are used in SD-OCT devices, the semiconductor based photo detectors that are used in SS-OCT devices can work at the 1000 $\mu$m and 1300 $\mu$m wavelength regimes and hence can acquire images from more light scattering types of tissue. The fact that SS-OCT devices dont need the mechanical movements of the reference mirror that happens in TD-OCT devices and the fact that they dont utilize the usually slow spectrometers used in SD-OCT devices, play a significant role in the superior imaging speed SS-OCT devices exhibit compared to the two previous iterations of the technology. Similar to SD-OCT, SS-OCT devices encode the interferometric information from different axial depths of the tissue under test at different frequencies. Due to the superior speed that SS-OCT devices exhibit compared to SD-OCT devices, two of the major players in the field, Carl Zeiss Meditec, and Topcon Corporation introduced their first ever SS-OCT products that were approved by the Food and Drug Administration (FDA) in the recent years. \\
\indent The introduction of ultra-fast swept source modalities such as Fourier domain mode lock (FDML) lasers, pushes the image acquisition speed in SS-OCT systems to a new realm of possibility. This superior imaging speed yields the possibility of acquiring 3D volumetric images that can be of great importance in different applications of OCT devices, namely elastography, doppler OCT, A-OCT, etc.,making the SS-OCT devices the trending iteration of the technology in the coming years. \\
\indent The purpose of this dissertation is to provide signal acquisition and processing algorithms that enable the SS-OCT devices to work at speed rates much higher than those that are currently available in the market.

\subsection{FD-OCT Formulation \cite{bookOCTFormula}} \label{FD_OCT Formulation}
In FD-OCT systems, the wavenumber dependant intensity of the interferometric output , formulated by Equation \ref{Equ.12}, needs to be recorded. In SD-OCT systems, the broadband interferometric output is captured by a wideband spectrometer and then is detected by an array of detectors each being set to record a certain component of the wavelength. In SS-OCT devices, however, the wavelength is swept in time and a semiconductor based (InGaAs) photo detector records the intensity of the interferometric output signal as a function of time. The goal is to develop a way to extract the sample reflectivity function $\sqrt{R_{Sn}}$. Using the inverse Fourier transformation properties below, 
\begin{align*}
x(k)\circledast y(k) & \xrightarrow{\mathcal{F}} X(z)Y(z) \\
cos(k(z-z_0))  \xrightarrow{\mathcal{F}} \frac{1}{2}\big[\delta(&z-z_0)+\delta(z-z_0)\big].
\end{align*}
The transformation is applied on Equation  \ref{Equ.12} and the formula below is obtained,
\begin{equation} \label{Equ.17}
\begin{split}
i_D(z)=\mathcal{F}^{-1}\{I_D(k)\}& = \frac{\rho}{4}\gamma(z)[R_R+R_{S1}+R_{S_2}+...+R_{Sn}] \\
&+ \frac{\rho}{2}\gamma(z)\bigg[\sum_{n\ne m=1}^{\infty}\sqrt{R_{Sm}R_{Sn}}\bigg(\delta\big(z\pm2(z_{Sm}-z_{Sn})\big)\bigg)\bigg] \\
&+ \frac{\rho}{2}\gamma(z)\bigg[\sum_{n=1}^{\infty}\sqrt{R_{R}R_{Sn}}\bigg(\delta\big(z\pm2(z_R-z_{Sn})\big)\bigg)\bigg].
\end{split}
\end{equation}
Simplifying Equation  \ref{Equ.17},
\begin{equation} \label{Equ.18}
\begin{split}
i_D(z)& = \frac{\rho}{4}\gamma(z)[R_R+R_{S1}+R_{S_2}+...+R_{Sn}] \\
&+ \frac{\rho}{2}\bigg[\sum_{n\ne m=1}^{\infty}\sqrt{R_{Sm}R_{Sn}}\bigg(\gamma(2(z_{Sm}-z_{Sn}))+\gamma\big(-2(z_{Sm}-z_{Sn})\big)\bigg)\bigg] \\
&+ \frac{\rho}{2}\bigg[\sum_{n=1}^{\infty}\sqrt{R_{R}R_{Sn}}\bigg(\gamma(2(z_{R}-z_{Sn}))+\gamma\big(-2(z_{R}-z_{Sn})\big)\bigg)\bigg].
\end{split}
\end{equation}

It can be assumed that $\gamma(z)$ has a Gaussian shape. Equation  \ref{Equ.17} suggests that for an artificial subject under test which is consisted of two differential particles that are located at $z_{S1}$ and $z_{S2}$, the A-scan $i_D(z)$ will consist of a Gaussian shaped spectrum $\gamma(z)$ located at DC that is caused by the first line in Equation  \ref{Equ.18}. For such subject under test, the A-scan also is consisted of the shifted version of the light spectrum located at $2(z_{S1}-z_{S2})$ and $-2(z_{S1}-z_{S2})$ that are caused by the second line in Equation  \ref{Equ.18}. The A-scan also is consisted of terms located by shifting the light spectrum to $2(z_{S1}-z_{R})$, $2(z_{S2}-z_{R})$, $-2(z_{S1}-z_{R})$, and $-2(z_{S2}-z_{R})$.\\
\indent The Gaussian shape of the function $\gamma(z)$ can blur out the quality of the tomographic image acquired as the tails of the Gaussian function can be mixed. The solution to solve that is to increase the bandwidth of the light , $\Delta k$, so that the function $\gamma(z)$ becomes narrower and the function $\sqrt{R_{Sn}}$ becomes more distinguishable . That is exactly the definition of axial resolution.\\
There are a few things that need to be taken into account when dealing with an A-scan.\\
\indent First and as Equation  \ref{Equ.18} suggests, there exists a term at DC with a rather significant gain , $\big[R_R+R_{S1}+R_{S2}+...+R_{Sn}\big]$. Due to the Gaussian nature of the term $\gamma(z)$, and the significant gain it experiences at DC, the tail of that Gaussian function might very well saturate all terms that are located at the positions determined by the cross correlation terms effectively making the recovery of $\sqrt{R_{Sn}}$ impossible. The DC gain term is dominated by the reflectivity of the reference mirror and can be estimated as below,
$$\big[R_R+R_{S1}+R_{S2}+...+R_{Sn}\big]\approx R_R.$$ Due to such approximation, and to eliminate the blurring effects of the tail of the Gaussian shape function, $\gamma(z)$, designers usually remove the sample under test and reconstruct the interferometric test from a mirror alone. That way, by negating the later interferometric pattern from the one that is resulted from the subject under test, the DC terms and their deteriorating effects could be minimized.\\
\indent The second point to be taken into account is the fact that the light spectrum function $\gamma(z)$ will be located at symmetric locations at $2(z_{Sn}-z_{R})$, and $-2(z_{Sn}-z_{R}).$ This phenomenon is understandable due to the real intensity of the light detected and the properties of the inverse Fourier transform in Equation \ref{Equ.17} \\
\indent Lastly and due to the fact that most auto correlation terms are located at around zero ($z_{Sn}\approx z_{Sm}$), the location of the reference mirror, $z_R$, should be picked with extra care so that the auto correlation terms are well away from the cross correlation ones.  The section below discusses the effect sampling speed has on the sensitivity of a Fourier Domain OCT system.
\section{Sensitivity Roll-off and Sampling Speed in FDOCT Systems}
The previous section of this dissertation explained the advantages of FD-OCT systems and also provided an overview of how the signal is captured and formulated and how the inverse Fourier transformation is applied to acquire the tomographic image. In reality, however, the physical components producing, and recording the interferometric data have limitations that should be taken into account \cite{bookOCTFormula}.\\
\indent First, lets us consider an SD-OCT system. In a SD-OCT system, the spectrometer as well as the line array camera used to capture the interferometric pattern have a certain spectral frequency resolution.\\
\begin{table}[t]
	\centering
	\caption{Effect of sampling time limit as well as the spectral resolution of the spectrometer in SD-OCTs and the resolution of the swept source in SS-OCT systems}	
	\label{Band}
    \begin{tabular}{*3c}   \toprule
    & Maximum one-sided & 6dB SNR  \\
    & 6dB SNR roll off & falloff point \\\midrule
    Wavenumber Unit & $\dfrac{\pi}{2\delta_{s}k}$ & $\dfrac{2ln(2)}{\delta_{r}k}$\\ \\
    Wavelength Unit & $\dfrac{\lambda_0^2}{4\delta_{s}\lambda}$ & $\dfrac{ln(2)}{\pi}\dfrac{\lambda_0^2}{\delta_r \lambda}$\\\bottomrule
    \hline
    \end{tabular}

\end{table}
Let us assume that the spectral resolution of the spectrometer is denoted as $\delta_r k$ which is the minimum spectral frequency distance there could be between two pixels of the spectrometer. In SS-OCT systems $\delta_r k$ denotes the minimum distance between two spectral frequencies of the sweep profile of the swept source.\\
\indent To model such phenomenon, the intensity of light that is detected by the spectrometer in the SD-OCT case and by the photo detector in the SS-OCT case is convoluted by a Gaussian function the full width half maximum (FWHM) of which equals $\delta_r k$. The new intensity of light then can be denoted by $\hat{I}_D(k)$ and equals,
\begin{equation} \label{Equ.19}
    \hat{I}_D(k) = I_D(k) \circledast e^{-\dfrac{4ln(2)k^2}{\delta_r k}},
\end{equation}
where $I_D(k)$ is what denoted by Equation  \ref{Equ.12}.\\
\indent The A-scan is then constituted by applying inverse Fourier transformation on $\hat{I}_D(k)$,
\begin{equation} \label{Equ.20}
    \hat{i}_D(z) = \mathcal{F}^{-1}\{\hat{I}_D(k)\}= i_{D}(z)\times e^{-\dfrac{\delta_r k^2 z^2}{4ln(2)}}.
\end{equation}
The exponential function shown by Equation  \ref{Equ.20} indicates a roll-off on the amplitude of the A-scan as the depth in the same equation increases. As such, the visibility of the system for particles of the tissue under test that are placed at deeper location falls. This is what is considered as the sensitivity roll-off and indicates the sample depth at which the intensity of light gets halved. It can be shown that the depth at which the intensity becomes half can be formulated as below,
\begin{equation} \label{Equ.21}
    \hat{z}_{6dB}= \frac{2ln(2)}{\delta_r k}= \frac{ln(2)}{\pi}\frac{\lambda_0^2}{\delta_r \lambda}
\end{equation}

Besides the resolution of the of spectrometer in SD-OCT systems and the sweep resolution of a swept source in SS-OCT systems, there exists another limitation that sets the maximum depth from which interferometric patterns are recognizable. As stated before, the interferometric pattern that is captured via the optical to electrical components need to be sampled and taken an inverse Fourier transformation upon. Similar to all discrete signals, the highest depth that can be distinguishable in an A-scan that is the result of applying an inverse Fourier transformation on the intensity of light received at the output of the OCT system can be shown to be a function of how fast the interferometric signal is being sampled.\\
\indent Let us assume that the entire spectral bandwidth of the incident light source is denoted by $\Delta k$ and each A-scan is sampled into $M$ equidistant points. In the case of a SD-OCT system, $M$ denotes the number of channels within the spectrometer and in an SS-OCT systems $M$ denotes the number of samples per A-scan. It can be trivially found that $\Delta K = M\delta_s k$. With the explanation above, the sampling intervals in the $z$ domain are denoted by $z_{min}=\dfrac{2\pi}{2\Delta k}$ and the maximum depth that can be sampled will be denoted by,
\begin{equation} \label{Equ.22}
    \pm z_{max}= \pm \frac{\pi}{2\times\delta_sk}=\pm \frac{\lambda_0^2}{4\delta_s \lambda}.
\end{equation}
\indent It can be inferred that for a fixed light source spectral bandwidth, the more $M$  becomes the higher the maximum imaging depth\textbackslash range gets. Table 1.1 summarizes the formulas for the maximum one-sided 6 dB SNR roll off as well as the 6dB SNR falloff point with the wavenumber and wavelength units. \\
\indent In this chapter, the OCT technology is explained, and formulated. The important specifications to explain an OCT system have also been explained. The next chapter will explain the advantages of the SS-OCT as the trending iteration of the technology and will explain the problems that this dissertation intends to solve.

%
%
%
%


\chapter{\uppercase{Problem Statement and the State of the Art}}
In Chapter. \ref{Chapter1}, the basic concepts of optical coherence tomography were introduced. Properties of OCT systems were also explained. The trend to improve the properties of OCT systems resulted in a chronological introduction of more novel iterations of the technology, namely TD-OCT, SD-OCT, and SS-OCT, respectively. As explained in the previous chapter, TD-OCT systems where first to be introduced. These devices mainly has been put out of use in industrial and clinical applications due to their insufficient acquisition speed. The slow speed is mainly attributed to the mechanical movements that are necessary to acquire tomographic images via these devices. \\
\indent FD-OCT systems on the other hand acquire the light echos from different depths of the sample under test all at once and do not require those mechanical movements necessary in TD-OCT systems; hence a better imaging acquisition speed and sensitivity. \\
\indent Imaging speed is the specification that drives OCT systems to newer realms of innovation and can be defined as the number of A-scan that can be acquired from the subject under test in one second. It is known that SS-OCT devices can achieve higher speeds compared to SD-OCT systems \cite{topcon} which in turn results in finer resolutions. The finer resolution relates to the fact that the higher imaging speed in SS-OCT systems can mitigate the deteriorating effects patients' eye movement has on the resolution of the system in the most common use of OCT technology, ophthalmology. Later sections of this chapter will identify the benefits of having higher image acquisition speed in different applications of SS-OCT systems.\\
\indent Another benefit of SS-OCT systems compared to SD-OCT systems can be attributed to the wavelength region SS-OCT systems can work at. It is reported that SS-OCT systems can operate at longer wavelengths (1300 nm) at which tissues show less light scattering properties; hence the capability of imaging tissues located at deeper layers of the ocular organ with SS-OCT devices \cite{topcon} .  \\
\indent It also has been reported \cite{topcon} that SS-OCT systems exhibit a better sensitivity compared to that of SD-OCT systems. Such phenomenon lets SS-OCT systems to acquire higher quality tomographic images from deeper layers of tissue all at once. Furthermore, the light that is projected  by the light sources that are used in SS-OCT systems are more often  working at wavelength at which the projected light is invisible making the involuntary movements of eye that minimal.\\
\indent Advantages mentioned above make the SS-OCT iteration of the technology defacto the one to be used in clinical and industrial applications in the future. Following sections of this chapter first talk about different light sources that are used in SS-OCT systems and then will explain what methods are used in the current state of the art OCT devices to overcome the shortcomings these light sources exhibit.
\section{Light Sources Used in SS-OCT Systems }
SS-OCT systems rely on a group of devices, circuits, and methods to perform the necessary optical operation and produce optical signals. SS-OCT systems then acquire those signals and process them to output a tomographic image corresponding to the object under test. In all these devices, there exists a low coherence light source that emits a narrowband spectral frequency light whose wavelength/wavenumber sweeps in time. These light sources are implemented via different methods that each exhibit unique characteristics suitable for different applications in which OCT devices are intended to be used. These light sources are referred to as lasers in the field. In recent years, novel developments in the laser technology have improved the speed at which these devices are capable of sweeping their wavelength. This phenomenon yields the possibility of having much faster OCT systems. While the conventional short cavity lasers are bound to have sweep rates in ranges of a couple of hundreds of thousands per second, newer vertical cavity surface emitting lasers (VCSEL) and Fourier domain mode locking (FDML) lasers can perform millions of sweeps per second. In fact, FDML lasers are shown to have no theoretical limit on how fast they can sweep their spectral frequency \cite{FDML}. Further, for such high sweep rates, lasers with low relative intensity noise (RIN) are required to compensate the limitations that photodetectors have to acquire such high-speed optical signals \cite{RIN}; hence the use of FDML lasers in these very high sweeping rate SS-OCT applications. These fast laser sweeps require very efficient signal processing and acquisition blocks to be able to output on-the-fly, high-resolution images or otherwise the overhead to perform all the necessary process cannot be afforded for such high imaging speeds. The device responsible for such data processing and acquisition is called an OCT engine and is used within the OCT device. Below, this dissertation explains the major drawback that both FDML and VCSEL light sources exhibit that need to be known and taken care of. 
\subsection{Lasers}
Unlike SD-OCT systems that use a wideband light source with low coherence length such as continuous wave (CW) super-luminescent LEDs (sLED), SS-OCT systems take advantage of more coherent and narrow bandwidth light sources or in short Lasers (Light Amplification by Stimulated Emission of Radiation). \\
\begin{figure}[!t]
	\centering
	\includegraphics[scale=1.1]{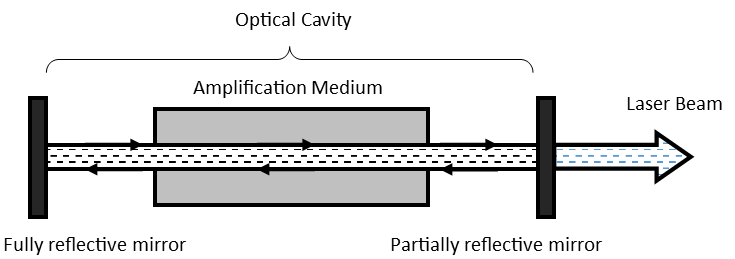}
	\caption{A systematic simplified schematic of how a laser generally works. Reprinted from \cite{laserfund}.}
	\label{laser}
\end{figure}
\indent The way all lasers work is by amplification of an emitted light that itself is the result of stimulating a semiconductor material. To explain more clearly, the reader should have some background knowledge of the electron energy state within a material and how receiving a specific amount of energy can elevate the electron from one energy state to another. The reader should also be familiar with the fact that an electron can loose a specific amount of energy and goes from a higher energy level to a lower one.\\
\indent Let us assume that an electron within an atom rests at its ground energy at a specific state. It can be shown \cite{laserfund} that a photon with an specific amount of energy can interfere with the electron and elevate it to a higher energy state should the energy difference between the ground energy level and the higher energy level equals the energy of the stimulating photon. The electron at the higher energy state then is rather unstable and can move to the lower energy level by a minuscule perturbation. The electron moving back to a lower energy level projects a photon with an energy that exactly equals the energy difference between the two energy levels.\\
\indent It can be shown that for a stacked array of stimulated atoms, a minuscule perturbation can emit one photon per one elevated electron per an atom. All these photons are coherent, all are projected at the same direction and have the same exact spectral frequency. Although such a beam projection matches the definition of a laser beam, stimulating an array of atoms may not be practical to begin with.\\
\indent One alternative to achieve such a coherent light source is by building an optical cavity that is bounded by two mirrors on the two sides. This way, a projected photon can get reflected back from the surface of the mirror and keep stimulating electrons to an elevated energy levels. Those elevated electrons then emit photon that keep being reflected back from the mirrors and continue building up a strong beam of the same spectral frequency coherent light. An avalanche of photons are created and is projected out of the laser cavity from one side of the laser through a mirror that is partially reflective and lets light pass through it (Figure \ref{laser}). \\
\indent The back and forth reflection of photons from the reflective surfaces on the borders of the cavity builds a standing electromagnetic wave the wavelength of which is an integer multiple of a value that itself is a function of the cavity length. The amplification medium within the cavity, acts as an optical filter and only lets one of the wavelengths to be selected; hence projecting a monochromatic coherent light. The optical filter can be implemented by a Fabry-Perot tunable Filter (FPF). In SS-OCT systems, there needs to be a mechanism that changes the projected wavelength in the time span of one A-scan.
\subsection{Semiconductor Lasers} \label{sweeplaser}
These types of lasers are similar to all other types where they generate and amplify coherent photons at their outputs. Semiconductor lasers use the excitation of an electrical current to merge the holes from a p-type semiconductor with the electrons from a n-type semiconductor. As the result of the electron annihilation (combination with the hole) a photon is produced. This amplification medium that produces the photons in a semiconductor laser is called a semiconductor optical amplifier (SOA). The avalanche of photons produced by the current excitation is trapped between two reflective surfaces and gets amplified. Depending on the type of laser used, the light beam is projected out of the laser cavity. Below, this dissertation explains two different laser types and the mechanism they use to sweep the spectral wavelength \cite{semilaser}.
\subsubsection{VCSEL Lasers Nonlinear Sweep}
Vertical cavity surface emitting lasers (VCSEL) \cite{VCSELOrg} are a specific type of semiconductor lasers that take advantage of some vertical distributed Bragg reflectors (DBR) located at the bottom and the top of an active layer which itself is constituted of specific material that produces a quantum well. An electrical excitation to electrical plates located at the bottom and the top of the active region causes the electron-hole combination to happen and the avalanche of photon production. Recent publication have reported repetition rates in the range of a couple of millions of wavelength sweep in a second \cite{VCSELspeed}.\\
\indent Tunable VCSEL lasers change the wavelength of emitting light with either of the two following mechanisms. \\
\indent First, these devices utilize the temperature-wavelength relationship \cite{VCSELTemp}. Different mechanisms can be used to change the temperature of the active region and the wavelength as a result \cite{VCSELtempchange}.\\
\indent The second way to change the emitted wavelength of the VCSEL laser is to change the length of the active region \cite{VCSELlengthchange}. The length change  usually occurs via some type of micro-electro-mechanical systems (MEMS). \\
\indent It should be noted that the control over the temperature of the active region is rather complicated. As the result of this phenomenon, the wavelength sweep as function of time is not linear and the wavelength of the output light of the VCSEL laser changes by different amounts in identical time spans within one spectral sweep of the laser 
\subsubsection{FDML Lasers Nonlinear Sweep}
\begin{figure}[!t]
	\centering
	\includegraphics[scale=0.45]{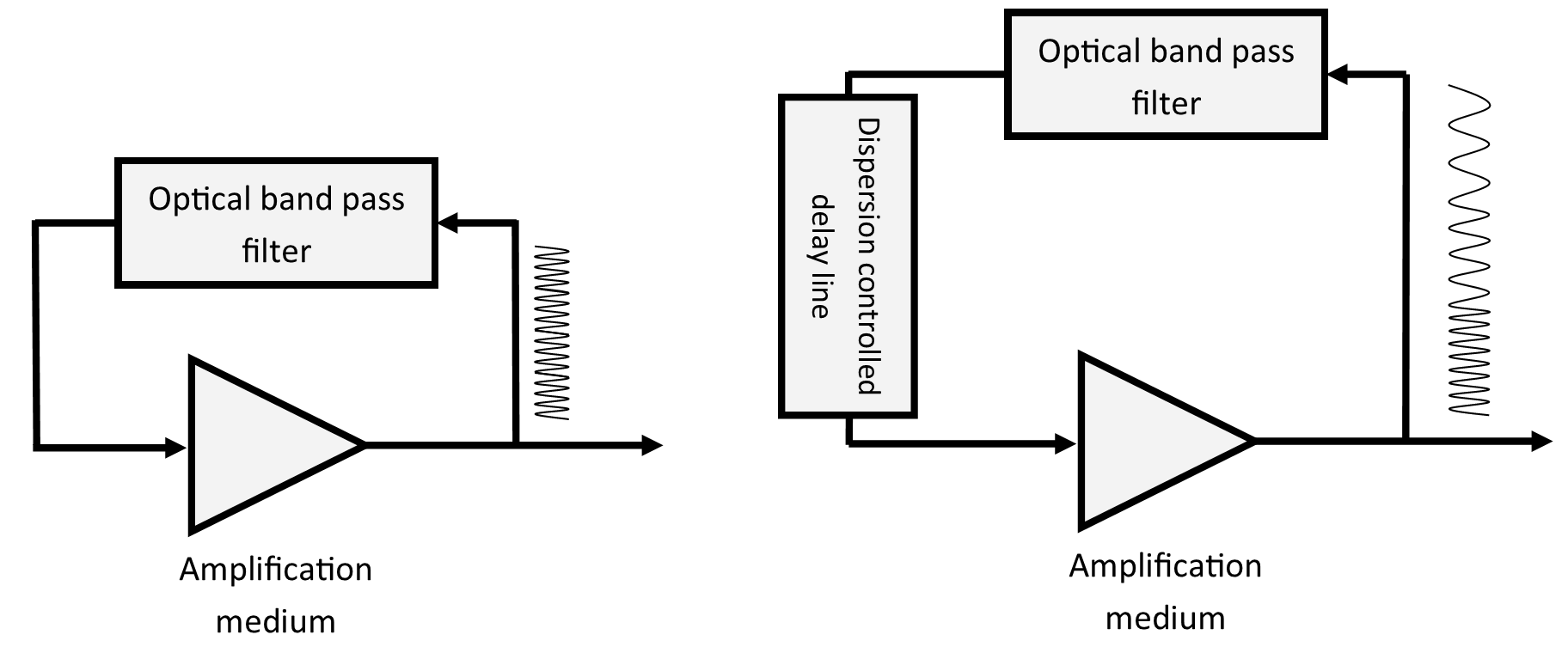}
	\caption{(Left) Standard tunable laser, The laser beam that is generated by the amplification medium is filtered by the optical bandpass filter; only one lasing spectral frequency exists in the loop. (Right) Basic FDML laser mechanism. A dispersion controlled delay line is added to the loop. All laser spectral frequencies are generated at once within the amplification medium. Reprinted from \cite{FDMLbasic}.}
	\label{FDML}
\end{figure}
Unlike typical lasers (Figure \ref{FDML}, left), in which an amplification medium is coupled with a bandpass optical filter, FDML lasers add a dispersion controlled delay line into the loop to better control the sweep speed and the coherence of light at any particular wavelength.\\
\indent In typical lasers, the optical filter emits one frequency at a time. When the center frequency of the bandpass filter is tuned, the lasing process needs to re-occur within the amplification medium at the newly determined spectral frequency; hence the existence of a trade of between the coherence of laser beam at any specific spectral frequency and the speed at which the laser can sweep the spectral frequency.\\
\indent In FDMLs, however, a dispersion controlled delay line is added to the loop to ensure that all the possible spectral frequency outcomes of the laser reach the amplification medium at the same time. As a result of such phenomenon, all spectral frequency components of a laser sweep happen at the same time in the amplification medium. To ensure such behavior, the propagation delay of the laser in the FDML loop (Figure \ref{FDML}, right) is designed to match the sweep time of the laser (the time it takes for the laser to cover all spectral frequencies that its intended to sweep across). Doing so, all spectral frequencies are effectively produced at the amplification medium at all time and the trade off mentioned before does not accrue; hence the ability to achieve much higher sweep rates.\\
\indent Similar to most semiconductor lasers, the FDML utilizes a Fabry-Perot filter to act as a bandpass optical filter. As mentioned before, these filters tune their center frequency by changing their internal cavity length by the means of some MEMS devices. The length change is rather difficult to control. Therefore, it is challenging to control the linearity of the spectral frequency sweep of these swept source FDML light sources.\\ \indent As SS-OCT systems rely on the spectral frequency sweep of lasers to produce interferometric patterns, it is important to have a sense of how the sweep profile affects the quality of the tomographic image produced and how the non-linear sweep can be estimated and accounted for. 
\section{The Non-linear Spectral Sweep of Lasers and Its Effect on the Quality of Tomographic Image}
An interferometric pattern in an SS-OCT system that is detected by a photo detector can be modeled by Equation \ref{Equ.12}. This signal needs to be taken an inverse Fourier transformation upon to produce an A-scan. In order to perform an inverse Fourier transformation, the interferometric patter (Equation \ref{Equ.12}) first needs to be sampled into the digital domain. \\
\indent For an SS-OCT system Equation \ref{Equ.12} can be revisited as below,
\begin{equation} \label{Equ.2.1}
\begin{split}
I_D(t)& = \frac{\rho}{4}S(k(t))[R_R+R_{S1}+R_{S_2}+...+R_{Sn}] \\
&+ \frac{\rho}{2}S(k(t))\bigg[\sum_{n\ne m=1}^{\infty}\sqrt{R_{Sm}R_{Sn}}cos\big(k(t)(z_{Sm}-z_{Sn})\big)\bigg]. \\
&+ \frac{\rho}{2}S(k(t))\bigg[\sum_{n=1}^{\infty}\sqrt{R_{R}R_{Sn}}cos\big(k(t)(z_R-z_{Sn})\big)\bigg],
\end{split}
\end{equation}
where the wavenumber\textbackslash wavelength sweeps in time. \\
\indent Any design of SS-OCT systems then have to sample Equation \ref{Equ.2.1} and take an inverse Fourier domain in the wavenumber ($k$) domain.\\
\indent As stated in Section \ref{sweeplaser}, most swept source lasers exhibit some sort of non-linearity when changing the wavenumber of their emitted light. A non-linear sweep means that the laser changes the wavenumber of the emitted light by different amounts in identical time spans within one spectral sweep of the laser. It is therefore obvious that sampling Equation. \ref{Equ.2.1} at equidistant time instances would result in sampling points at which $k$ has not changed by identical amounts. To explain more clearly, a uniformly sampled SS-OCT interferometric pattern is modeled as below, \\
\indent For an SS-OCT system, Equation \ref{Equ.12} can be formulated as below,
\begin{equation} \label{Equ.2.2}
\begin{split}
I_D(\Delta D)& = \frac{\rho}{4}S(k(\Delta D))[R_R+R_{S1}+R_{S_2}+...+R_{Sn}] \\
&+ \frac{\rho}{2}S(k(\Delta D))\bigg[\sum_{n\ne m=1}^{\infty}\sqrt{R_{Sm}R_{Sn}}cos\big(k(\Delta D)(z_{Sm}-z_{Sn})\big)\bigg]. \\
&+ \frac{\rho}{2}S(k(\Delta D))\bigg[\sum_{n=1}^{\infty}\sqrt{R_{R}R_{Sn}}cos\big(k(\Delta D)(z_R-z_{Sn})\big)\bigg],
\end{split}
\end{equation}
where $\Delta$ is the constant sampling interval at which the interferometric pattern is sampled. Different values of  $k(\Delta D)$ for different $D$s in Equation \ref{Equ.2.2} are not apart from each other by identical values. That is due to the non-linear sweep of the swept source laser that is indeed used in SS-OCT devices.\\
\indent As the inverse Fourier transformation occurs in the \textit{k-domain} and not the time domain, applying the inverse Fourier transformation on the uniformly sampled interferometric pattern (shown by Equation \ref{Equ.2.2}) can not demodulate the value of $\sqrt{R_{Sn}}$ and therefore the tomographic pattern can not be distinguished. \\
\indent \textbf{The main contribution of this dissertation is to propose ways to sample the interferometric pattern in Equation \ref{Equ.12} at exact time instances at which the wavenumber has changed by identical values.}\\
\indent Let us assume that the entire span that the laser sweeps its spectral frequency across is shown by $[k_0,k_{end}]$.  Assuming that $k_i$ is an element of a set of wavenumbers denoted by $\textbf{K}_{linear}$ described as below,
\begin{equation}\label{Equ.2.3}
    \textbf{K}_{Linear}=\bigg\{k_i\big|k_0<k_i<k_{end}, k_i=k_0+i\times\delta_s k, i\in\mathbb{N}\bigg\},
\end{equation}

where $\delta_s k$ is a value that two equidistant wavenumbers are apart by. The goal is to find time instances $t_c$ at which the changing value of $k(t)$ hits any of the predefined values of $k_i$ and sample the interferometric signal (Equation \ref{Equ.12}) at those time values. The aforementioned process is called spectral calibration in this dissertation.\\
\indent It is obvious that the wavenumber sweep profile in a SS-OCT system needs to somehow be extracted and then the  effects of its non-linear sweep be accounted for. Auxiliary interferometers such as Mach-Zehnder interferometers (MZI) or Fabry-Perot interferometer (FPI) are used in an OCT systems for that exact purpose. 
\section{Fabry Perot Interferometers, FPI}
\begin{figure}[!t]
	\centering
	\includegraphics[scale=0.28]{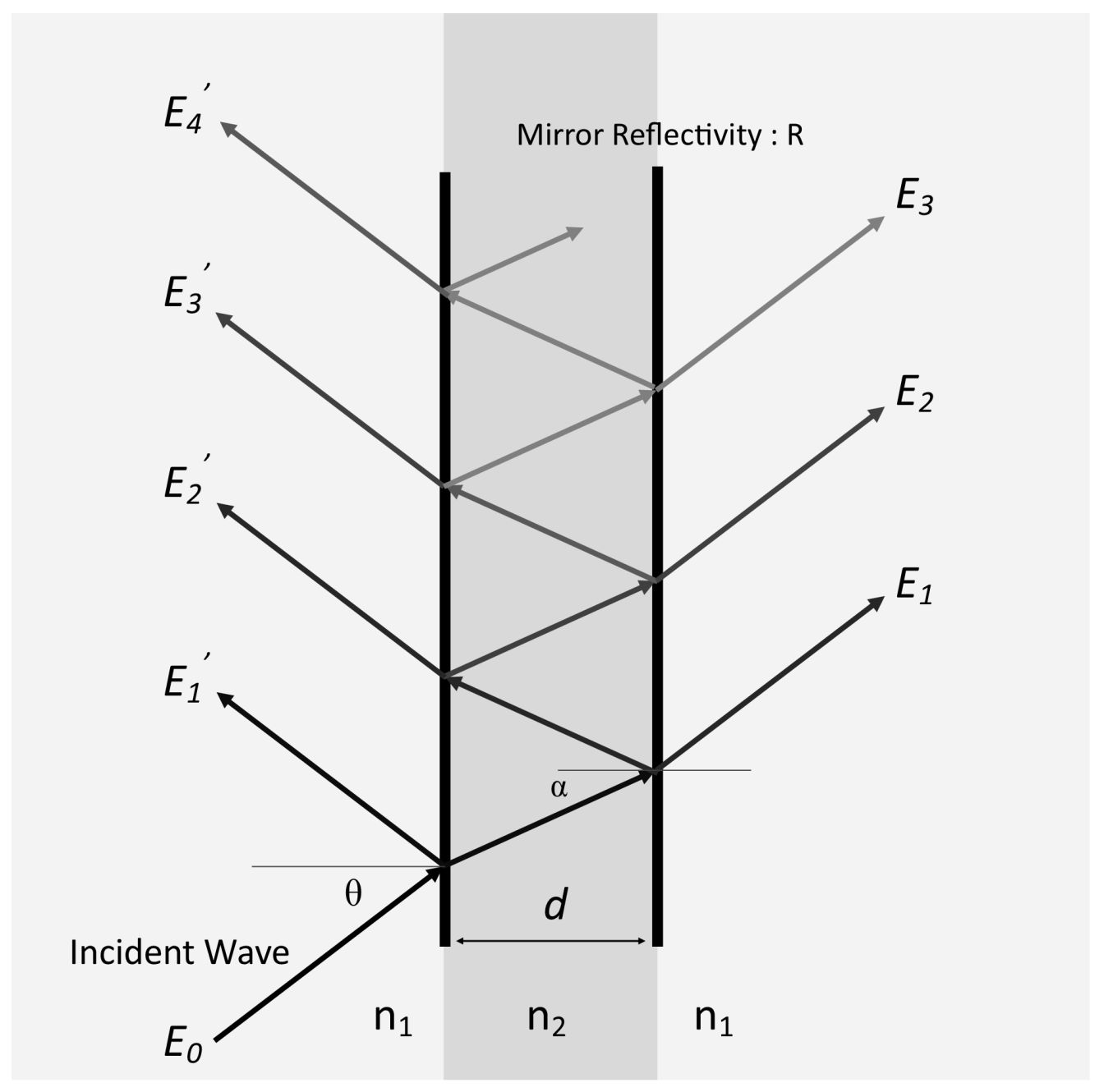}
	\caption{Generic representation of a FPI.}
	\label{FPI}
\end{figure}
Fabry-Perot interferometer (FPI) is a type of interferometer that can be used to determine the wavelength of the incident light that is coupled to the interferometer \cite{FPI}. In an FPI, an incident light is projected to the interferometer. It first hits  a partial reflector that reflects part of the light and transmits the rest. The transmitted light gets dispersed in the new medium with the light refraction index $n_2$. The light then hits the second partial reflector; the light then partially reflects back from the surface of the second partial reflector and partially transmits to the third medium with a refraction index equaling that of the first medium (Figure \ref{FPI}).  The back and forth of light continues to happen as Figure \ref{FPI} suggests. As a result, light beams $E_1$, $E_2$, $E_3$, $...$ are transmitted to the third medium. A specific mechanism is then used to interfere and add up light beams in the third medium. It can be shown that the total intensity of light at the output of the FPI can be represented as,
\begin{equation}\label{Equ.2.4}
    I_{T}=\frac{I_0(1-R)^2}{(1-R)^2+4Rsin^{2}(\delta)} = \frac{I_0}{1+Fsin^{2}(\delta)} ,
\end{equation}
where $F=\dfrac{4R}{(1-R^2)}$, and $\delta = \dfrac{2\pi ndcos(\alpha)}{\lambda}$.
If the incident light to a FPI is the light projected from a swept source, the intensity of light at the output of the FPI, $I_T$, changes as the wavelength ($\lambda$) and $\delta$ as the result change. Because the way $\lambda$ is embedded in the formula (Equation \ref{Equ.2.4}) rather complicated, other interferometers such as Mach-Zehnder interferometers (MZI) are used in OCT systems to distinguish the sweep profile of an OCT system.\footnote{Changing $d$ in Equation \ref{Equ.2.4}, changes the intensity of the output light for an specific wavelength. A specific  value for $d$ maximizes the intensity of light for specific wavelength; therefore the interferometer can be used as an optical filter.}
\section{Mach-Zehnder Interferometers, MZI}
\begin{figure}[!t]
	\centering
	\includegraphics[scale=0.48]{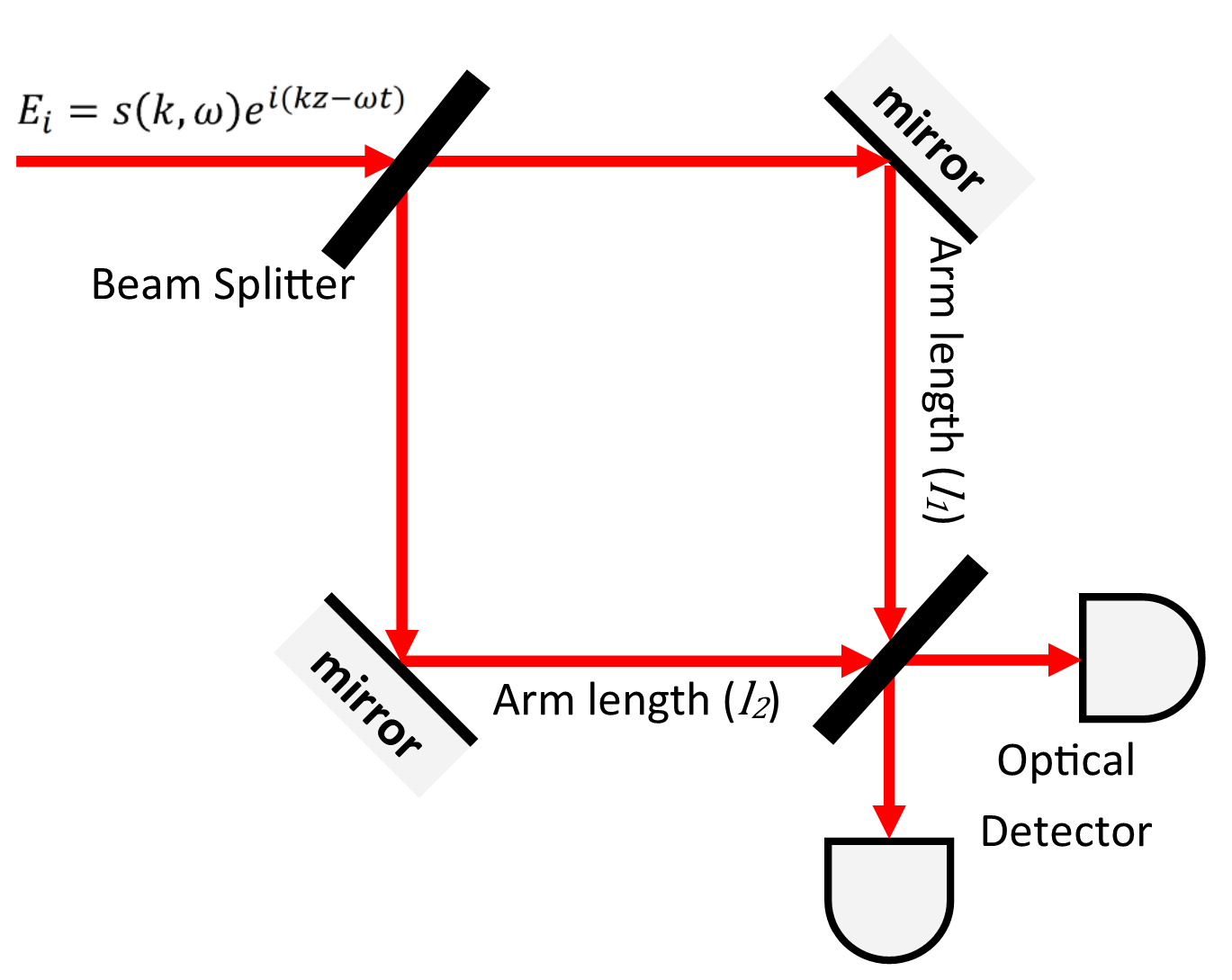}
	\caption{Generic representation of a MZI.}
	\label{MZI}
\end{figure}
Mach-Zehnder interferometers are widely used in SS-OCT systems to estimate the nonlinear wavenumber sweep that occurs in those systems. In a MZI (Figure \ref{MZI}) the incident light is projected upon a beam splitter. The light is split into two beams with equal powers, both are projected to different mirrors and re-coupled at a second beam coupler. Two light beams are then projected out of the second coupler and are received by two optical detectors \cite{MZI}.\\
\indent Research \cite{MZI} shows that there exits a specific relationship between the phase of photons that have traveled from the top or the bottom path as shown by Figure \ref{MZI}.\\
\indent Let us assume that the length of the top path in the MZI shown in Figure \ref{MZI} is denoted by $l_1$ and that of the bottom path by $l_2$. \\
\indent The incident light $E_i$, passes through the top beam splitter and experiences a phase shift that equals $\dfrac{2\pi l}{\lambda}$, where lambda is the wavelength of the light that is coupled into the interferometer and  $l$ is the path length that the light travels within the splitter. The light that gores through the top path experiences a total phase shift of $\dfrac{2\pi l_1}{\lambda}$ and two phase shifts each equaling $\pi$ before it is received by the optical detector at the right hand side of the MZI configuration as shown by Figure \ref{MZI}; hence a total phase shift equaling $2\pi+\dfrac{2\pi l_1}{\lambda}+\dfrac{2\pi l}{\lambda}$.\\
\indent Similar to the top, path the light beam traveling in the bottom path experiences the same value of phase shift, $\dfrac{2\pi l}{\lambda}$ when is passed through the splitter. Passing through the lower path, the light experiences another phase shift that equals $\dfrac{2\pi l_2}{\lambda}$. Being received by the optical detector on the right hand side of the configuration shown by Figure \ref{MZI} the light in the lower path experiences two reflections from the top splitter and the bottom mirror; hence an additional phase shift of $2\pi$. The total phase shift that the light experinces in the bottom path equals $2\pi+\dfrac{2\pi l_2}{\lambda}+\dfrac{2\pi l}{\lambda}$. The phase difference between the two paths is denoted by $\theta$ and equals,
\begin{equation} \label{MZIPhase}
    \theta = (2\pi+\dfrac{2\pi l_2}{\lambda}+\dfrac{2\pi l}{\lambda})-(2\pi+\dfrac{2\pi l_1}{\lambda}+\dfrac{2\pi l}{\lambda})=\dfrac{2\pi (l_2-l_1)}{\lambda} = k(l_2-l_1).
\end{equation}
Having calculated the phase difference between the light beams, the intensity of light being detected by the photo detector is formulated as,
\begin{equation} \label{MZIFormula}
    I_{MZI} = CS(k)cos\big(k(l_2-l_1)\big),
\end{equation}
where $C$ is a constant and  $S(k)$ encodes the power spectral dependence of the light source. \\
\indent As seen in Equation \ref{MZIFormula}, the wavenumber of the incident light is embedded in the argument of the MZI output. It is obvious that demodulating the wavenumber sweep from a MZI signal is much easier compared to that of a FPI. Therefore, there exists an auxiliary MZI in OCT systems to capture the wavenumber sweep profile of the light source and mitigate the deteriorating effects its non-linear sweep causes. 
\section{Calibration}
\indent The major challenge to acquire high-quality, high-speed images that all SS-OCT systems face is related to a fundamental limitation that all widely adopted spectral frequency sweeping lasers exhibit; the spectral frequency sweep is a non-linear function of time. The wave number sweeps by different amounts at different time spans during the entire time of a single sweep. Then, converting the acquired optical signals that are passed through the object under test into the electrical domain and sampling them on uniformly-spaced time intervals into the digital domain requires extra processing to be able to compensate for the non-idealities the non-linear wavenumber sweep causes as ideally the optical signals need to be sampled at times that the wavenumber has changed by identical amounts \cite{bookOCTFormula}.\\
\indent Conventionally, all SS-OCT systems utilize an additional interferometer to act as a reference to be able to reverse the unwanted effects that the nonlinear sweep of the wave-number causes. In this dissertation, we call the output of the reference interferometer as the calibrating signal and the signal that passes through the object under test as the interferometric signal. While the calibrating signal is almost unanimously implemented by the previously introduced MZI, the interferometric signal is the result of using a Michelson interferometer and passing light through the object under test. While the wavenumber sweep information is embedded in the calibrating signal, the interferometric signal entails the tomographic information needed to be acquired from the object under test. The MZI signal is sampled into the digital domain, mathematical operations are applied on the sampled signal and a calibrating profile ultimately is produced that is used to sample the interferometric signals at time stamps at which the wavenumber of the swept source laser is changed by identical values. This process is called spectral calibration in this dissertation.
\subsection{Calibration Based on Zero Crossing Detection}
\begin{figure}[!t]
	\centering
	\includegraphics[scale=0.37]{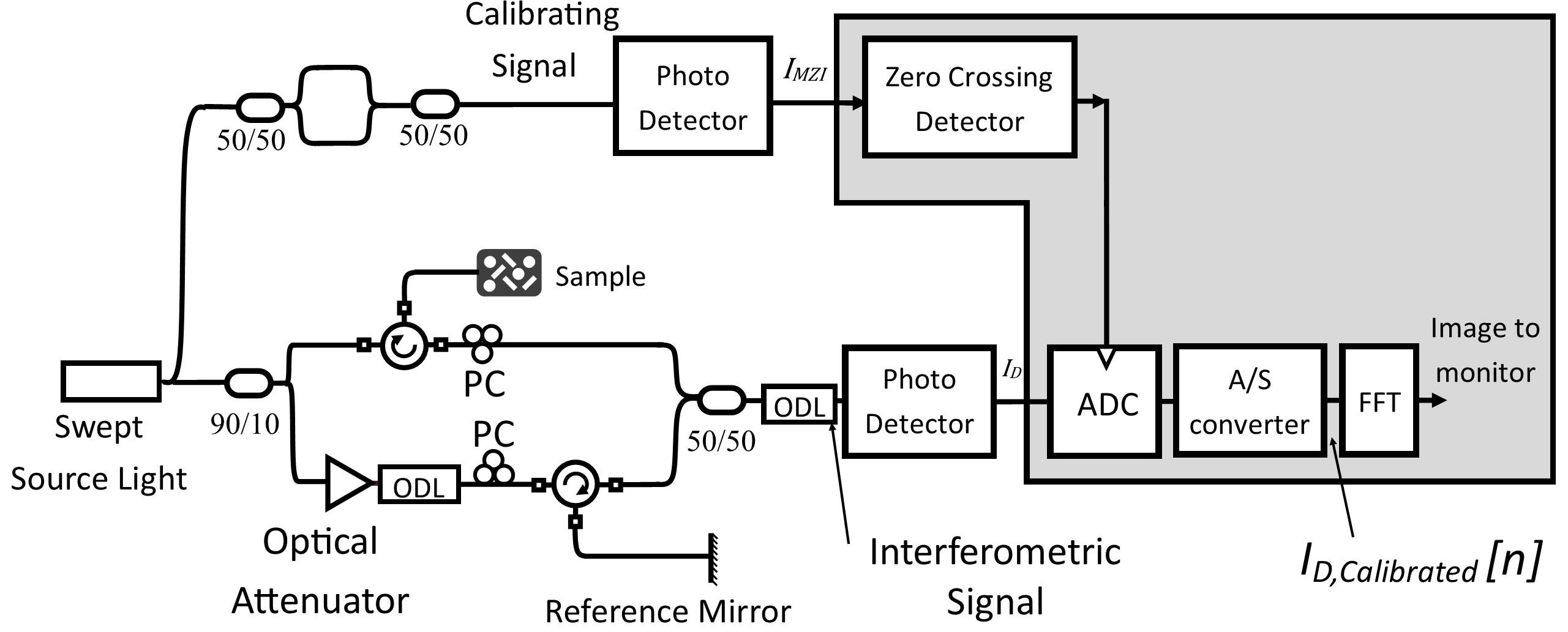}
	\caption{Conventional OCT System based on sampling on the zero-crossings of the MZI signal.}
	\label{Zerocrossingconfig}
\end{figure}
\begin{figure}[!t]
	\centering
	\includegraphics[scale=0.52]{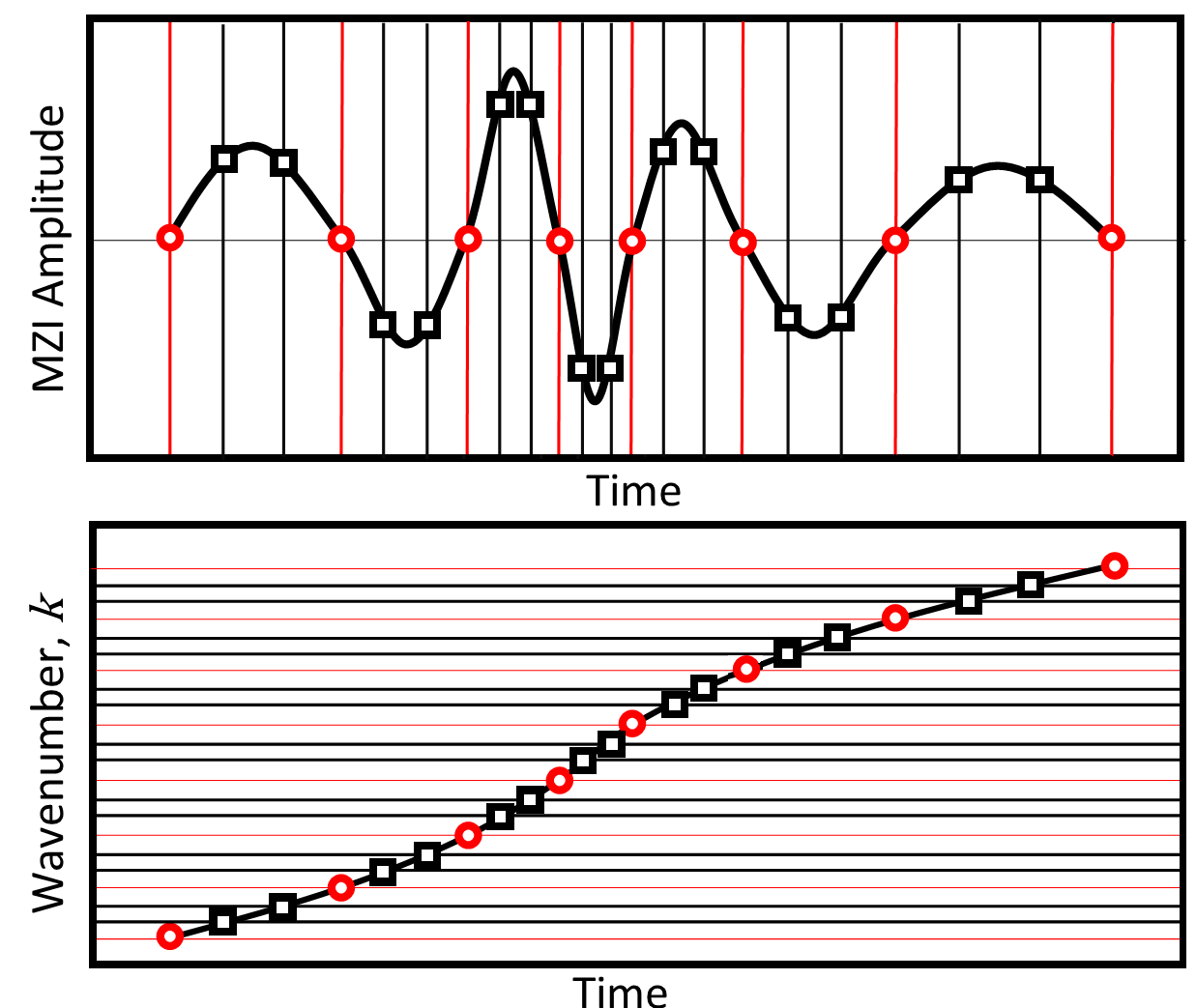}
	\caption{(Top) the calibrating signal, MZI signal, (bottom) the wavenumber, k, associated with calibrating signal. Circles and squares are sampling instances at which the interferometric signal gets sampled.}
	\label{Zerocrossing}
\end{figure}
In this method (Fig. \ref{Zerocrossingconfig}), the calibrating signal (MZI) is used as a wavenumber clock (k-clock) to sample the interferometric signal at correct timestamps at which the wavenumber has swept by identical amounts. The way this methodology works is to use the calibrating signal (MZI) and sample the interferometric signal at time instances at which the calibrating signal crosses zero. At any of these timestamps, the value of the parameter $k(t)$ has increased from the previous sampling stamp by a constant value of $\dfrac{\pi}{l_2-l_1}$; hence sampling the interferometric signal at time instances at which the value of k has increased by identical amounts. These time instances are denoted by $t_{zero_{cn}}$.
\begin{equation}
    I_{MZI}(t)= CS(k)cos\big((l_2-l_1)k(t)\big)=0\rightarrow t_{zero_{cn}}=\frac{n\pi+\frac{\pi}{2}}{l_2-l_1}.
\end{equation}
\indent The major problem with such implementation is the limited number of zero crossings in a particular calibrating signal causes a limited number of sampled points from the interferometric signal per scan that are readily available to apply further processing to acquire a tomographic image. It can be proven mathematically \cite{bookOCTFormula} that the maximum distance ($z_{max}$, imaging range/depth) that the object under test can have from the optical setup used in the SS-OCT system before the acquired image becomes indistinguishable is a function of the number of points the interferometric signal is sampled by. As such, restricting the sampling points to the time stamps at which the calibrating signal has crossed zero indeed limits the imaging range which limits the utilization of such SS-OCT systems in applications where higher ranges are required such as meter range optical coherence tomography \cite{meterrange} or in certain areas of ophthalmology where imaging the entire depth of the anterior segment of the eye is necessary \cite{anterior}. The solution to such problem is to be able to sample the interferometric signal at time instances in between the zero crossings of the calibrating signal which by definition is not possible in this solution for the spectral calibration of SS-OCT systems.  Figure \ref{Zerocrossing} shows a typical MZI signal (top) and the laser wavenumber sweep as function of time ($k(t)$) that is embedded in its argument (Equation \ref{MZIFormula}). While the zero crossing time instances are only limited to a few, more instances could be added to obtain sufficient number of sampling points. As Figure \ref{Zerocrossing} (bottom) shows, all sampling instances are picked at times that the wavenumber undergoes identical changes in time.
\subsubsection{Detecting Zero Crossing with Analog Electronic Circuits}\label{zerocrossinganalog}
\begin{figure}[!t]
	\centering
	\includegraphics[scale=0.53]{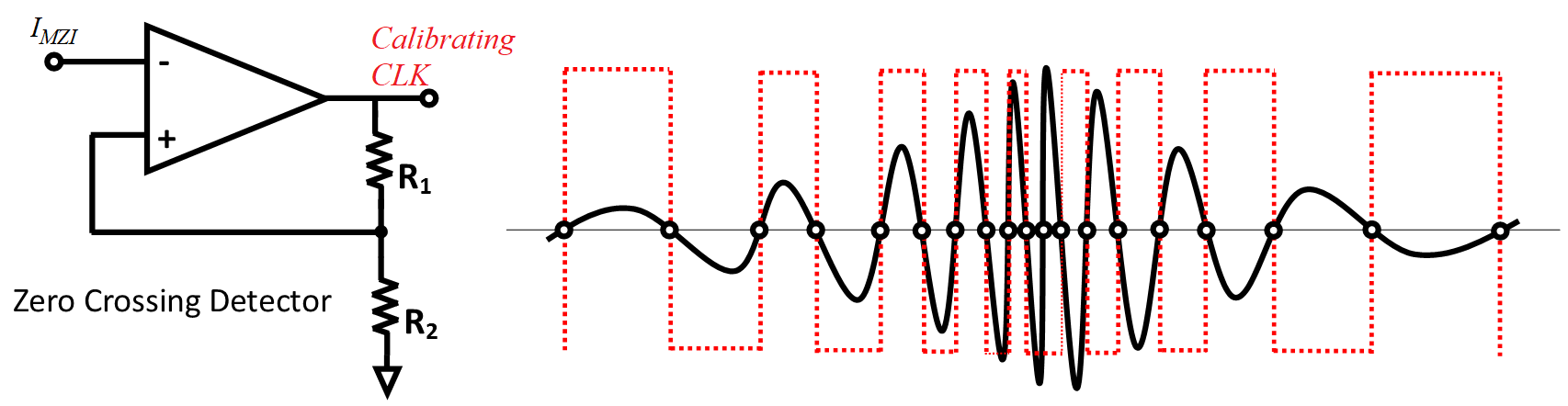}
	\caption{Detecting the zero crossings of a calibrating signal (MZI signal) via electronic circuits. The configuration on the left is realized by putting an OpAmp in a positive feedback configuration ($R_1>>R_2$).}
	\label{zerocrossinganalog}
\end{figure}
In this approach \cite{zerocrossingdigital}, the calibrating signal (MZI signal, $I_{MZI}$) is fed to an electronic circuit configuration. The electronic configuration is built by putting an operational amplifier (OpAmp) in a positive feedback configuration. In this configuration ($R_1>>R_2$), the voltage value at the positive pin of the OpAmp is always set at around zero. It also can be observed that the output of the OpAmp is clamped to either $+V_{CC}$ or $-V_{CC}$ due to the existence of a positive feedback. Let us assume that the output of the OpAmp configuration is set at $-V_{CC}$; it can be understood that the voltage on the positive pin is around very small values close to zero and any crossing of the input signal ($I_{MZI}$) will trigger the configuration and sets the output signal to $V_{CC}$. The same analogy can be comprehended in the opposite direction when the output is set at $V_{CC}$ and the input signal $I_{MZI}$ crosses zero towards negative voltages. This circuit effectively produces a Calibrating CLK with positive and negative edges at the zero crossings of the MZI signal ($I_{MZI}$). The calibrating clock then can be used to sample the interferometric signal $I_D$ at time instances at which the laser has changed the wavenumber by identical values. 
\subsubsection{Detecting Zero Crossing with Digital Circuits}
Some innovative ways also have been reported in the literature to use the calibrating signal ($I_{MZI}$) and produce more sampling points at which the laser has changed the wavenumber by identical amounts. 
\begin{figure}[!t]
	\centering
	\includegraphics[scale=0.85]{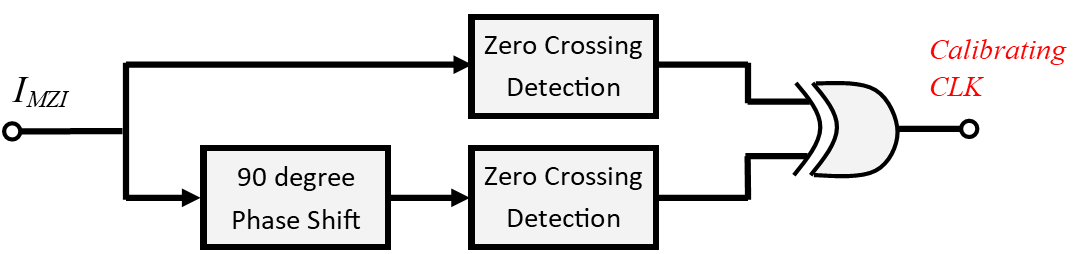}
	\caption{Producing a calibrating clock using digital circuitry. Reprinted from \cite{zerocrossingdigital}.}
	\label{zerocrossingdigital}
\end{figure}
\\ \indent As Figure \ref{zerocrossingdigital} suggests, the calibrating signal $I_{MZI}$ will be subject to a 90 degree phase shifter. At the output of this block, $I_{MZI,90}$, the maximums and minimums of the signal are aligned with the zero crossings of the original signal, $I_{MZI}$. Both signals ($I_{MZI}$, and $I_{MZI,90}$) are subject to two different zero crossing detector blocks. Doing so, the top path (Figure \ref{zerocrossingdigital}) produces rising\textbackslash falling edges on zero crossings of the calibrating signal, $I_{MZI}$. The bottom path, however, produces rising\textbackslash falling edges at the maximums and minimums of the calibrating signal, $I_{MZI}$. Using a X-Or gate at the output of the two zero crossing blocks, a calibrating clock with rising\textbackslash falling edges at the zero crossings and minimums and maximums of the calibrating signal, $I_{MZI}$ is produced. It can be shown that the the wavenumber of the laser has changed by identical values that equal $\dfrac{\dfrac{\pi}{2}}{(l_2-l_1)}$ between subsequent maximums, zero crossings, and minimums of the calibrating signal, $I_{MZI}$. This approach can produce twice as much sampling point compared to the system proposed in Section \ref{zerocrossinganalog}; hence, increasing the range the SS-OCT system can be increased. The drawback of this algorithm relates to the excess hardware used and the limited improvement it results in terms of the number of sampling points available in an A-scan. 
\subsection{Calibration Based on Using Oscilloscopes in Meter Range SS-OCT}
\begin{figure}[!t]
	\centering
	\includegraphics[scale=0.9]{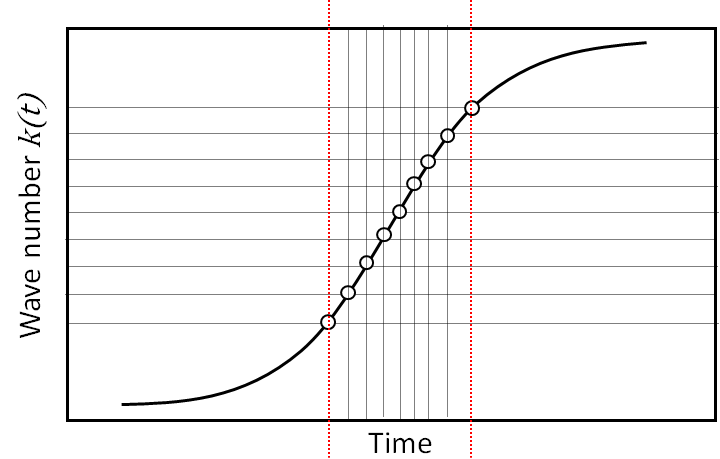}
	\caption{A typical wavenumber sweep in a SS-OCT system. The middle range shows a part of the spectral sweep that is fairly linear.}
	\label{meter}
\end{figure}
The maximum axial depth ($z_{max}$) an OCT system is capable of acquiring images from is a function of how close the wavenumber distance is between two subsequent samples of an A-scan. The less the wavenumber distance between two subsequent samples of an A-scan gets, the higher the imaging range becomes. Recently, there has been a desire to acquire tomographic images from subjects that have dimensions in the range of meters \cite{meterOCT,meterOCT1}.\\ 
\indent In order to achieve such gigantically large imaging ranges, each A-scan needs to be sampled at very high speeds such that the specific wavenumber sweep span that the laser causes is divided into more points and the distance between two subsequent wavenumbers is minimized. In these meter range SS-OCT application, a relatively low sweeping rate laser is used. The interferometric signal that is produced is then sampled with oscilloscopes that are capable of sampling an A-scan at rates in tens of billions of samples per second. Assuming that the laser sweeps the spectral frequency at a rate close to a couple of hundreds of thousands of sweeps per scan, such a fast sampling rate oscilloscope can pick close to a millions of samples within an interferometric pattern that is produced in an A-scan. \\
\indent  The caveat to such approach relates to the inherent uniform sampling capability of oscilloscopes that are used in this approach. It is obvious that sampling via an oscilloscopes can only occur at a uniform sampling period; hence, the need to have a laser that sweeps the spectral frequency accurately linearly. \\
\indent At such high imaging range specifications (meters), any minuscule spectral sweep nonlinearity can deteriorated the quality of image acquired. It should be noted that due to physical movements required to produce a spectral sweep within a swept source laser, having an entire linear spectral frequency sweep in one sweep of the laser is rather physically impossible. Non-linearities are observed at the sweep profile ($k(t)$) specially at the beginning and the end of the spectral sweep span as Figure \ref{meter} suggests. The solution to spectral calibrate the interferometric signal is to discard the samples of the signal at the beginning and the end of the spectral sweep and sample the interferometric signal only at time spans during which the wavenumber is being changed rather linearly (Figure \ref{meter}).\\
\indent As such, other techniques need to be introduced that are capable of sampling an interferometric signal non-uniformly. That way it is possible to sample the interferometric signal at the beginning and the end of a spectral frequency sweep and increase the quality of tomographic images produced.
\subsection{Calibration Based on Interpolation, Resampling, and Lookup Table}
\begin{figure}[!t]
	\centering
	\includegraphics[scale=0.278]{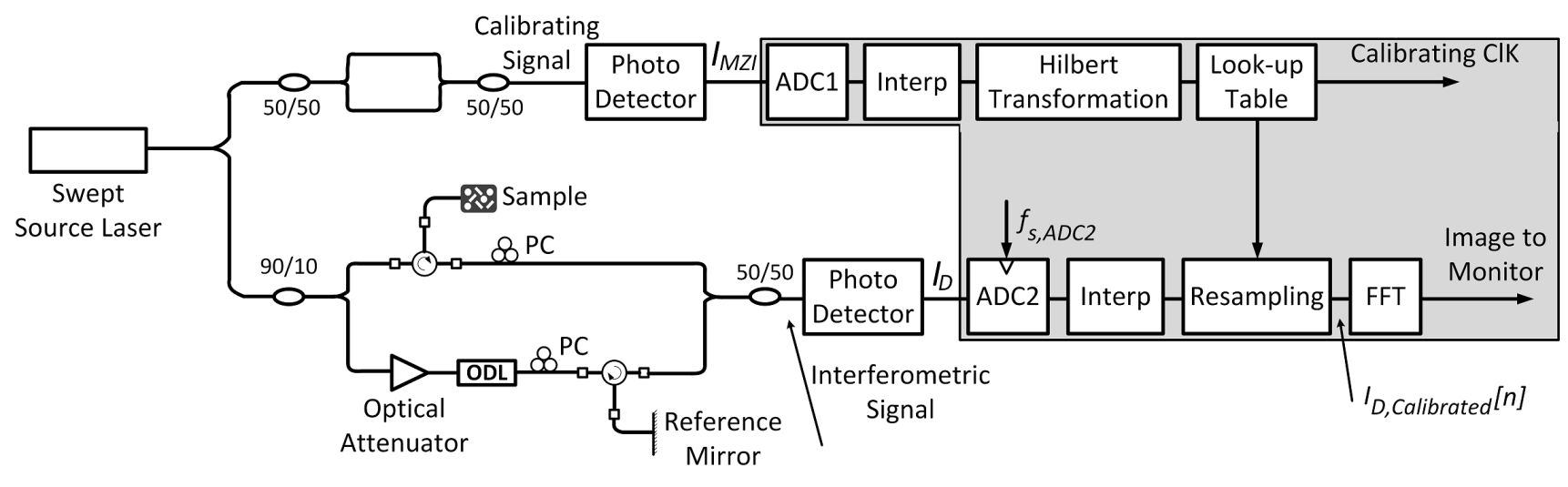}
	\caption{Conventional OCT system based on the Hilbert Transformation of the MZI signal and interpolation and resampling of the interferometric signal.}
	\label{hilbertconfig}
\end{figure}
Similar to the previous spectral calibration technique, both the MZI and the interferometric signal are used in the manner that Fig. \ref{hilbertconfig} suggests. In this technique, both signals are first sampled and recorded into the digital domain. The MZI signal then goes through a demodulation technique that often is realized by a Hilbert transformation (after being interpolated). The Hilbert transformation can demodulate the MZI argument $k[n]$. By using a lookup table, the interpolated values of $k[n]$ are then compared against a predefined set of equidistant values of the wavenumber ($\{k_{{Hilbert}_{cn}}\}$). That is how the correct time instances are chosen to resample the interferometric signal. Interpolation has to be carried out on both signals to increase the digital time resolution so that the resampling time instances are picked more accurately. The sample indices that are picked in the resampling block correspond to those that are the closest to the values saved in the look-up table. \\
 \indent Unlike the Zero Crossing Based Spectral Calibration, this technique enables the designer to pick the number of the equidistant wavenumber values. However, a realistic implementation of the Hilbert Transformation is proved to be neither the most computationally efficient \cite{IpDFT} nor can it account for noise \cite{biocas}. This method is often carried out in post processing and on a recorded data. Further, in high repetition rate OCT applications, or those where a high imaging range \cite{meterrange} is desired, achieving a real-time on-the-fly realization of Fig. \ref{hilbertconfig}, becomes extraordinarily challenging due to the digital domain clock frequency and bit resolution specifications.

This approach relies on applying a Hilbert transformation on the calibrating signal (MZI signal) \cite{Hilbert} to extract the sweep profile of the laser used in an SS-OCT device, $k(t)$. The produced calibrating and interferometric signals ($I_D, I_{MZI}$) are processed by a photodetector that converts them into the electrical domain. Both these signals are then sampled by data acquisition units (DAQ) and all the necessary signal processing is carried out in the digital domain. First, the digitally sampled calibrating signal ($I_{MZI}$) passes through a Hilbert transformation,
\begin{equation} \label{MZIHilbert}
    I_{MZI,H} = Hilbert(I_{MZI})
\end{equation}
This signal then is used in conjunction with the signal $I_{MZI}$ in some arithmetic process to produce the wavenumber signal in the digital domain. The process to acquire the wavenumber signal, $k$, is called demodulation in this text.
\begin{equation} \label{k[n]}
   k[n]=tan^{-1}\bigg(⁡\dfrac{I_{MZI,H}[n]}{I_{MZI}[n]}\bigg)    
\end{equation}                                                          
It should be noted that the arctan arithmetic is rather process intensive in the digital domain. Extra circuitry also is needed to unwrap the parameter $k[n]$ as the arctan arithmetic produces only output values in the main trigonometric circle, $(-\dfrac{\pi}{2},\dfrac{\pi}{2})$. The interpolated wavenumber profile signal is then used to find an array of wavenumber values, $\textbf{K}_{Linear}=[0,k_{1},k_{2},k_{3},…,k_{m}]$, that are equally distributed in the entire wavenumber span that the swept source laser sweeps in. Each of the components of the array $\textbf{K}_{Linear}$ then correspond to time indices that the interferometric signal needs to be sampled at. The acquired interferometric signal as well undergoes an interpolation step and the samples of it corresponding to those associated with the components of the array $\textbf{K}_{Linear}$ are then picked for further processing. The problems with such approach stem from first, the lack of efficiency that the Hilbert Transformation exhibits as well as fundamental structural limitations that the configuration in Fig. \ref{hilbertconfig} suffers from. Below, two different methods are described using which the Hilbert transformation gets applied.
\subsubsection{Software Implementation of Hilbert Transformation}
In the most widely accepted application of Hilbert transformation, the digitally acquired calibrating signal $I_{MZI}$ is received by pieces of software such as MATLAB or LABView. These pieces of software have predefined programs that can perform a Hilbert transformation by applying a Fourier transformation on the sampled signal. The negative frequencies of the resulted signal is then then zeroed out and the newly produced signal is applied an inverse Fourier transformation upon. It can be proven that the arithmetic operation mentioned above can produce the signal $I_{MZI,H}$ if applied on the signal $I_{MZI}$.\\
\indent Although close to all SS-OCT devices use the mentioned approach to acquire the signal $I_{MZI,H}$, It can be proven that applying the Fourier transformation on the entirety of the signal $I_{MZI}$ and then applying an inverse Fourier transformation is rather heavily hardware intensive and not efficient at all.  

\subsubsection{Hardware Implementation of Hilbert Transformation}
Recently, there have been new reports in the literature in which more \cite{Hilbert} innovative ways of applying a Hilbert transformation is reported. It can be proven that the signal $I_{MZI,H}$ can be calculated by applying a Finite Impulse Response (FIR) filter to the signal $I_{MZI}$ after it is sampled into the digital domain. \\
\indent Although the hardware based implementation of the Hilbert transformation is much more efficient than the software based one, the following chapter introduces new techniques that surpass the Hilbert Transformation in terms of efficiency. Although Hilbert transformation provides a mathemtical solution to demodulate the parameter $k(t)$ from $I_{MZI}$ in Equation \ref{MZIFormula}, it is neither arithmetically the optimum nor does it account for unwanted non-idealities such as noise. Therefore, the need to introduce new demodulation techniques (extracting $k(t)$ from $I_{MZI}$ in Equation \ref{MZIFormula}) that are more efficient. The next chapter introduces several demodulation techniques to optimize the spectral calibration of SS-OCT devices.

\chapter[DEMODULATION TECHNIQUES]{DEMODULATION TECHNIQUES \footnote[1]{Reprinted with permission from "A novel continuous time ternary encoding based SS-OCT calibration" by Amir Tofighi Zavareh, Oscar Barajas, and Sebastian Hoyos, 2016. 2016 IEEE Biomedical Circuits and Systems Conference (BioCAS), pp 5-8, Copyright [2016] by Institute of Electrical and Electronics Engineers (IEEE).\\\\
\indent Reprinted with permission from "An efficient estimation algorithm for the calibration of low-cost SS-OCT systems" by Amir Tofighi Zavareh, Oscar Barajas, and Sebastian Hoyos, 2017. 2017 IEEE 14th International Symposium on Biomedical Imaging (ISBI 2017), pp 1169-1172, Copyright [2017] by Institute of Electrical and Electronics Engineers (IEEE).\\\\
\indent Reprinted with permission from "The Spectral Calibration of Swept-Source Optical Coherence Tomography Systems Using Unscented Kalman Filter" by Amir Tofighi Zavareh, and Sebastian Hoyos, 2018. 2018 IEEE Biomedical Circuits and Systems Conference (BioCAS), pp 1-4, Copyright [2018] by Institute of Electrical and Electronics Engineers (IEEE).\\\\
\indent Reprinted with permission from "Towards an on-chip signal processing solution for the online calibration of SS-OCT systems" by Oscar Barajas, Amir Tofighi Zavareh, and Sebastian Hoyos, 2017. 2017 IEEE International Symposium on Circuits and Systems (ISCAS), pp 84-87, Copyright [2017] by Institute of Electrical and Electronics Engineers (IEEE).
}}
\indent So far, the poor imaging range resulted from the zero crossing approach as well as an extensively high processing overhead and lack of efficiency in the arithmetic processes done by a Hilbert transformation as well as some system level drawbacks on the calibration method based on interpolation and re-sampling has been explained as the reason why to employ other demodulation techniques and spectral calibration mechanisms.\\
\indent This chapter proposes alternative demodulation techniques and expresses the advantages they propose over the conventional Hilbert transformation. The calibrating signal, $I_{MZI}$, is first sampled in to the digital domain via conventional ADCs. Ultra-fast DAQ units can be used to capture the calibrating signal into the digital domain at ultra-fast rates \cite{FASTADC}. Once the signal $I_{MZI}$ is sampled into the digital domain by the means of ultra-fast A/Ds, the demodulation techniques presented will be applied on the sampled calibrating signal. Below, this dissertation will introduce 4 different newly introduced demodulation techniques each suitable for different applications and compares them against the state of the art Hilbert transformation based demodulation.
\section[Demodulation Based on the Square Law Envelope Detection]{Demodulation Based on the Square Law Envelope Detection }

\begin{figure}[!t]
	\centering
	\includegraphics[scale=0.50]{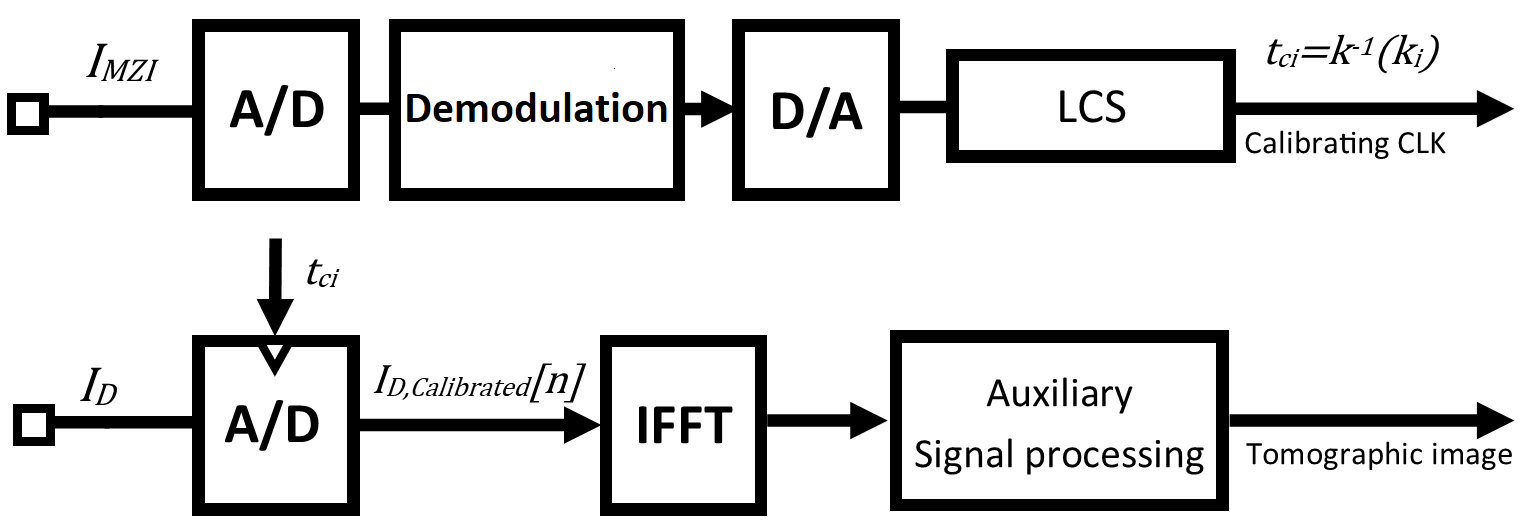}
	\caption{The calibration method proposed in this dissertation. $k_i$ represents wavenumbers in the entire spans of the spectral sweeps of the swept source laser that are equally distanced.}
	\label{demodulation}
\end{figure}

This demodulation technique \cite{biocas} intends to provide a very simple digital arithmetic to acquire the spectral sweep profile ($k[n]$) of a calibrating signal (MZI Signal). Predefined thresholds can be used to set up a level crossing sampler (LCS, will be discussed in detail in Chapter. 4) to calculate non-uniform wavenumber equidistant sampling points. As  Figure \ref{demodulation} shows, these time instances are used to sample the interferometric signal, $I_D$. Let us assume that the entire full scale of $k[n]$ is divided into equidistant instances denoted by $k_i$; $t_{ci}$ then would be the time instance in which $k$ passes $k_i$. The LCS creates a calibrating CLK with positive edges at all instances denoted by $t_{ci}$. It should be noted that the signal $I_D$ contains the reflectivity information of the tissue under test and should be sampled at time instances ($t_{ci}$) at which the wave-number sweep has changed by identical values; hence, the calibrating CLK is used to sample the interferometric signal and produce the signal $I_{D,Calibrated}[n]$. This signal is then taken an inverse Fourier transform on to produce a tomographic image. Further necessary signal processing algorithms can be done to yield a high quality tomographic image. The difference between the calculated time instances can be dynamically changed by software, effectively modifying the number of sampling points that can be calculated from the calibrating signal (MZI). These sampling instances are used to sample the interferometric signal at correct time stamps (calibration process). In the current spectral calibration state of the art, the change in the number of sampling points readily available to sample the interferometric signal can either happen by some physical changes in the MZI (changing the path length in a MZI) in the zero crossing detection approach or by changing the oversampling factor in the interpolation, lookup table, and resampling based technique. Both these approaches are heavily process intensive and in cases might deem impossible. \\
\indent In the demodulation technique based on the squared law envelope detection, the MZI signal will be sampled and it is amplitude will be equalized. By applying a simple square law envelope detection, the envelope of the calibrating signal ($I_{MZI}$) is calculated and the calibrating signal is then divided by the calculated envelope signal.
\subsection{Configuration and Methodology}

\begin{figure}[!t]
	\centering
	\includegraphics[scale=0.45]{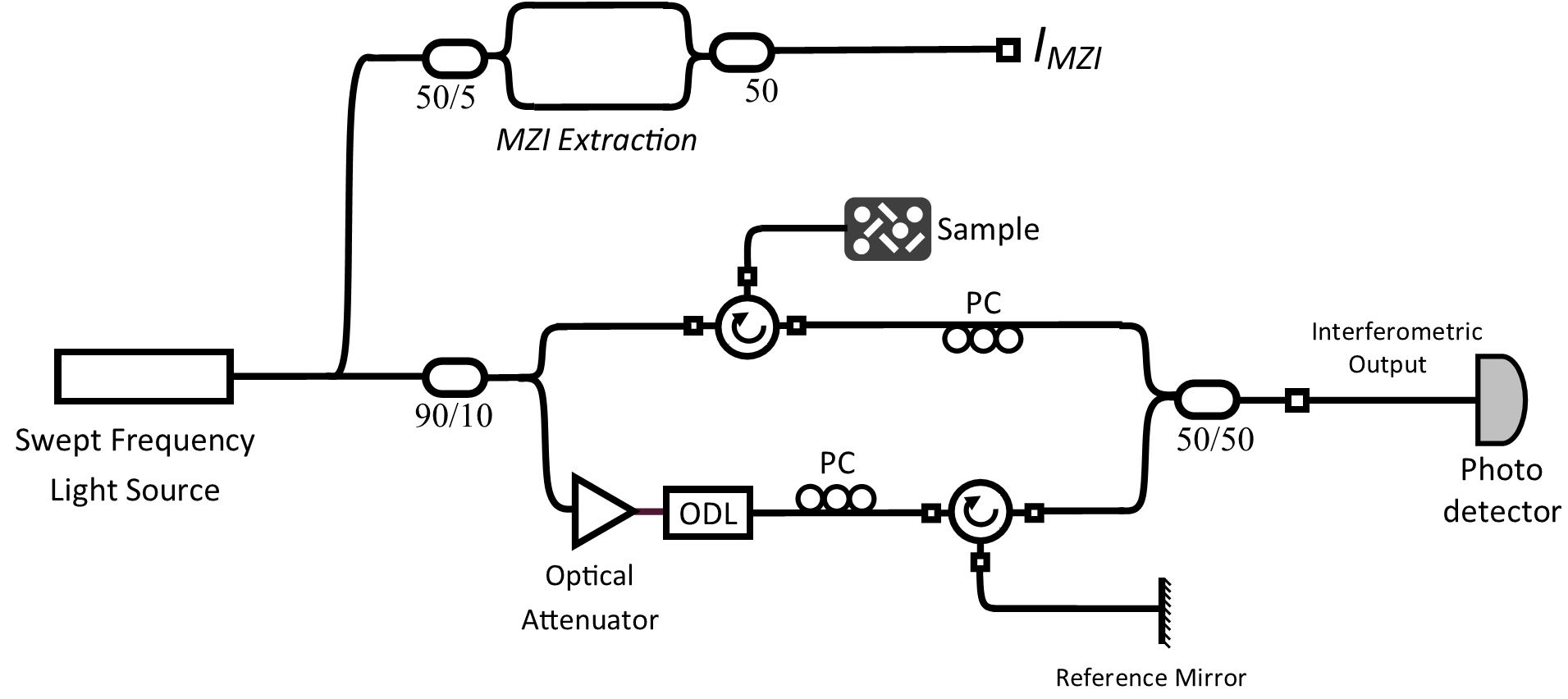}
	\caption{The full optical configuration used to acquire the calibrating and the interferometric signal.}
	\label{configuration}
\end{figure}

The interferometric signal, acquired by the use of a Michelson interferometer, is received via the utilization of a  broadband  Swept  source  (ESS,  Exalos)  that works at center wavelength around 1310 nm  and exhibits a sweeping rate equaling 150-kHz. The calibrating signal ($I_{MZI}$) is acquired using a 14-Bit 250 MS/s data acquisition module (NI5761, National Instruments). As Figure \ref{configuration} shows, a MZI is used in conjunction with the normal Michelson interferometer that passes light through the tissue under test. The MZI signal is used as a calibrating signal; it is sampled as Figure \ref{demodulation} suggests and is used to produce a calibrating clock that is then used to sample the interferometric signal at time instances at which the spectral wavenumber of the swept laser has changed by identical values.
\subsubsection{Amplitude Equalization}
Amplitude equalization is a widely adopted technique that is used in several applications specifically for the demodulation of an amplitude modulated (AM radio) signal. Amplitude demodulation is typically used to detach a low frequency amplitude information from a high frequency carrier it is modulated around. Although different techniques can be used to perform the demodulation, a very simple scheme is utilized in this dissertation that cuts the complexity of the conventional Hilbert transformation by several factors.  \\
\indent Let us assume that the signal $x_{BB}(t)$ is the low frequency (baseband) amplitude signal that is modulated to a high frequency carrier, $Acos(\omega_c t)$. A constant value, $k_a$ determines the degree at which the baseband signal is modulated to the carrier. It also should be assumed that the baseband signal has a frequency spectrum that is bounded to a certain limited frequency, $f_{BB,high}$, which means that $$X_{BB}(f)=0~for~f>f_{BB,high},$$ where $X_{BB}(f)$ is the frequency component, in the Fourier transformation, of the signal $x_{BB}$. It should be noted that a MZI signal typically looks like an amplitude modulated chirp, hence can be written as,
\begin{equation} \label{Equ.3.1}
I_{MZI}(t) = C[1+k_{a}m(t)]\cos{(\omega_{c}t)},
\end{equation}
which can be easily attributed to the theoretical MZI signal in Equation \ref{MZIFormula} where,
\begin{equation} \label{Equ.3.2}
\begin{split}
I_{MZI}(t) = CS(k)&\cos{\big((l_2-l_1)k(t)\big)},\\
CS(k) = &C[1+k_{a}m(t)].
\end{split}
\end{equation}
In the squared law envelope detection based technique, the MZI signal first sampled and then squared in the digital domain,
\begin{equation} \label{Equ.3.3}
I_{MZI}^{2}[n] =\frac{1}{2}A_{c}^{2}[1+k_{a}m[n]]^2 \newline +\frac{1}{2}A_{c}^{2}[1+k_{a}m[n]]^2\cos(2\omega_{c}n),
\end{equation}
As seen in Equation \ref{Equ.3.3}, the first term is a DC term and the squared of the amplitude of the MZI signal to be detected. The second term in Equation \ref{Equ.3.3} is a sinusoidal function located at a carrier frequency at $2\omega_{c}t$. It is obvious that by applying a lowpass filter that rejects the sinusoidal function the DC term can be retrieved.
\begin{equation} \label{Equ.3.4}
LPF\{I_{MZI}^{2}[n]\} =\frac{1}{2}A_{c}^{2}[1+k_{a}m[n]^2.
\end{equation}
The above signal then can be taken a square root upon and the result would be, 
\begin{equation} \label{Equ.3.5}
I_{MZI,Demodulated} =\frac{\sqrt{2}}{2}A_{c}[1+k_{a}m(t)].
\end{equation}
\begin{figure}[t]
\centering
\subfloat{%
\includegraphics[width=0.85\textwidth]{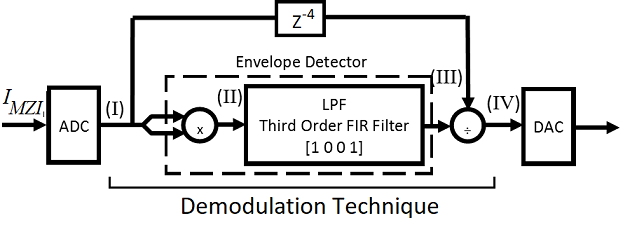}
\label{Envelope1}}
\quad
\subfloat{%
\includegraphics[width=.35\textwidth]{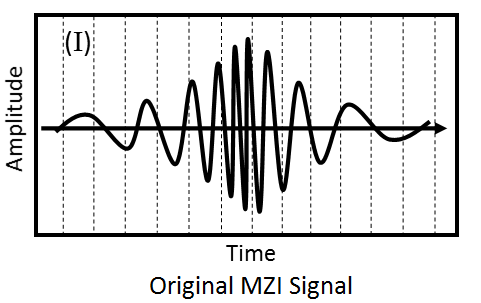}
\label{Envelope2}}
\quad
\subfloat{%
\includegraphics[width=.35\textwidth]{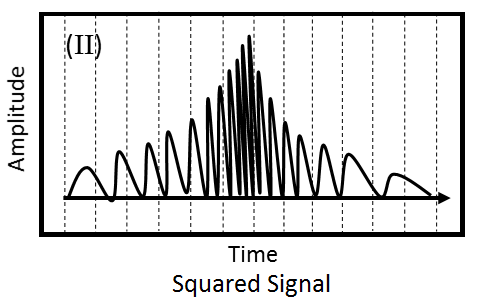}
\label{Envelope3}}
\quad
\subfloat{%
\includegraphics[width=.35\textwidth]{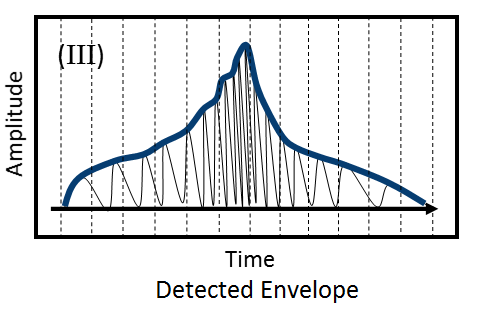}
\label{Envelope4}}
\quad
\subfloat{%
\includegraphics[width=.35\textwidth]{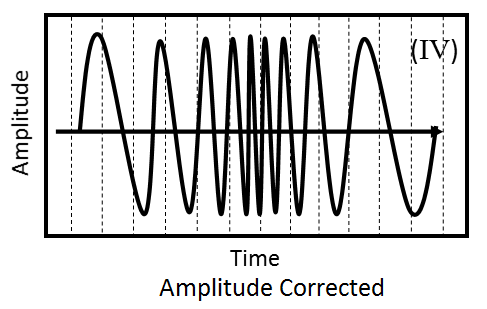}
\label{Envelope5}}
\caption{Envelope detection process from a continuous time perspective. Bottom Figures correspond to points I-IV in the top graphic, respectively.}
\label{EnvelopeDetection}
\end{figure}
\indent The lowpass filter is implemented via a very simple 4-tap FIR filter that cuts the complexity by a factor of more than 4 compared to similar demodulation techniques when implemented by a Hilbert transformation \cite{Hilbert}. This improvement in the complexity of the systems is achieved without any scarification of the axial resolution as the most important specification that quantifies the effectiveness of the demodulation and the spectral calibration. It should be noted that without a calibration process, the tomographic image is in distinguishable and the axial resolution is gravely compromised. Figure \ref{EnvelopeDetection} shows how the envelope detection demodulation technique is embedded within the overall demodulation scheme shown by Figure \ref{demodulation} and how the detected envelope is used to divide the MZI signal to yield an amplitude equalized signal. 
The $z^{-4}$ delay is used to delay compensate for the delay that the envelope detection FIR filter causes. As Figure \ref{demodulation} shows, the equalized amplitude signal is then passed through a digital to analog converter (DAC) to the analog domain. A level crossing sampler is then used to calculate the correct sampling time instances at which the spectral frequency of the laser has changed by identical values. 
\subsubsection{Level Crossing Sampler and Its Application in the Envelope Detection Demodulation Technique} \label{3.1.1.2}
\begin{figure}[t]
\centering
\subfloat{%
\includegraphics[width=0.85\textwidth]{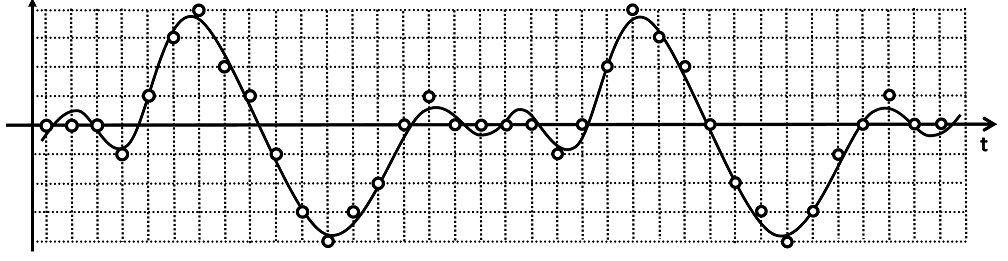}
\label{LCSConcept1}}
\quad
\subfloat{%
\includegraphics[width=.85\textwidth]{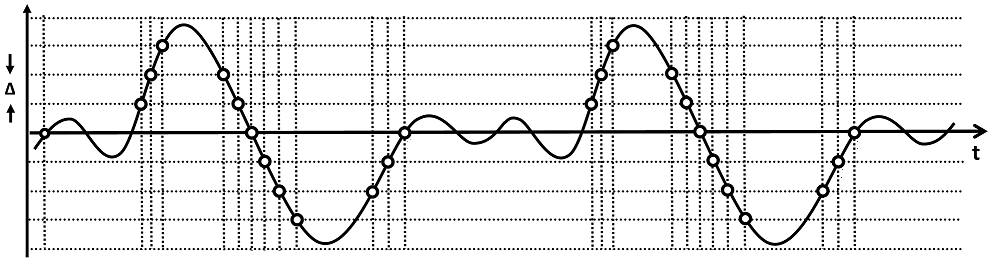}
\label{LCSConcept1}}
\caption{(Top) A generic signal being sampled by a uniformly sampling ADC. (Bottom) A generic signal sampled by a level crossing sampler, LCS. As seen the sampling points are non-uniformly distributed and the sampling pints are when the amplitude of the signal hits predefined values. Reprinted from \cite{tcas1}.}
\label{LCSConcept}
\end{figure}
\begin{figure}[!t]
	\centering
	\includegraphics[scale=0.8]{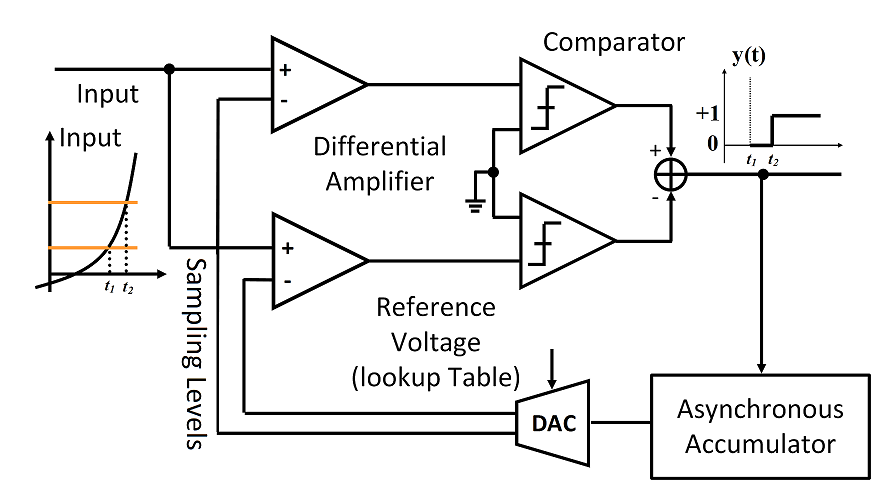}
	\caption{A possible realization of level crossing sampler. Reprinted from \cite{tcas1}.}
	\label{LCS}
\end{figure}

Unlike regular uniformly sampling A/Ds, level crossing samplers employ innovative ways to sample a signal when its amplitude hits a predefined value. Figure \ref{LCSConcept} shows the difference there is between sampling a signal with regular A/D (Figure \ref{LCSConcept}, top) and sampling a generic signal by a LCS. Although In Figure \ref{LCSConcept} (bottom), the predefined amplitude values are equally distanced from one another, the amplitude levels that the sampling act happens at can be non-uniformly distributed across the entire time span of the signal that the LCS is applied on. One realization of a LCS is suggested by Figure \ref{LCS}. The next chapter in this dissertation will introduce a circuit level realization of the above system level introduction. \\
\indent Figure \ref{LCS} is constituted of differential amplifiers with the input signal connected to the positive pin of both, and two sampling levels connected to the negative pins of both differential amplifiers. The two sampling levels produce a window that the input signal always lays in between. If the input hits the top value of the window, the first differential amplifier triggers the comparator and the output clamps to the logic value $+1$. The output then activates the asynchronous accumulator and increases the value it outputs by one logic value. The DAC then decodes the digital value to two analog voltages that re-define the sampling levels that the LCS introduced in Figure \ref{LCS} work with. Once that happens, the output of the LCS falls back to zero as the value of the input falls between the predefined window once it gets updated. In case the input voltage hits the bottom side of the predefined window, the entire LCS acts in the opposite direction and changes the predefined window in the opposite direction. The entire LCS, as shown by Figure \ref{LCS}, produce falling\textbackslash rising edges when the input voltage hits predefined values suggested by a look up table that the decoder (DAC) in Figure \ref{LCS} works with.\\
\indent Once the amplitude of the MZI signal is equalized, a parameter $M_C$ is defined that divides the unit trigonometrical circle that the argument of the MZI signal travels in as the swept source sweeps the value of the spectral frequency, $k(t)$, into $M_C$ equidistant angles.

\begin{figure}[!t]
	\centering
	\includegraphics[scale=1.0]{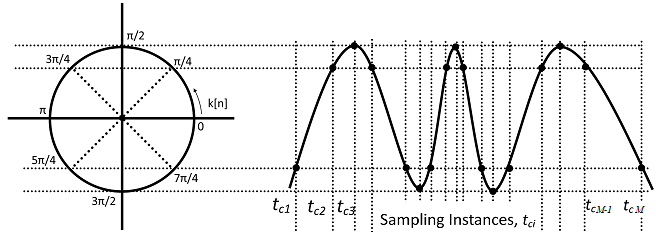}
	\caption{A representation of the LCS is applied on the amplitude equalized MZI signal in the squared law envelope detection based demodulation technique.}
	\label{Samplingenvelope}
\end{figure}

\indent In Figure \ref{Samplingenvelope}, an amplitude equalized MZI signal is represented (right). The trigonometrical circle on the left depicts how it is divided into eight equidistant angles and how those angles correspond to amplitude values on the amplitude equalized MZI signal. It should be clear that the amplitude equalization is necessary to yield a fixed look up table that the encoder (DAC) in the LCS suggested by Figure \ref{LCS} can work with.
\subsection{Results}
\begin{figure}[t]
\centering
\subfloat{
    \includegraphics[width=.45\textwidth]{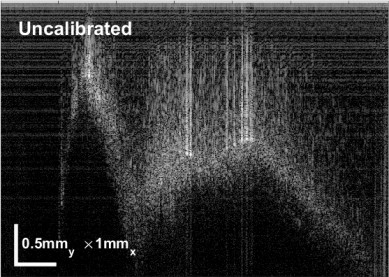}}
\subfloat{
    \includegraphics[width=.45\textwidth]{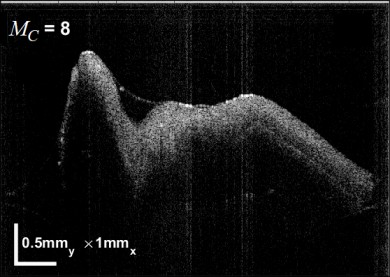}}\\
\subfloat{    
     \includegraphics[width=.45\textwidth]{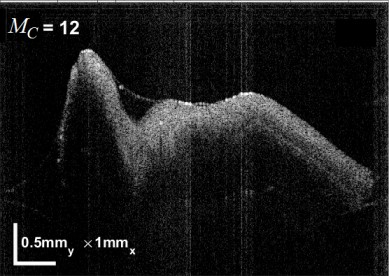}}
\subfloat{
    \includegraphics[width=.45\textwidth]{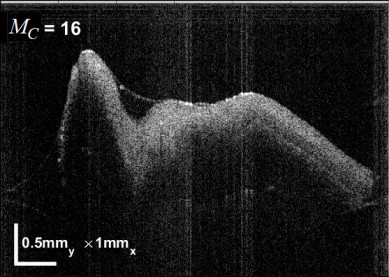}}
    \caption{The results of calibration procees being performed on the quality of images produced. (Top left) Not calibrated. (Top left) Calibrated, $M_C=8$. (Bottom left) Calibrated, $M_C=12$. (Bottom right) Calibrated, $M_C=16$.}
\label{envelopetomographic}  
\end{figure}

The simulations to evaluate the performance of this technique was carried out in SIMULINK. The axial resolution performance was calculated by acquiring 500 A-scan from reflective mirrors at 3 different depth locations. At each depth, the ADC resolution that the MZI signal was sampled by was swept across 4 different values, 8, 10, 12, 14.\\
\begin{figure}[t]
\centering
\subfloat{%
\includegraphics[width=.44\textwidth]{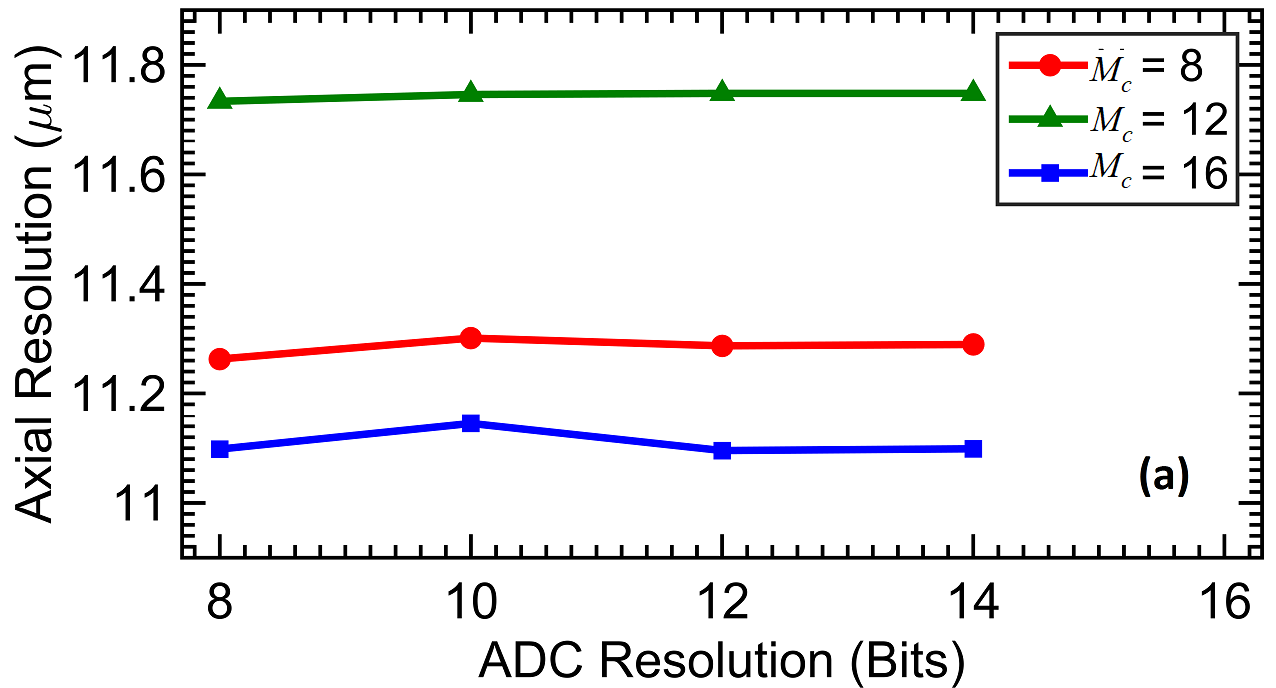}
\label{Envelope2}}
\quad
\subfloat{%
\includegraphics[width=.44\textwidth]{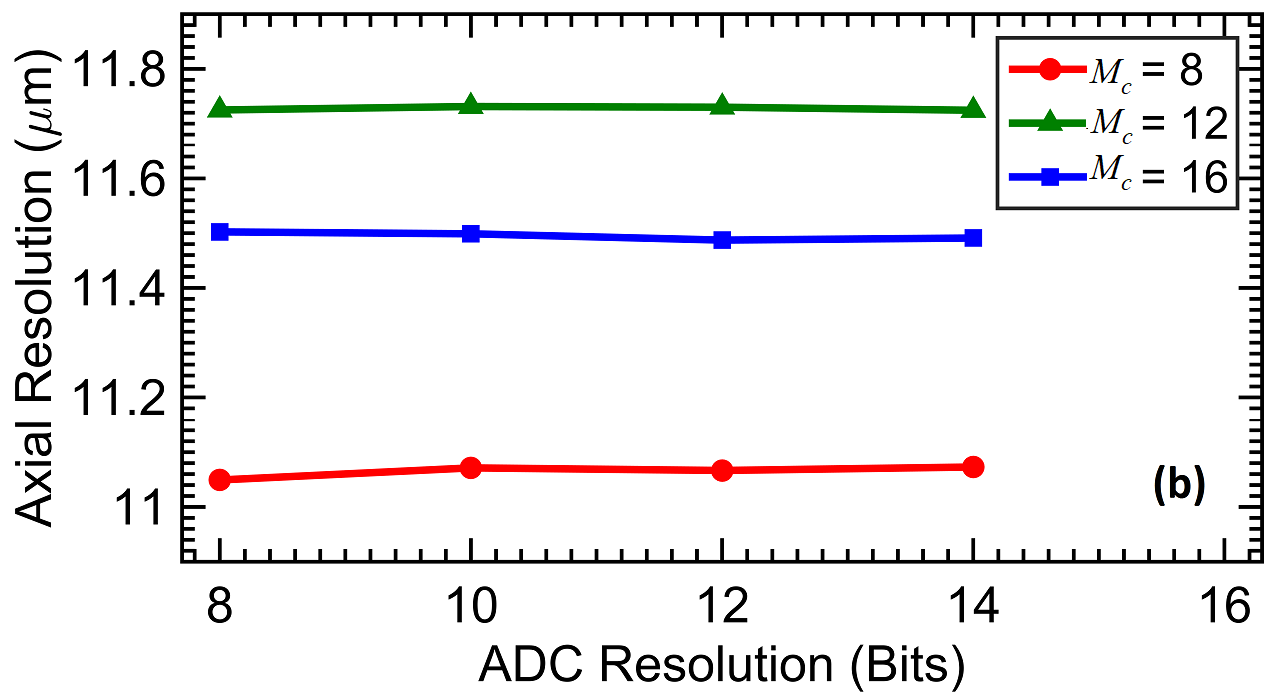}
\label{Envelope3}}
\quad
\subfloat{%
\includegraphics[width=.44\textwidth]{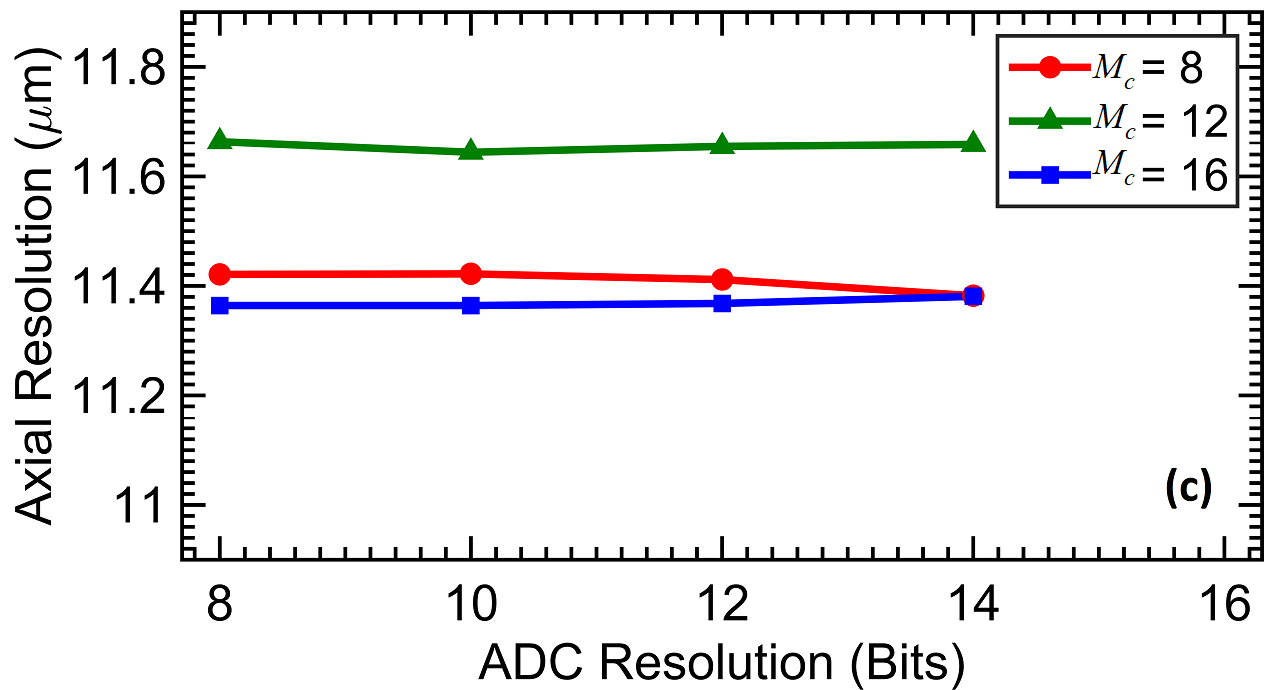}
\label{Envelope4}}
\quad
\subfloat{%
\includegraphics[width=.44\textwidth]{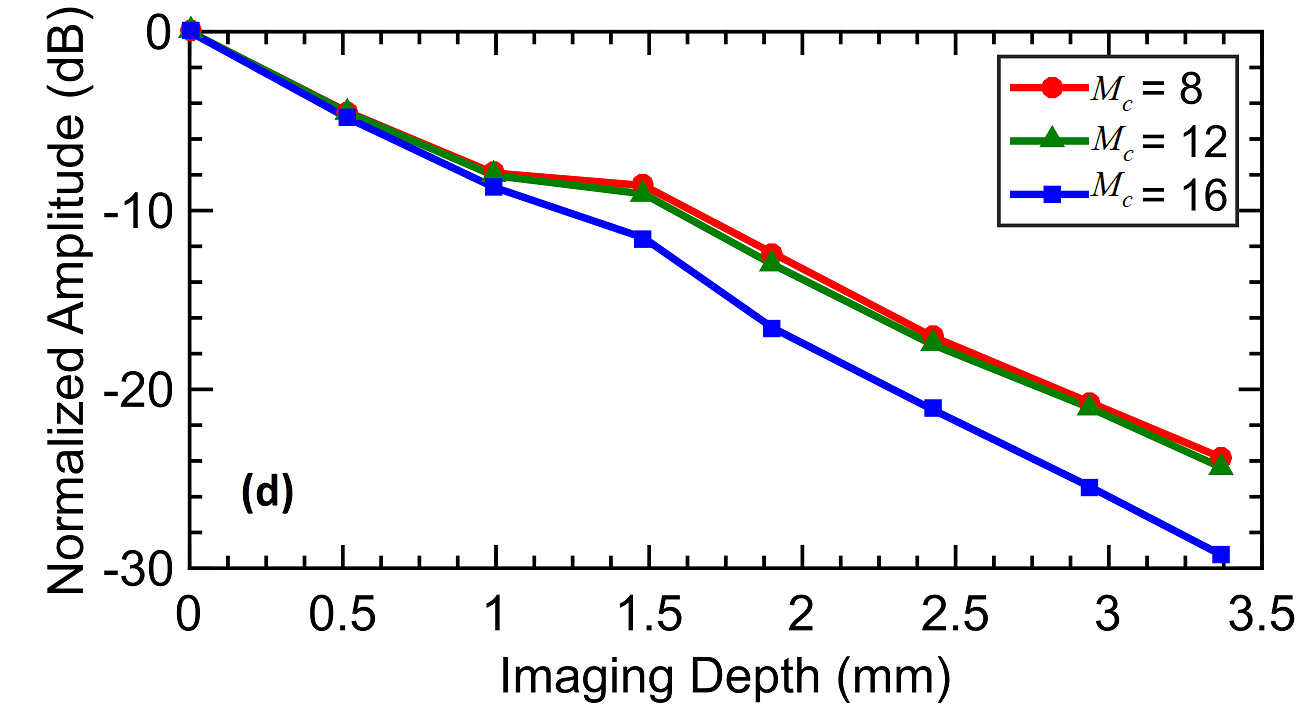}
\label{Envelope5}}
\caption{Axial Resolution for different ADC resolutions sampling the MZI signal. (Top left, top right, bottom left) Axial resolution for three axial depths that the reflective mirror is located at, 500 $\mu$m, 1000 $\mu$m, 1500 $\mu$m, respectively. $M_c$ is the parameter that the trigonometrical circle is divided by. (Bottom right) Sensitivity roll off.}
\label{EnvelopeDetectionResult}
\end{figure}
\indent Results shown in Figure \ref{EnvelopeDetectionResult} show that the axial resolution remains fairly constant across bit resolutions that is attributed to the ADC that samples the MZI signal. This simulation implies that regardless of the resolution of the ADC, the axial resolution can be remained constant. The literature \cite{ADCpower} shows that cutting the resolution of an ADC can reduce the energy it burns by a significant amount. The obtained results show that by reducing the number of bits, the power consumption of the system can be cut significantly.  \\
\indent Another observation that can be drawn from Figure \ref{EnvelopeDetectionResult} is that the axial resolution remains constant regardless of the value of $M_C$ (number of samples per cycle of the MZI signal). By increasing the number of samples per A-scans, images from deeper layers of tissue can be obtained. In this proposed approach, the number of samples per A-scan is defined in software without physical modifications of the Mach-Zehnder Interferometer. These physical modifications are sometimes rather impossible to accomplish. Figure \ref{EnvelopeDetectionResult} (bottom right) shows the sensitivity roll off of the systems for three values of $M_C$.\\
\indent This proposed demodulation technique exhibits much less complexity compared to that of previously reported demodulation work related to Hilbert transformation \cite{Hilbert}. Despite the significant reduction of complexity (17 taps FIR filter in \cite{Hilbert} to 4 taps FIR filter in this proposed algorithm) results are on par with what the Hilbert transformation technique can achieve in terms of axial resolution. 
\begin{table}[!t]   \label{Axialenvelope}
\centering
\caption{Comparison between the axial resolution of the demodulation method based on the envelope detection and spectral calibration based on Hilbert transformation. Proposed Method (Top). Hilbert transformation (Bottom). All quantities are in $\mu$m. 
$[cycle^{-1}]$ indicates the value of $M_C$ and equals the number of sampling time instances that would be available  to sample the interferometric signal from one cycle (period) of the MZI signal.} 
 \label{Axialenvelope}
\begin{tabular}[!h]{cc}
 \label{Axialenvelope}
\begin{tabular}{*4c}    \toprule
\emph{$M_C$} & \multicolumn{3}{c}{\emph{Depth ($\mu$m)}}   \\\midrule
    & 514  & 998  & 1481    \\ 
 8  [$Cycle^{-1}$]   & 11.28 & 11.07 & 11.38 \\ 
	 12 [$Cycle^{-1}$] & 11.74 & 11.72 & 11.68 \\
 16 [$Cycle^{-1}$] & 11.09 & 11.49 & 11.38 \\\bottomrule 
 \hline
 \label{Axialenvelope}
\end{tabular}
\\ \\ 
\begin{tabular}{*4c}    \toprule
\emph{Quasi $M_C$} & \multicolumn{3}{c}{\emph{Depth ($\mu$m)}}   \\\midrule
    & 514  & 998  & 1481    \\ 
 8 [$Cycle^{-1}$]    & 11.10& 11.10 & 11.09 \\ 
12 [$Cycle^{-1}$] & 11.11 & 11.11 & 11.14 \\
 16 [$Cycle^{-1}$] & 11.13 & 11.12 & 11.12 \\\bottomrule 
 \hline
 \label{Axialenvelope}
\end{tabular}
\label{Axialenvelope}
\end{tabular}

\label{Axialenvelope}
\end{table}
As Table. 3.1 shows, the results obtained by the demodulation technique based on envelope detection and those obtained by the Hilbert based demodulation technique are tantamount; hence, the demodulation techniques can be interchangeably used, albeit have the benefit of the complexity reduction when the former is used. \\
\indent This dissertation further shows (Figure \ref{envelopetomographic}) the effect of the calibration process by applying the demodulation process based on envelope detection on the quality of tomographic images if or if not the calibration process is applies.

\subsection{Conclusion}
This proposed technique utilizes an innovative, computationally cost-effective and less energy consuming spectral calibration by introducing an envelope detection demodulation technique and combining it by a system level LCS.  The MZI signal is used as calibrating signal to produce a calibrating clock that can sample the interferometric signal that is resulted from a Michelson interferometer and being passed through the subject under test. While the computational complexity of the technique is significantly lower than those reported by performing Hilbert transformation, the viability of the demodulation technique was validated by measuring the axial resolution resulted from the proposed technique as well as the quality improvement of tomographic images that is achieved after the demodulation and calibration process is being applied.
\section[Demodulation Based on Extended Kalman Filtering, EKF]{Demodulation Based on Extended Kalman Filtering, EKF}

Kalman filtering is often used to estimate the inner state parameters of a state space model only by knowing the output signals they produce. Multiple research in the literature \cite{noiseOCT1,noiseOCT2,noiseOCT3,noiseOCT4} show the importance of reducing the noise level present in multiple applications of OCT devices. Although reducing the noise is always desirable in all signal processing algorithms and processes, the conventional demodulation technique based on applying Hilbert transformation does not denoise the signals that the transformation is applied to by any means. The Kalman filter, on the other hand, relies on a specific signal processing algorithm in which the value of the additive noise on both the calibrating signal and the interferometric signal can be estimated and accounted for. The noise originates from all non-ideal optical and electrical components present in the system. The novel and innovative Kalman based technique proposed \cite{EKF} provides a solution to use lower cost optical and electrical components in SS-OCT devices as the Kalman filter, unlike the Hilbert transformation, can account for the noise present in the system. It should be noted that the Kalman filter is applied to the calibrating signal ($I_{MZI}$), as Figure \ref{demodulation} suggests to produce the wave number sweep, $k[n]$, as a function of time. The Kalman filter denoises the MZI signal and extracts the signal $k[n]$ which is fed to the proceeding block of the system. Clearly, not accounting for this noise disturbs the values of $k[n]$ and makes the calibration process less efficient. This deterioration causes a higher noise level in the Interferometric signal when it is recalibrated by the values obtained for $k[n]$ if the Kalman filter is not applied and conventional Hilbert techniques used instead. 
\subsection{Configuration and Methodology}
The optical configuration shown in Figure \ref{configuration} is used to produce the calibrating signal $I_{MZI}$ and the interferometric signal that entails the reflectivity information of the tissue under test and is produced by using a Michelson interferometer. The MZI signal is then used as Figure \ref{demodulation} shows to produce a calibrating clock that samples the interferometric signal at correct time instances (Figure \ref{demodulation}). As explained before, the MZI signal can be modeled as $I_{MZI} = CS(k)\cos\big((l_2-l_1)k(t)\big)$. The goal of applying a Kalman filter is to extract the parameter $(l_2-l_1)k(t)$ and $k(t)$ as the result to invert the deteriorating effects its nonlinear change causes. 
\subsection{Estimation}
To estimate the argument of the MZI signal as formulated by Equation  \ref{MZIFormula}, an extended Kalman filter is applied. Due to some specific particularities that a MZI signal exhibits, some extra caution need to be taken into account when applying the EKF to the MZI signal. The first particular property of the MZI signal relates to its non-monotonic sweep of the argument of the MZI signal. Most commercially available lasers, sweep the parameter $k(t)$ in a way that the MZI signal exhibits a lower frequency at the beginning and the end of the time span in which the sweep happens. The second challenge relates to the non-constant amplitude of a MZI signal where the amplitude of the signal is low at the beginning and the end and more in the middle in the time span of one scan of the laser. The third relates to the highly noisy nature of the argument of the MZI signal requiring extra care to be taken into account. As stated above, a Kalman filter is applied on a state space model to estimate its inner products knowing the values of the output signals of the state space model.
\subsubsection{State Space Model}
A non linear state space model can be written as,
\begin{equation} \label{nonlinearstatespace}
\begin{split}
   \textbf{X}[n] = f(\textbf{X}[n-1]&,\textbf{N}[n-1]),\\
   \textbf{Y}[n]=h(\textbf{X}[n]&,\textbf{W}[n]),
\end{split}
\end{equation}
where \textbf{X} is the state space matrix, \textbf{Y} is the output, \textbf{N} is the state noise, and \textbf{W} is the output noise.
\subsubsection{State Equations} \label{section3.2.2.1}
Let us define the state matrix \textbf{X} as below,
\begin{equation} \label{stateequation}
  \textbf{X}[n] = \big[A[n] ~~ k[n] ~~ k^{(1)}[n] ~~ k^{(2)}[n]~~ k^{(3)}[n] \big],
\end{equation}
where A[n] represents the changing amplitude of a MZI signal and $k[n]$ the argument of the MZI signal. $k^{(i)}[n]$ represents the $i^{th}$ discrete derivative of the parameter $k[n]$. Based on the previously shown representation of a MZI signal in Equation  \ref{MZIFormula}, a sampled MZI signal can be shown as below,
\begin{equation} \label{sampledMZIKalman}
  y[n] = A[n]cos(k[n])+\textbf{W}[n] = A[n]cos(\sum_{i=l}^{M}b_{i}n^{i})+\textbf{W}[n],
\end{equation}
where $b_i$s are some constant coefficients that represent the polynomial that best fits the shape that the swept laser sweeps the spectral frequency in one scan. $\textbf{W}[n]$ is the noise there is when the MZI signal is being recorded and then sampled into the digital domain. Using the definition of the discrete signal derivative where,
\begin{equation} \label{derivative}
  k^{i}[n]=k^{i-1}[n]-k^{i-1}[n-1],
\end{equation}
the state equation can be written as below,
$$\textbf{X}[n] = \textbf{F}\textbf{X}[n-1] + \textbf{N}[n-1] =$$
\begin{equation} \label{matrixEKF}
  \begin{bmatrix}
    1 & 0 & 0 & 0 & 0\\
    0 & 1 & \frac{1}{1!} & \frac{1}{2!}  & \frac{1}{3!}\\
    0 & 0 & 1 & \frac{1}{1!} & \frac{1}{2!}\\
    0 & 0 & 0 & \frac{1}{1!}  & 1 \\
    0 & 0 & 0 & 0  & 1\\
  \end{bmatrix}\textbf{X}[n-1]+\begin{bmatrix}
    n_{A}[n-1]\\
    n_{k}[n-1] \\
    n_{k^{(1)}}[n-1]\\
    n_{k^{(2)}}[n-1]\\\
    n_{k^{(3)}}[n-1]\\
  \end{bmatrix}\\.
\end{equation}
In the above equation, the amplitude of the MZI signal is modeled as what is known as \textit{random walk}. It is assumed that the value of the amplitude is different between two consequent samples of the MZI signal by a random stationary Gaussian noise the value of which equals $n_A[n]$. The parameter $n_{k}[n]$ is defined to model the noise that of the signal $k[n]$. In Equation  \ref{matrixEKF},
\begin{equation} \label{derivative}
  n_{k^{(i)}}[n]=n_{k^{(i-1)}}[n]-n_{k^{(i-1)}}[n-1].
\end{equation}
\indent Subsequently, the noise covariance matrix can be represented as below,
\begin{equation}  \label{covariancematrix}
{\textbf{V}}=Cov(\textbf{N})=
  \begin{bmatrix}
    \sigma^{2}_{n_{A}} & 0 & 0 & 0 & 0\\
    0 & \sigma^{2}_{n_{k}} & \sigma^{2}_{n_{k}} & \sigma^{2}_{n_{k}} & \sigma^{2}_{n_{k}}\\
    0 & \sigma^{2}_{n_{k}} & 2\sigma^{2}_{n_{k}} & 3\sigma^{2}_{n_{k}} & 4\sigma^{2}_{n_{k}}\\
    0 & \sigma^{2}_{n_{k}} & 3\sigma^{2}_{n_{k}} & 6\sigma^{2}_{n_{k}} & 10\sigma^{2}_{n_{k}} \\
    0 & \sigma^{2}_{n_{k}} & 4\sigma^{2}_{n_{k}} & 10\sigma^{2}_{n_{k}} & 20\sigma^{2}_{n_{k}}\\
  \end{bmatrix},
\end{equation}
where $\sigma^{2}_{n_{k}}$ is the variance of the noise $n_{k}$ and $\sigma^{2}_{n_{A}}$ is the variance of the noise $n_{A}$.
\subsubsection{Observation Equations}
\begin{figure}[t] 
\centering
\subfloat{
    \includegraphics[width=.47\textwidth]{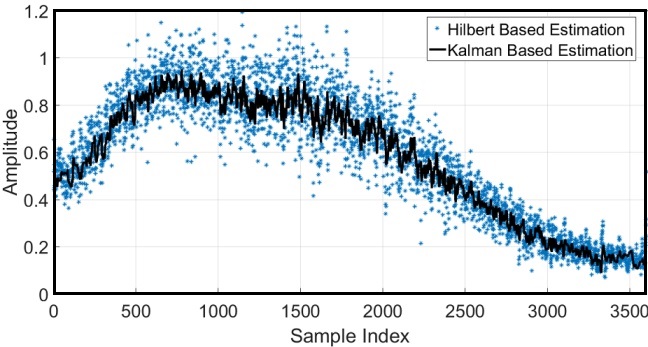}}
\subfloat{
    \includegraphics[width=.48\textwidth]{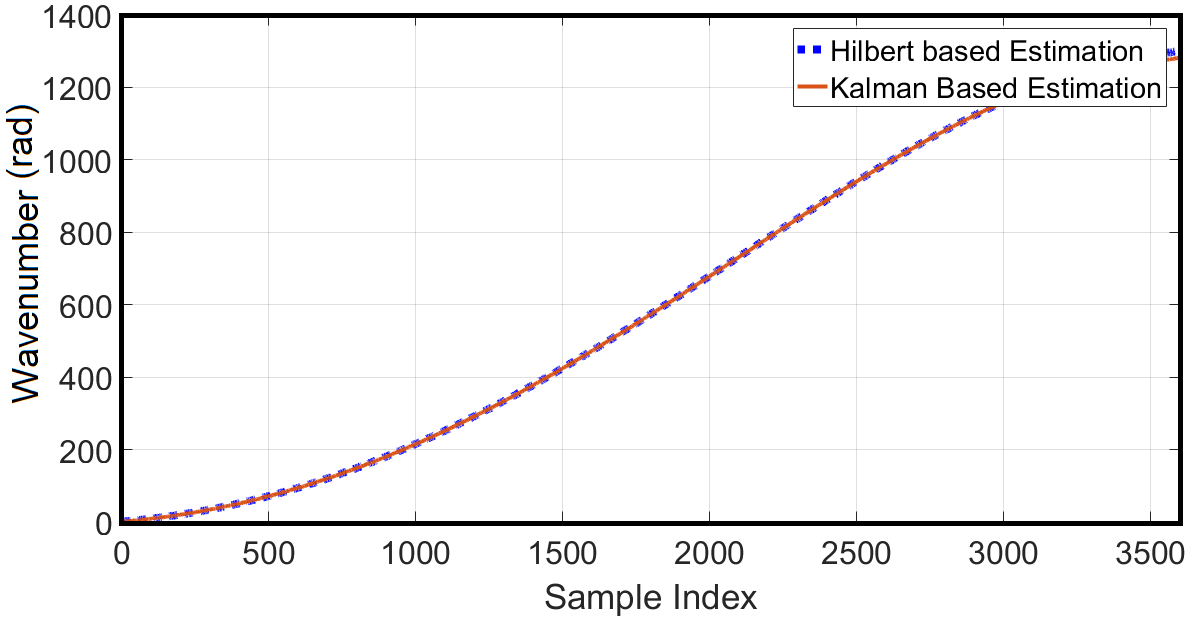}}
  \caption{The estimation of the amplitude and the argument (instantaneous phase) of the MZI signal. (Left) Amplitude parameter extraction. (Right) MZI argument extraction.}
\label{EKFphaseandamplitude}   
\end{figure}

To construct the observation equations, the state vector should be related to the output signal that is measured. It is obvious that the measurable quantity is the value of the calibrating signal ($I_{MZI}$) and not its argument ($(l_2-l_1)k(t)$) nor its amplitude $CS(k)$. Therefore, there needs to be an equation that relates the state vector element to the measurable quantity, $I_{MZI}$. This equation inevitably would be a nonlinear function and can be formulated as below,
\begin{equation} \label{observequat}
  \textbf{Y}[n]=h(\textbf{X}[n],\textbf{W}[n])=x_{1}[n]cos(x_{2}[n])+\textbf{W}[n],
\end{equation}
where $\textbf{W}[n]$ is a Gaussian, stationary, white noise with $E\{\textbf{W}[n]\textbf{W}[n+k]\} =  \dfrac{\sigma_w^2}{2}  \delta[k].$ \\
\indent The main premise of an EKF is to linearize the nonlinear functions denoted by $f$ and $h$ in Equation  \ref{nonlinearstatespace} by the use of a Jacobian matrix. As $f$ is already a linear function, $h$ can be linearized using the equation below,
\begin{equation} \label{Jacobian}
  \textbf{H}[n]=[h_{ij}]_{1\times{5}}=\frac{\partial {h}_i}{\partial x_{j}}\bigg|_{\hat{x}^{-}_{n}}.
\end{equation}
In the equation above, $\hat{x}^{-}_{n}$ is the priori state estimate of the state space vector and used as Algorithm 1 suggests. $\textbf{H}[n]$ can is calculated as follows,
\begin{equation} \label{jacobianHlinearized2}
\begin{bmatrix}
cos(\hat{x}^{-}_{2}[n])   & -\hat{x}^{-}_{1}[n]sin(\hat{x}^{-}_{2}[n])  & 0 & 0 & 0 
\end{bmatrix}.
\end{equation}
Obtaining Equation  \ref{jacobianHlinearized2} completes the construction of the state space model. At this point, Algorithm 1 can be applied to estimate the values of the state space vector including the argument of the MZI signal and as the result, invert the effects that the nonlinear sweep profile of the swept source laser causes. The introduction of the state and measurement noise makes it possible to utilize lower cost optical and electronic components and mitigate the deteriorating effects the noise they introduce causes. Figure \ref{EKFphaseandamplitude} shows how the EKF is capable of estimating the values of the amplitude and the argument of the MZI signal, and $k[n]$ as the result.

\begin{algorithm}[t]
\SetAlgoLined
\KwResult{$\hat{\textbf{x}}_n$ is the estimate of the state matrix $\textbf{X}$.}
Initialization:\\
{$\textbf{\textit{P}}_{0} = {\textbf{\textit{I}}}$~  (Identity~ matrix).}\\
 $\hat{\textbf{\textit{x}}}_{0}$ = Initial~ estimate~ of~ the~ state~ space~ model.\\
\For{n = 1 : length($I_{MZI}$)}{
 Time update:\\
$\hat{\textbf{\textit{x}}}^{-}_{n} = \textbf{\textit{F}}\hat{\textbf{\textit{x}}}_{n-1}$\\

${\textbf{\textit{P}}}^{-}_{n} = \textbf{\textit{F}}\textbf{\textit{P}}_{n-1}\textbf{\textit{F}}^{T}+\textbf{\textit{R}}_\textbf{\textit{V}}$\\
 Measurement Update:\\
$\textbf{\textit{K}}_{n} = {\textbf{\textit{P}}}^{-}_{n}{\textbf{\textit{H}}}^{T}_{n}\bigg(\textbf{\textit{H}}_{n}\textbf{\textit{\textit{P}}}^{-}_{n}\textbf{\textit{H}}_{n}^{T}+\sigma^{2}_{w}\bigg)^{-1}$.\\
 $\hat{\textbf{\textit{x}}}_{n} = \hat{\textbf{\textit{x}}}^{-}_{n} + \textbf{\textit{K}}_{n}(\textit{\textbf{Y}}[n]-\textbf{\textit{Y}}_{n}|_{\hat{\textbf{\textit{x}}}^{-}_{n}}).~~~ \textbf{\textit{Y}}_{n}|_{\hat{\textbf{\textit{x}}}^{-}_{n}} = \textbf{\textit{H}}_{n}\hat{\textbf{\textit{x}}}^{-}_{n}. $ \\
$\textbf{\textit{P}}_{n} = (\textbf{\textit{I}}-\textbf{\textit{K}}_{n}\textbf{\textit{H}}_{n})\textbf{\textit{P}}^{-}_{n}$}
 \caption{Extended Kalman Filter to \textit{k}-estimate the MZI instantaneous phase.}
\end{algorithm}
\subsection{Producing a Calibrating Clock} \label{sec3.2.3}
\begin{figure}[!t]
	\centering
	\includegraphics[scale=1.3]{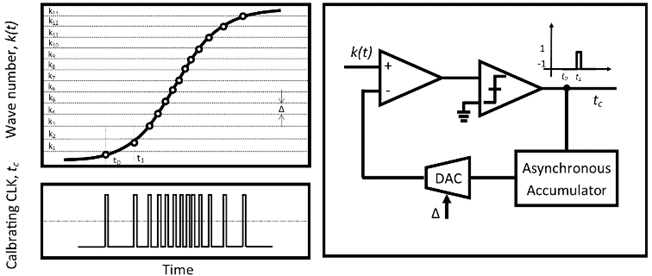}
	\caption{(Left Top) A typical wavenumber sweep caused by the swept source laser. (Left bottom) The calibrating CLK produced. (Right) Circuitry used to implement a level crossing sampler.}
	\label{LCSEKF}
\end{figure}

\indent Once the wavenumber sweep, $k[n]$, is extracted by the Kalman based demodulation technique mentioned before, it is fed to a digital to analog convertor (D/A) (as Figure \ref{demodulation} shows) and its output, $k(t)$, is fed to a level crossing sampler (Figure \ref{demodulation}). This level crossing sampler is comprised of a differential amplifier at the front end. The input of the LCS is connected to the positive pin of the differential amplifier.  Every time the wavenumber sweep, $k(t)$, passes one of the specific predefined values, $[k_1,k_2,k_3,…,k_m ]$, the differential amplifier triggers the comparator shown in Figure \ref{LCSEKF}. The comparator sets the output of the entire LCS to one. Once that happens, the asynchronous accumulator and the DAC, that is shown in \ref{LCSEKF}, change the predefined value of threshold voltage connected to the negative pin of the differential amplifier, i.e., $k_i$ changes to $k_{i+1}$; the output of the LCS drops to zero, producing a pulse every time $k(t)$ passes one of the predefined values $k_i$. For a non-linearly sweeping $k(t)$, these pulses effectively produce a non-uniform calibrating CLK that can be used to sample the interferometric signal at correct time stamps at which the wavenumber has changed linearly. It should be noted that all $k_i$ are equally spaced on the entire span that the swept source laser sweeps the wavenumber at. 
\subsection{Results}
\begin{figure}[t]
\centering
\subfloat{
    \includegraphics[width=.45\textwidth]{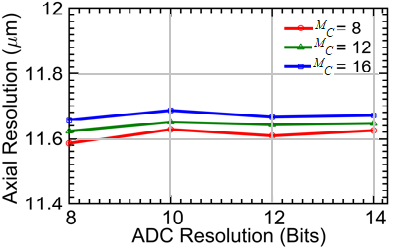}}
\subfloat{
    \includegraphics[width=.45\textwidth]{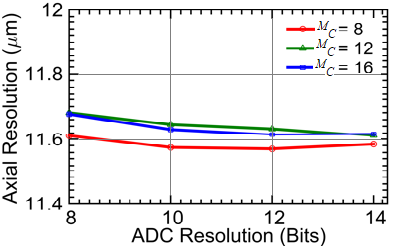}}\\
\subfloat{    
     \includegraphics[width=.45\textwidth]{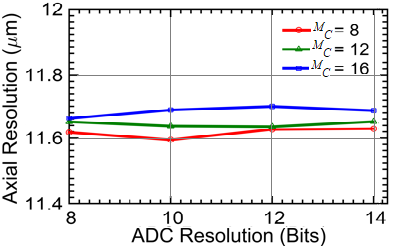}}
\subfloat{
    \includegraphics[width=.45\textwidth]{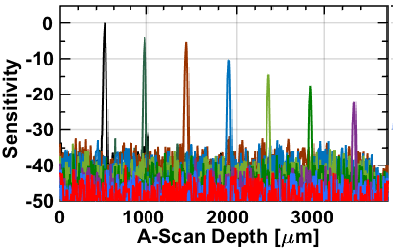}}
    \caption{Axial Resolution measurments for three different axial dpeths of the subject under test (Top left) Axial Depth 1448$\mu$m, (Top right) Axial Depth 1921$\mu$m, (Top left) Axial Depth 2361$\mu$m, across 500 realizations as a function of ADC Resolution (8bits-14bits) and N (Bottom right): sensitivity roll-off ($20log_{10},dB$) across all depths.}
\label{EKFResults}  
\end{figure}
\begin{figure}[t]
\centering
\subfloat{
    \includegraphics[width=.43\textwidth]{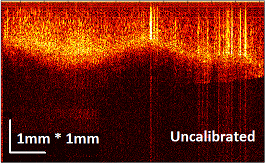}}
\subfloat{
    \includegraphics[width=.43\textwidth]{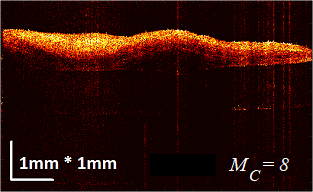}}\\
\subfloat{    
     \includegraphics[width=.43\textwidth]{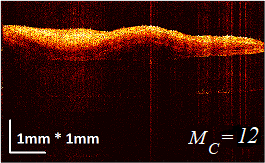}}
\subfloat{
    \includegraphics[width=.43\textwidth]{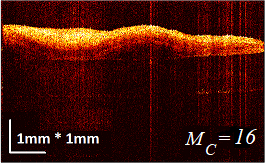}}
    \caption{The quality of the tomographic image produced by the proposed calibration technique based on EKF. Minimal change is observed when $M_c$ is changed}
\label{tomogEKF}  
\end{figure}

The performance of the proposed demodulation and clock extraction technique is evaluated by applying the technique on 500 different A-scans acquired from a mirror that is used as the sample under test. Similar to the previously reported demodulation technique (envelope based), the resolution of the A/D that samples the MZI signal sweeps across four different values, 8, 10, 12, 14. Similar to the previous demodulation technique, the parameter $\Delta$ (shown in Figure \ref{LCSEKF}) changes so that the number of sampling points calculated per each cycle of the MZI signal is changed. Similar to the envelope detection based demodulation technique, the numbers picked for the number of samples per cycles were chosen to be 8, 12, 16. Similar to the previous method, parameter $M_C$ denotes the description above. Figure \ref{EKFResults} shows the results for the experiments explained above. As observed, the axial resolution is on par with those reported in the previous section and to those that could be achieved by the Hilbert transformation based technique (Figure \ref{EKFResults}). Figure \ref{EKFResults}, (bottom right) shows the sensitivity roll off that was observed in this technique. \\
\indent Further, this dissertation examines the results of performing the calibration process based on the EKF demodulation technique and the LCS on the quality of the tomographic images produced. Figure \ref{tomogEKF} shows that the choice of $M_C$ has minimal effect of the axial resolution , and the quality of image as a result. However increasing the number $M_C$ can result in advantages such as increased imaging range as explained before. 

\subsection{Conclusion}
In this section, a newly developed demodulation technique is presented. In this technique, an EKF is applied on the calibrating signal to extract the argument of the MZI signal and as spectral sweep profile of the swept source laser used in the SS-OCT configuration. The acquired sweep profile then drives a LCS to produce a calibrating clock to sample the interferometric signal at correct time instances. The benefit of using an EKF relates to its ability to estimate, account for, and mitigate the unwanted effects it cause. Therefore, letting the designer use lower cost electronic and optical devices that potentially introduce more noise. 
\section[Demodulation Based on Unscented Kalman Filtering, UKF]{Demodulation Based on Unscented Kalman Filtering, UKF }

\indent In this section, another version of Kalman filtering, unscented Kalman filtring (UKF), is used in the context of SS-OCT spectral calibration \cite{UKF}. Similar to the EKF, this technique estimates the argument of the calibrating signal (MZI) by only measuring the value of the same signal. Compared to the EKF, the UKF exhibits a definite improvement in terms of the capability of hardware implementation. The UKF works more robustly for more non-linear spectral sweeps of the laser making it more suitable to be used for a wider application of SS-OCT systems. By utilizing this well-established techniques in innovative ways, in the context of SS-OCT spectral calibration, these devices can quickly move into a more affordable imaging modality able to reach an ever wider audience, by allowing device manufacturers to use more cost effective components that potentially suffer from more non-idealities such as noise.
\subsection{Configuration and Methodology}
The optical configuration used in this demodulation technique is similar to the two previously reported demodulation technique in this dissertation. The calibrating signal ($I_{MZI}$) is again sampled into the digital domain so that the UKF can be applied on it. The interferometric signal as well is produced via the means of a Michelson interferometer as explained in previous sections of this dissertation. Similar to the EKF, the UKF is used to extract the argument of the MZI signal and the spectral sweep profile of the swept source laser as the result. This sweep profile then drives a LCS to produce a calibrating clock to sample the interferometric signal at the correct time instances (Figure \ref{demodulation}).  
\subsection{Estimation}
\subsubsection{State Space Model}
Similar to the EKF, the UKF is capable of estimating the inner states of a state space model by only measuring its output signals. Similar to the EKF, a non linear state space mode (Equation  \ref{nonlinearstatespace}) is used to model the process.
\subsubsection{State Equations}
Similar to the EKF, the MZI signal in this approach can be modeled as below, 
\begin{equation} \label{sampledMZIUKalman}
  y[n] = A[n]cos(k[n])+\textbf{W}[n],
\end{equation}
where the parameters above are those represented in the previous sections of this dissertation (Section \ref{section3.2.2.1}). Due to the specific shape that the spectral sweep of the laser exhibits, $k[n]$ can be modeled as below,
\begin{equation} \label{phiUKF}
k[n] = a_3(nT)^3 + a_2(nT)^2 + a_1(nT)^1,
\end{equation}
where $T$ is the sampling period that the MZI is being sampled with and $a_3$, $a_2$, and $a_1$ are some constant coefficients to represent the polynomial that best fits the spectral  sweep of the the swept source laser. The parameters $a_{1-3}$ as well as the amplitude of the MZI signal constitute the state vector,
\begin{equation} \label{statevectorUKF}
\textbf{X}[n] = \big[A[n]~~~ a_3[n]~~~ a_2[n]~~~ a_1[n]\big].
\end{equation}
Obviously, the state space dimension equals four and is the length of the matrix in Equation  \ref{statevectorUKF}.
Similar to the EKF, the amplitude is modeled by a random walk where it is assumed to have a constant amplitude that changes by a white Gaussian stationary noise $n_{A}[n]$ from one sample to the other. $a_{1-3}$ being constant, the state equation can be modeled as below,
\begin{equation} \label{stateequU}
\textbf{X}[n] = \textbf{I}_{N}\textbf{X}[n-1] + \textbf{G}\textbf{N}[n-1],
\end{equation}
where $\textbf{I}_N$ is the identity matrix and \textbf{N} is the state noise as explained in previous sections of this dissertations. It is assumed that the variance of the amplitude noise is denoted by $\sigma^{2}_A$. The noise on $a_{1-3}$ are modeled by a Gaussian white stationary noise the variance of which can be denoted by $\sigma^{2}_{a_1}$, $\sigma^{2}_{a_2}$, and $\sigma^{2}_{a_3}$, respectively. It is obvious that these noises are uncorrelated and independent and the state noise variance can be modeled as,
\begin{equation} \label{covariancematrixU}
\begin{split}
\textbf{N}[n]&= [n_A[n] ~~~n_{a_{3}}[n] ~~~n_{a_{2}}[n] ~~~n_{a_{1}}[n]],
 \\
  \textbf{V} &= diag\{\sigma^2_{n_{A}}~~~\sigma^2_{a_{3}}~~~\sigma^2_{a_{2}}~~~\sigma^2_{a_{1}}\}.\\
 \end{split}
\end{equation}
In Equation.\ref{covariancematrixU}, $diag\{.\}$ represents a diagonal matrix with $\{.\}$ being its diagonal.  
\subsubsection{Observation Equations}
The observed MZI signal then can be formulated as below,
\begin{equation} \label{observationUkal}
\textbf{y}[n]= I_{MZI}= h(\textbf{X}[n],\textbf{W}[n])= \textbf{X}_1[n]\bigg[sin\bigg(\textbf{X}_3[n](nT)^3+\textbf{X}_2[n](nT)^2+\textbf{X}_1[n](nT)^1\bigg)\bigg]
\end{equation}

\begin{figure}[!t]
	\centering
	\includegraphics[scale=0.80]{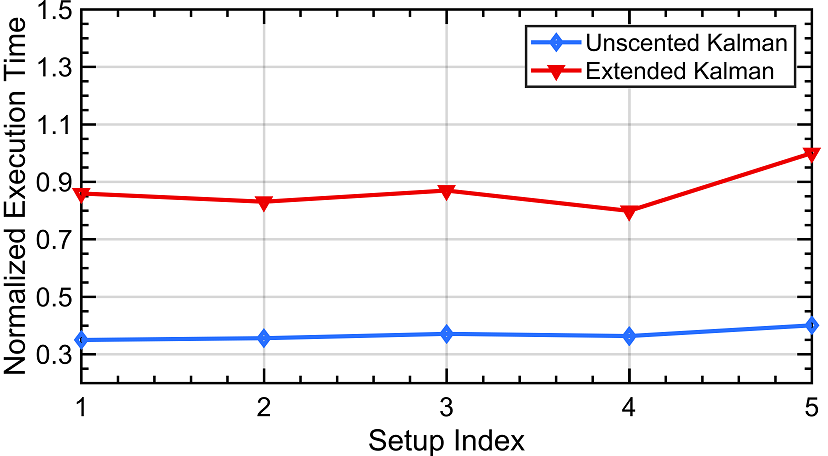}
	\caption{The Execution time comparison between the EKF and the UKF for five different data sets extricating the MZI argument.}
	\label{EKFVUKF}
\end{figure}

\indent Developing equation above, Algorithm 2 can be applied to extract the argument of the MZI signal.\\ 
\indent There are a multitude of benefits UKF has over the EKF presented in previous sections of this dissertations. It can be proven that while the EKF is able to capture the non-linearities of random processes up to the first order, the UKF is able to capture it up to the third order.
\begin{algorithm}[t]
\SetAlgoLined
\caption{The Unscented Kalman Filter (UKF) used for spectral calibration of SS-OCT systems.}
 Initialization:\\
 $\textbf{\^{X}}_0 = \mathbb{E}\{\textbf{X}_0\}, ~ \textbf{P}_{X_0} =  \mathbb{E}\{(\textbf{X}_0-\textbf{\^{X}}_0)(\textbf{X}_0-\textbf{\^{X}}_0)^T\}$ \\
\While{$n \le length ~(I_{MZI}[n])$}{
1) Sigma Point Calculations:\\
$\boldsymbol\chi_{n-1} = [\textbf{\^{X}}_{n-1} ~~~ \textbf{\^{X}}_{n-1}+\gamma\sqrt{\textbf{P}_{n-1}}~~~\textbf{\^{X}}_{n-1}-\gamma\sqrt{\textbf{P}_{n-1}}]$\\
2)	Time update equations:\\
$\boldsymbol\chi_{n|n-1} = f(\boldsymbol\chi_{n-1})$\\
$\textbf{\^{X}}_{n} = \sum_{i=0}^{2L} {w_{i}^{(m)}\boldsymbol\chi_{i,n|n-1}}$\\
$\textbf{P}_{\textbf{X}_n}^{-} = \sum_{i=0}^{2L} {w_{i}^{(c)}(\boldsymbol\chi_{i,n|n-1}}-\textbf{\^{X}}_n)(\boldsymbol\chi_{i,n|n-1}-\textbf{\^{X}}_n)^T+\textbf{R}_v$\\
3) Measurement update equations:\\
$y_{n|n-1} = h(\boldsymbol\chi_{n-1})$\\
$\textbf{\^{y}}_{n} = \sum_{i=0}^{2L} {w_{i}^{(m)}y_{i,n|n-1}}$\\
$\textbf{P}_{\textbf{y}_n} = \sum_{i=0}^{2L} {w_{i}^{(c)}(y_{i,n|n-1}}-\textbf{\^{y}}_n)(y_{i,n|n-1}-\textbf{\^{y}}_n)^T$\\
$\textbf{P}_{\textbf{xy}_n} = \sum_{i=0}^{2L} {w_{i}^{(c)}(\boldsymbol\chi_{i,n|n-1}}-\textbf{\^{X}}_n)(y_{i,n|n-1}-\textbf{\^{y}}_n)^T$\\
$\textbf{K}_n = \textbf{P}_{\textbf{xy}_n} \textbf{P}_{y_n}^{-1}$\\
$\textbf{\^{X}}_{n} = \textbf{\^{X}}_{n}^{-} + \textbf{K}_n(I_{MZI}[n]-\textbf{\^{y}}_{n})$\\
$\textbf{P}_{\textbf{\^{X}}_n} = \textbf{P}_{\textbf{X}_{n}}^{-} - \textbf{K}_n\textbf{P}_{{y}_{n}}\textbf{K}_n^{T}$\\
4) Index update\\
$n= n+1$
}
\end{algorithm}
The ability of the UKF to capture higher degrees of non-linearity, make this approach much more suitable to be used for systems in which lower priced lasers with higher degrees of non-linear spectral sweep are used. Further, the UKF, unlike the EKF, does not need the calculation of a Jacobian matrix of the non-linear functions at every time stamps which tends to be erroneous when dealing with these algorithms. Moreover, the UKF presented in this section has reduced the problem dimension compared to the previously reported EKF. As Figure \ref{EKFVUKF} shows, the calibration execution time is cut by half when the UKF is used compared to the same when an EKF is used, making the UKF more suitable for the hardware implementation in real time applications.\\ 
\indent This dissertation uses the scaled UKF to ensure that the posteriori covariance matrix ($\textbf{P}_k$) remains positive definite at all time. Algorithm 2 is used to estimate the state vector ($\textbf{X}$). The estimation is denoted by $\textbf{\^{X}}$. In this algorithm, $w_0^{(m)}=\frac{\lambda}{\lambda+L}$, $w_0^{(c)}=\frac{\lambda}{\lambda+L}+(1-\alpha^{2}+\beta)$, $w_i^{(m)} = w_i^{(c)} = \frac{1}{2(L+\lambda)}$, $\lambda = \alpha^{2}(L+\kappa)-L$ and $\gamma = \sqrt{L+\lambda}$, where $L$ is the dimension of the state space model, $\alpha$ is an arbitrarily positive small number. $\beta$, and $\kappa$ are positive numbers. For Gaussian random variables, the optimum $\beta$ is set at two. It also can be shown that $\textbf{P}_k = \sqrt{\textbf{P}_k}\sqrt{\textbf{P}_k}^T$, where  $\sqrt{\textbf{P}_k}$ is the lower triangular matrix of the Cholesky decomposition applied on the matrix $\textbf{P}_k$. $\textbf{P}_{X_0}$ is the initial state posteriori covariance matrix and is assumed to be an identical matrix. $\textbf{X}_0$ is the initial estimation of the state matrix.
\subsection{Producing a Calibrating Clock}
Once the parameter $k[n]$ is extracted, the calibrating clock is retrievied as exactly suggested by Section \ref{sec3.2.3}.
\subsection{Results}
Once Algorithm. 2 is applied on the MZI signal, the state vector is estimated and the values of $A[n]$ and $a_{1-3}$ are found. Figure \ref{UKFResults} shows that the values of $a_{1-3}$ converge to steady values and the spectral frequency sweep of the SS-OCT system, as suggested by Equation  \ref{phiUKF} can be calculated. Figure \ref{UKFResults} shows that the value of the MZI argument (phase) is accurately estimated using the UKF. 

\begin{figure}[t]
\centering
\subfloat{
    \includegraphics[width=.47\textwidth]{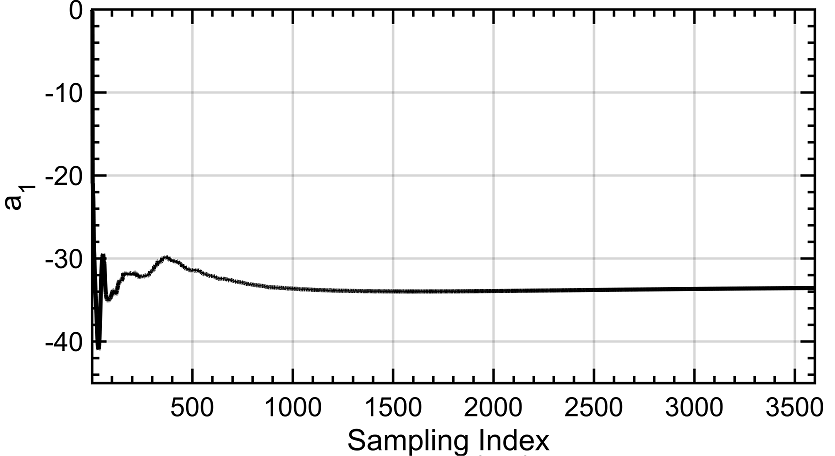}}
\subfloat{
    \includegraphics[width=.47\textwidth]{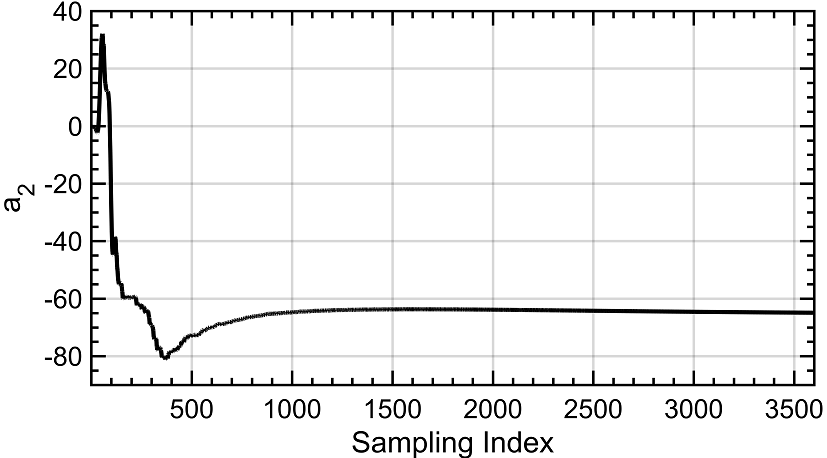}}\\
\subfloat{    
     \includegraphics[width=.47\textwidth]{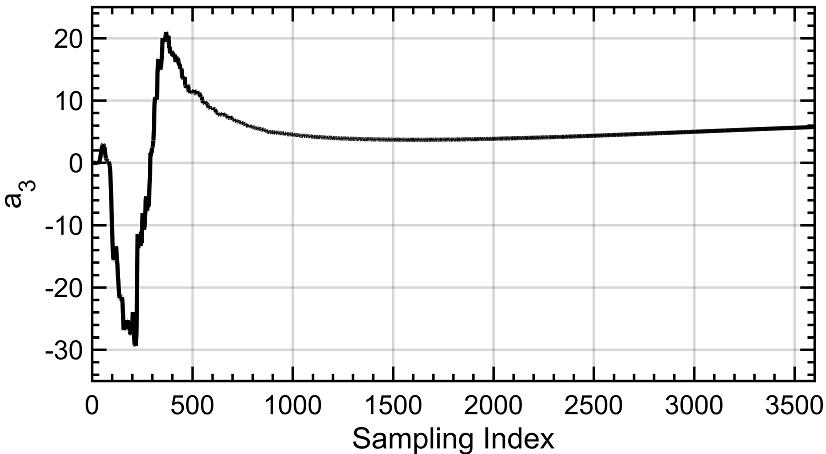}}
\subfloat{
    \includegraphics[width=.47\textwidth]{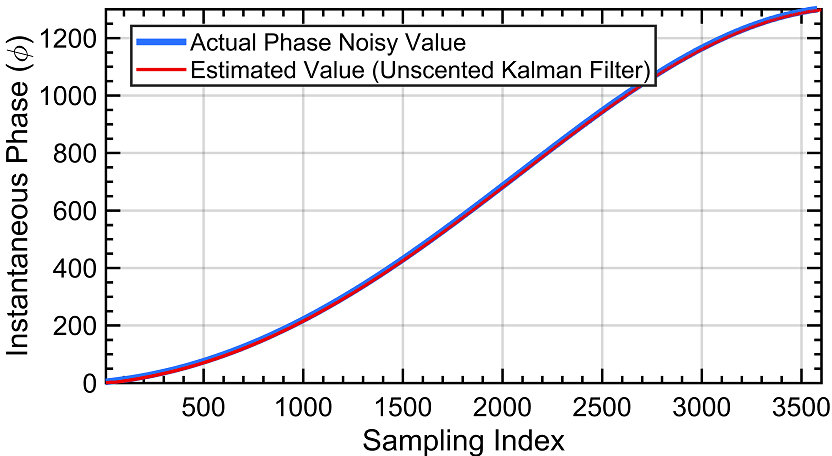}}
    \caption{The performance evaluation of the UKF algorithm. (Top left, top right, bottom left) The convergence of the parameters $a_{1-3}$. (Bottom Right) Value Estimation of the phase (argument) of the MZI signal.}
\label{UKFResults}  
\end{figure}

\begin{figure}[t]
\centering
\subfloat{
    \includegraphics[width=.47\textwidth]{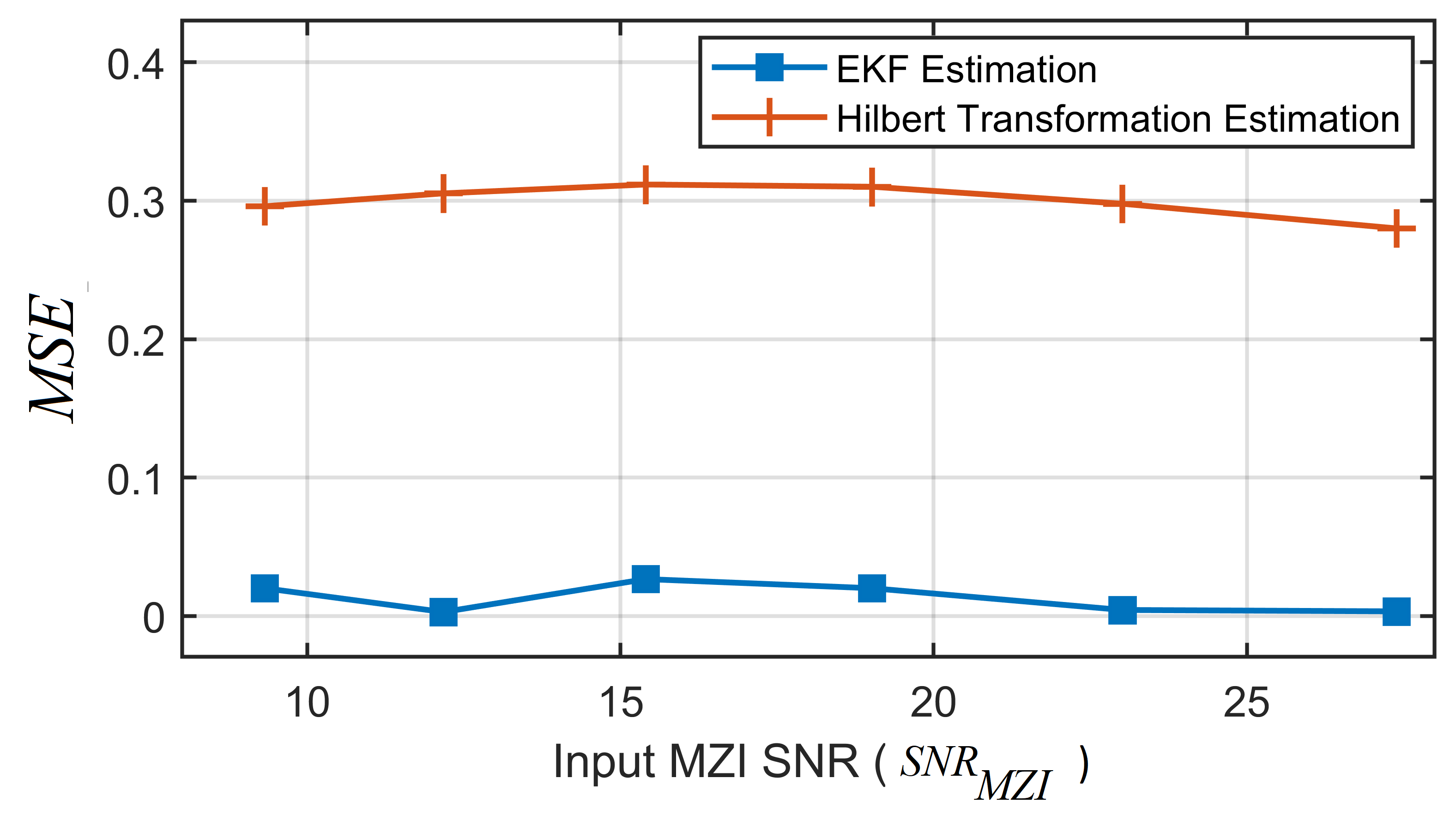}}
\subfloat{
    \includegraphics[width=.47\textwidth]{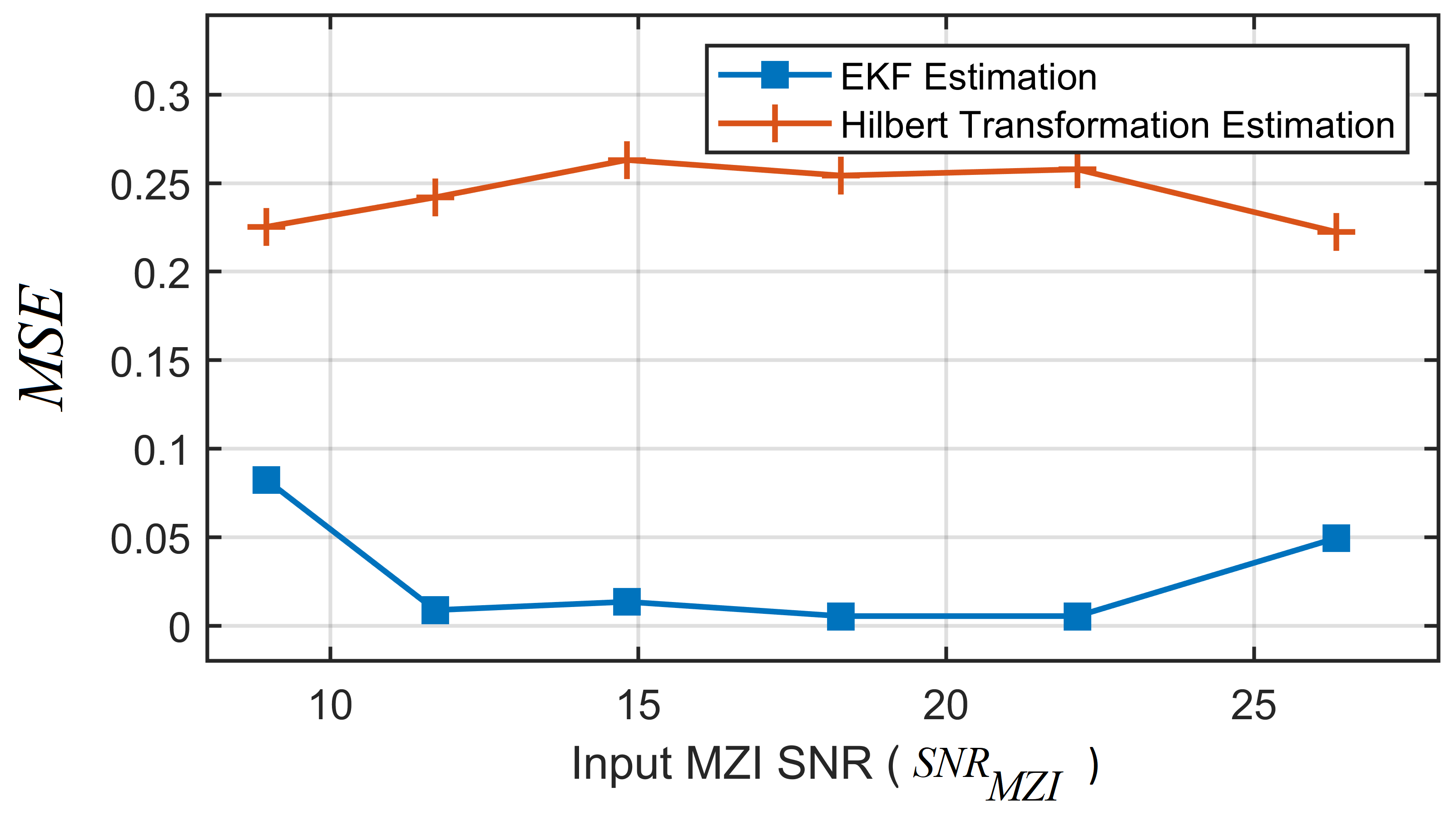}}\\
\subfloat{    
     \includegraphics[width=.47\textwidth]{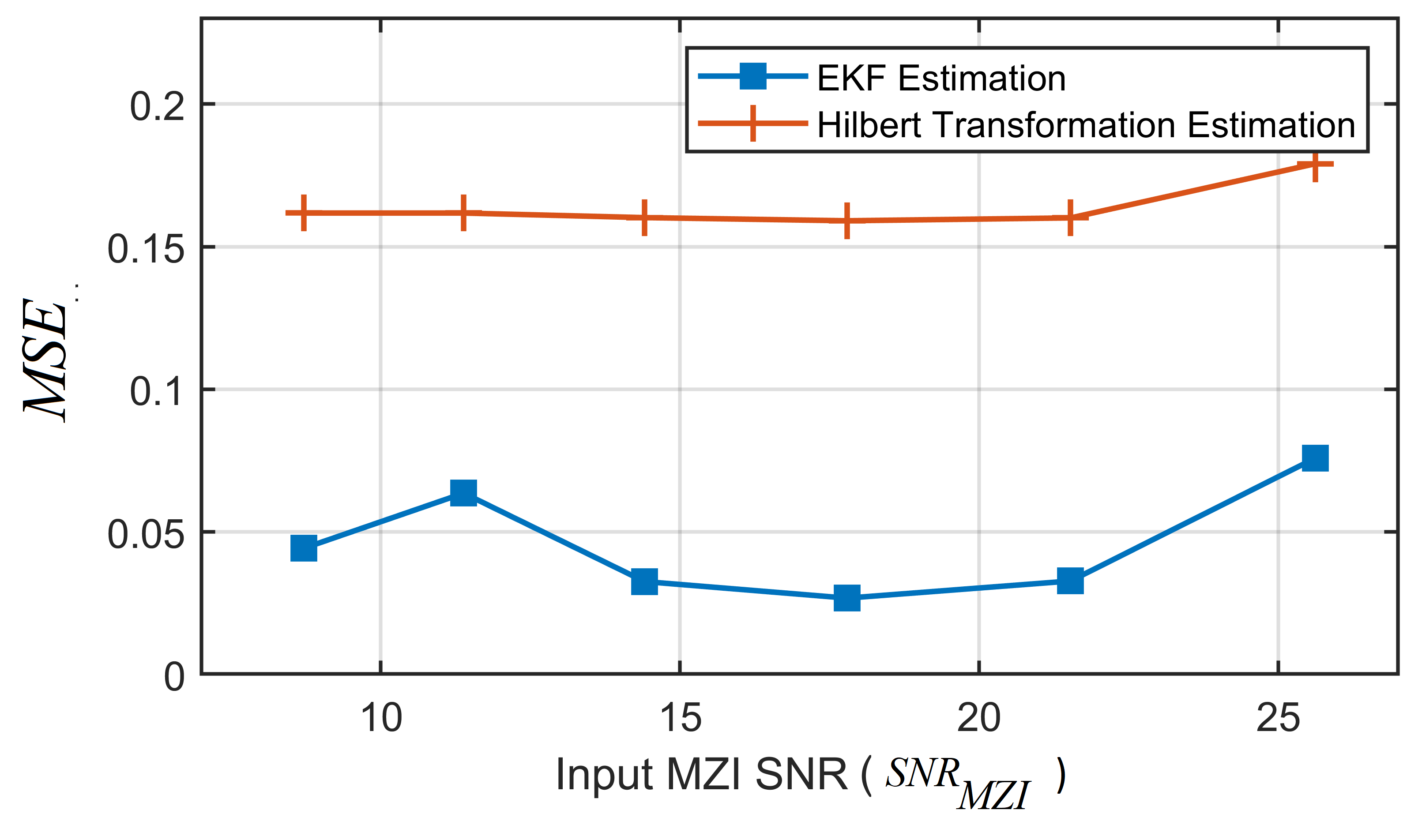}}
    \caption{Interferometric mean square error comparison. (Top left) the mirror depth is 500 $\mu$m, noise is added to the MZI signal. (Top right) The mirror depth is 1000 $\mu$m, noise is added to the MZI signal. (Bottom) The mirror depth is 1500 $\mu$m, noise is added to the MZI signal.}
\label{SNRplots}  
\end{figure}

\indent Unlike the EKF, the elements within the state space vector defined in the UKF converge to constant values. Such property makes the elements of all matrices in Algorithm 2 more homogeneous giving the designer the ability to pick a shorter bit width when implementing the algorithm. Such bit width reduction gives the hardware implementation a significant advantage in terms of the complexity, throughput and amount of digital resources that need to be consumed. The above mentioned advantage along side the inherent advantages of the UKF such as the reduction of the state dimension from five to four gives the UKF a clear advantage over the EKF.\\
\indent Finally, the effect of the calibration process is examined in terms of the axial resolution and tomographic image quality enhancement (Fig. \ref{UKFResults}).\\
\begin{figure}[t]
\centering
\subfloat{
    \includegraphics[width=.4\textwidth]{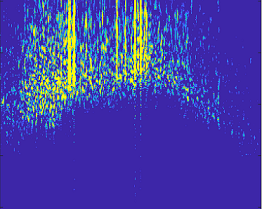}}
\subfloat{
    \includegraphics[width=.4\textwidth]{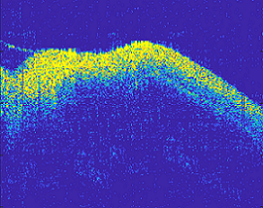}}\\
\subfloat{    
     \includegraphics[width=.65\textwidth]{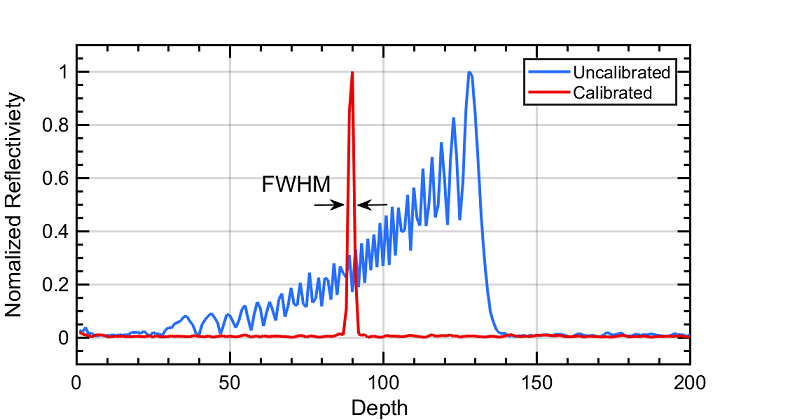}}
    \caption{The performance evaluation of the UKF algorithm. (Top left) Tomographic image without the calibration process being applied. (Top right) Tomographic image with the calibration process being applied. (Bottom) Axial resolution enhancement in the FWHM sense.}
\label{UKFResults}  
\end{figure}
\indent The results of the applying the Kalman filtering compared to its counterpart Hilbert transformation based calibration was evaluated in noisy environments. 
 In the first simulation, the viability of the denoising and the argument extraction is tested by applying the EKF. White Gaussian noise with different variances ($\sigma^2_{N_{MZI}}$) are added to the MZI signal and the extended Kalman filter is used for the k parameter demodulation. The EKF is applied on the noisy input MZI signal ($I_{MZI}+W_{MZI}[n],~var(W_{MZI}[n])= \sigma^2_{N_{MZI}}$) to extract $k[n]$. The extracted $k[n]$ is then passed through a DAC and a level crossing sampler to create the non-uniform calibrating clock to sample the interferometric signal at the correct time instances. For a uniformly reflective optical surface (mirror) under test, the interferometric $MSE$ is simulated when the noise is added only to the MZI signal once when the EKF is applied and once when it is not and a Hilbert transformation based \textit{k}-estimation is used instead. Simulation proves that the application of the Kalman Filter can significantly enhance the interferometric $MSE$ even in the presence of additive noise on the MZI signal. Fig. \ref{SNRplots} quantifies how the proposed EKF algorithm can enhance the MSE of the calibrated interfrometric signal in three different scenarios where the subject under test is located at three different depths ($Z_R$) of 500 $\mu m$, 1000 $\mu m$, and 1500 $\mu m$, respectively. It should be noted that $MSE$ quantifies the error between the calibrated interferometric signal when noise is added to the MZI signal and when it is not. \\
 \begin{equation}
 SNR_{{MZI}} = \frac{\text{Input MZI Power}}{\sigma^2_{N_{MZI}}}\\
 \end{equation}

\begin{figure}[t]
\centering
\subfloat{
    \includegraphics[width=.47\textwidth]{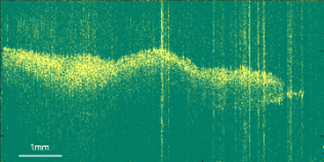}}
\subfloat{
    \includegraphics[width=.47\textwidth]{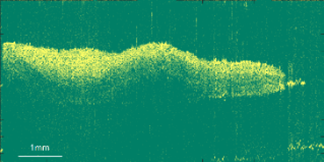}}
    \caption{(Left) Kalman filtering is not applied, (Right) Kalman filtering is applied.}
\label{MZIKalmanTomo}  
\end{figure}

Figure \ref{MZIKalmanTomo} shows a scenario when some additive noise is added to the calibrating MZI signal. The tomographic image then is reconstructed in two cases when the signal $k[n]$ is once retrieved by the conventional demodulation techniques (Hilbert transformation, Figure \ref{MZIKalmanTomo}, Left) and once when the Kalman filtering is used to demodulate the calibrating signal, Figure \ref{MZIKalmanTomo} (Right). 
\subsection{Conclusion}
Another demising technique based on the application of an unscented Kalman filtering was demonstrated. The demodulation process and the the production of the calibrating clock were introduced and the effects of calibrations and its importance were evaluated. At last, the performance of a Kalman based demodulation was compared to that of a Hilbert based demodulation in noisy environments.
\section[Demodulation Based on Interpolated Discrete Fourier Transformation, IpDFT]{Demodulation Based on Interpolated Discrete Fourier Transformation, IpDFT }

\begin{figure}[!t]
	\centering
	\includegraphics[scale=0.35]{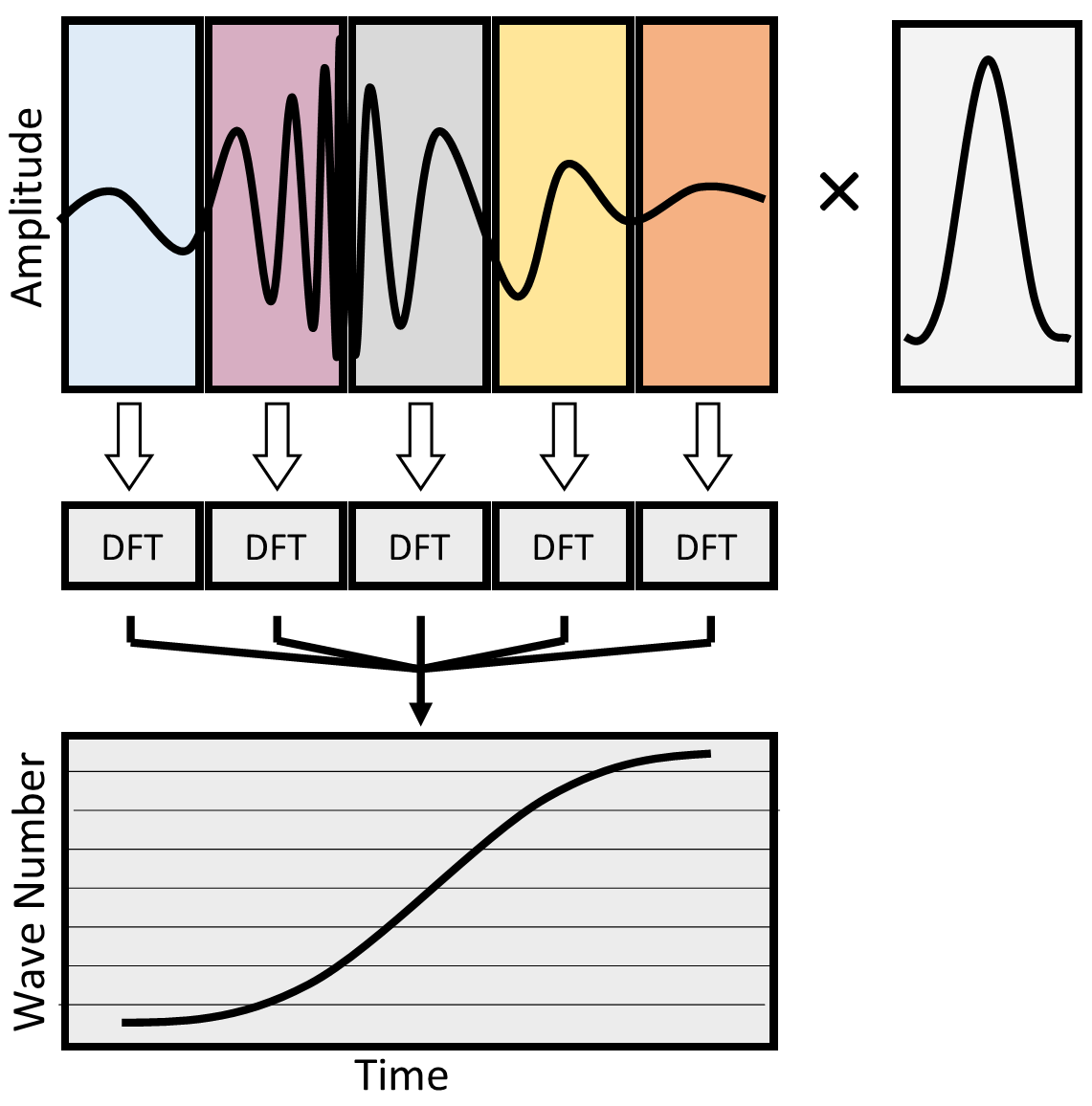}
	\caption{IpDFT algorithm implementation diagram, bottom right shows a split-radix FFT butterfly diagram.}
	\label{howipdft}
\end{figure}

IpDFT usually is used in power electronics to estimate the instantaneous parameters of energy transmission lines or to estimate the parameters that an electrical motor or generator is operating on in a very efficient and fast manner. In this algorithm, the sampled calibrating signal is received by the system block performing the IpDFT technique \cite{IpDFT}. Blocks of the sampled MZI signal with a data length of $P$ are then captured and a discrete Fourier transform (DFT) is applied to them. As a result of this arithmetic computation, three parameters, $Y_K^*$,$Y_{K-1}*$, and $Y_{K+1}^*$ are calculated. $Y_K^*$ is the maximum bin within any block that the DFT is being applied on and $Y_{K-1}^*$, and $Y_{K+1}^*$ are the adjacent bins. In each block then, some extra arithmetic is applied to calculate the instantaneous parameters of the calibrating signal, $I_{MZI}$. A split radix FFT algorithm is used on each block of the MZI signal for a better efficiency. A window (hamming, hanning, etc.) gets multiplied to each block of the calibrating signal for better accuracy of instantaneous parameter estimation. Figure \ref{howipdft} shows how the IpDFT algorithm is applied on different blocks of the calibrating signal. Research \cite{IpDFT} shows that the proposed IpDFT based demodulation technique requires much less arithmetic calculations compared to the conventional Hilbert based one making it more suitable for ultra-fast applications where a large amount of data needs to be processed in a shorter span of time. \\
\subsection{A Discussion on Different Windows that Can Be Used}

IpDFT is a well-established, precise technique to calculate the temporal parameters of a signal without the conventional problems, such as spectral leakage that applying a normal discrete Fourier transformation on a signal causes. To do so, windows of the MZI signal with specific length $P$ are acquired and are multiplied to a appropriate window as Figure \ref{howipdft} suggests.\\
\indent There are multiple points to be taken into account when picking a specific window. In general, a window is characterized by the main lobe width of its spectral shape and the attenuation of its side lobes. The narrower the former is and the more the attenuation, the better the window becomes in terms of instantanous parameter estimation.\\
Choosing a high order Rife-Vincent Class I (RVCI) window, causes minimal estimation error due to the large attenuation that can be observed on their side lobes. However, a longer block of the data in the time domain should be taken a DFT upon to yield those accurate results. On the other hand, applying a Kaiser-Bessel and\textbackslash or Dolph-Chebyshev results in a more inherent estimation error while making it possible to apply the DFT on shorter block of the signal, cutting the arithmetic count significantly.
\subsection{Configuration and Methodology}
The optical configuration is the same as discussed in previous sections of this dissertation. The MZI and the interferometric signal are produces as described by Figure \ref{configuration} and the calibration mechanism is applied as Figure \ref{demodulation} shows. This section introduces a new way of demodulating the argument of the calibrating signal (MZI). 
\subsection{Estimation}
Knowing that a MZI signal can be modeled as an amplitude modulated chirp, and assuming that it has a constant frequency and amplitude (with an exponential decay), in each temporal block that the DFT is applied to (Figure \ref{howipdft}), the signal in each block can be modeled as,
\begin{equation} \label{MZIIPDFT}
\begin{split}
  y[n] = Acos&(\omega_{0}n+\theta)e^{-dn} =\\ \frac{A}{2}\big(e^{-dn}e^{j(\omega_{0}n+\theta)}&+e^{-dn}e^{-j(\omega_{0}n+\theta)}\big),
\end{split}
\end{equation}
where $A$ represent the constant amplitude of the signal in each block, $\omega_0$ represents the constant frequency in each block and $e^{-dn}$ represents the changing nature of the amplitude of the MZI.\\
\indent Applying the DFT on the first element of the second line in Equation  \ref{MZIIPDFT} results in,
\begin{equation} \label{DFTapplied}
 DFT\{e^{-dn}e^{j(\omega_{0}n+\theta)}\} = e^{j\theta}\dfrac{1-e^{(j\omega_{0}-j\omega_{k}-d)P}}{1-e^{j\omega_{0}-j\omega_{k}-d}},
\end{equation}
the DFT of the entire Equation  \ref{MZIIPDFT} can be approximated as,
\begin{equation} \label{DFTappliedall}
Y_{k} \approx \dfrac{A}{2}\bigg(e^{j\theta}\frac{1-\lambda^{P}}{1-\lambda e^{-j\omega_{k}}}\bigg),  ~~~~      0\le\omega<\pi,
\end{equation} 
where $\lambda = e^{-d+j\omega_{0}}$. As explained before, $Y_K^*$ denotes the highest bin value within each block, and $K$ the bin index it corresponds to.\\
\indent There are two methods introduced here to estimate the instantaneous frequency in each temporal blocks that the DFT is applied to.
\subsubsection{Betrocco-Yoshida (BY-2) Algorithm}
Let us assume that the following highest bin in the calculated DFT after $Y_K^*$ is $Y_{K-1}^*$ followed by $Y_{K+1}^*$ and then {$Y_{K-2}^*$}. It can be proven that $\lambda$ (as shwon by Equation  \ref{DFTappliedall}) can be calculated as,
\begin{equation} \label{eq2}
 \lambda = e^{j\omega_{k}}\dfrac{1-R}{e^{-j2\pi/P}-Re^{-j2(2\pi/P)}},
\end{equation}
where,
\begin{equation} \label{eq2}
 R= \frac{Y^{*}_{k-2}-2Y^{*}_{k-1}+Y^{*}_{k}}{Y^{*}_{k-1}-2Y^{*}_{k}+Y^{*}_{k+1}}.
\end{equation}
In the case that the highest bin is followed by $Y_{K+1}^*$, $Y_{K-1}^*$ and $Y_{K+2}^*$,
\begin{equation} \label{eq2}
 \lambda = e^{j\omega_{k}}\dfrac{1-R}{e^{-j2(2\pi/P)}-Re^{-j2\pi/P}},
\end{equation}
where,
\begin{equation} \label{eq2}
 R= \frac{Y^{*}_{k-1}-2Y^{*}_{k}+Y^{*}_{k+1}}{Y^{*}_{k}-2Y^{*}_{k+1}+Y^{*}_{k+2}}.
\end{equation}
\indent Once $\lambda$ is calculated, the temporal frequency within each block, $\omega_0$ can be calculated.
\subsubsection{RVCI Windows (Order O)}
Let us assume that the window multiplied to each temporal block of the MZI signal is denoted by $w_n$. Each block of the MZI signal then , after the window multiplication, can be modeled as,
\begin{equation} \label{RVCIMZI}
 y[n]=w_{n}x_{n}=w_{n}Acos(\omega_{0}n+k)e^{-dn}=\bar{w}_{n}Acos(\omega_{0}n+k).
\end{equation}
Taking the DFT, $R_1$ and $R_2$ can be calculated as below, 
\begin{equation} \label{eq2}
\begin{split}
 R_{1}&= \frac{\|Y^{*}_{k+1}\|^2}{\|Y^{*}_{k}\|^2}  \approx \frac{(\delta+O)^2+D^2}{(\delta-O-1)^2+D^2}, \\
  R_{2}&= \frac{\|Y^{*}_{k-1}\|^2}{\|Y^{*}_{k}\|^2}  \approx \frac{(\delta-O)^2+D^2}{(\delta+O+1)^2+D^2},
 \end{split}
\end{equation}
where $D$ = $\dfrac{dP}{2\pi}$. Knowing $R_1$ and $R_2$, $\delta$ can be calculated to be,
\begin{equation} \label{eq2}
\delta =-\frac{2M+1}{2}\frac{R_{1}-R_{2}}{2(O+1)R_{1}R_{2}-R_{1}-R_{2}-2O},
\end{equation}
where $O$ is te order of the RVCI window applied. With all the equations obtained above $d$ and $\omega_0$ can be calculated as below,
\begin{equation} \label{eq2}
\begin{split}
 d&= \frac{2\pi}{P} \sqrt{\frac{(\delta+O)^{2}-R_{1}(\delta-O-1)^2}{R_{1}-1}},~~~~~~\delta>0, \\
  d&= \frac{2\pi}{P} \sqrt{\frac{(\delta-O)^{2}-R_{2}(\delta+O+1)^2}{R_{2}-1}},~~~~~~\delta<0,
 \end{split}
\end{equation}
and,
\begin{equation}
    \omega_0=(k+\delta)\frac{2\pi}{P}.
\end{equation}
A RVCI window with $O$ being equal zero is called a rectangular window.
\subsection{Producing a Calibrating CLK}
Once the parameter $\omega_o$ is extracted for each temporal block of the MZI signal, the spectral sweep profile of the laser is build by building a function that is constituted of lines with slopes being equal the temporal frequency at each temporal block. This newly generated function resembles the argument of the MZI signal over time ($k[n]$ in Section \ref{sec3.2.3}). The calibrating clock is retrieved as exactly suggested by the same Section \ref{sec3.2.3}.
\subsection{Computational Complexity Analysis}
This section aims to provide a thorough analysis of how this demodulation technique compares to the state of the art demodulation technique, Hilbert transformation, and proves that the proposed demodulation technique cuts the arithmetic complexity by a significant factor.\\
\begin{table}[t] \label{HilVIpDFTcomp}
\centering
\caption{Arithmetic count comparison between a Hilbert FIR filter based demodulation and an IpDFT based demodulation }

\begin{tabular}{c c c c c}    \toprule
\emph{Techniques} & \multicolumn{2}{c}{\emph{Hilbert}} & \multicolumn{2}{c}{\emph{IpDFT [5]}}  \\\midrule
    &\multicolumn{1}{c|}{Order} & $H$&  \multicolumn{1}{c|}{Obs Window Length}& $P $    \\ \midrule
 \# Add / per output & \multicolumn{2}{c}{$H$} & \multicolumn{2}{c}{$P(\frac{4}{3}log_{2}(P)-\frac{8}{9})-\frac{1}{9}(-1)^{log_{2}(P)}$}\\ 
	\# Mult / per output& \multicolumn{2}{c}{$H$} &  \multicolumn{2}{c}{P($\frac{2}{3}log_{2}(P)-\frac{19}{9})+\frac{1}{9}(-1)^{log_{2}(P)}$}\\
 \# Total Arithmetic  & \multicolumn{2}{c}{2$HN$} & \multicolumn{2}{c}{$\frac{L}{P}(2Plog_{2}P-3P$)}\\\bottomrule 
 \hline

\end{tabular}

\end{table}
\indent The complexity comparison between the IpDFT technique and the Hilbert based technique can be broken down into two different front. First aspect of complexity relates to the core algorithms themselves and the second relates to the auxiliary signal processing algorithms that need to be used in conjunction with the core algorithm to achieve the calibration process. It is obvious that for a FIR based Hilbert transformation to work there needs to be $H$ multiplications and $H$ additions to produce one sample of its output signal $I_{MZI,H}$. Assuming that the MZI signal is comprised of $L$ different samples, the total number of arithmetic operations would equal $2HL$. To perform the DFT on each temporal block of the MZI signal in the IpdFT based technique, certain addition and multiplications need to happen before a DFT result is achieved. Table 3.3 shows the number of multiplication and addition necessary to perform the DFT. Assuming the total $L$ samples of the MZI signal is divided into blocks with the length of $P$, there would be $\dfrac{L}{P}$ blocks that the DFT operation would be applied on. The total number of arithmetic count for an IpDFT based technique can be shown \cite{IpDFT} to equal $2Llog_{2}P-3L$. As reported in the literature \cite{Hilbert}, the minimum number set for $H$ can equal 17. Knowing that in realistic scenarios a wise $log_2P$ can not never exceed a value that equals five or six, it is obvious that the total arithmetic count in an IpDFT technique is significantly lower than that of the Hilbert transformation. Further, the latency of a FIR filter can be shown to equal $H$ while that of the IpDFT is $log_2P$ giving it another significant advantage in terms of real time implementation.

\begin{table*}[t]

\centering
\caption{Computational complexity comparison between auxiliary blocks necessary to yield the spectral sweep of a spectral sweep light source. $M(n)$ stands for the complexity of a multiplication algorithm. The multiplication algorithm complexities are discussed in Table 3.4. }
\begin{tabular}[!b]{cc}
\begin{tabular}{ c c c c}    \toprule
 \multicolumn{2}{c}{\small \emph{Hilbert}} & \multicolumn{2}{c}{\small \emph{IpDFT}}  \\\midrule
    \multicolumn{2}{c|}{\small Tangent Inverse}&  \multicolumn{2}{c}{\small Max Bin Det}     \\ \midrule
 {\small Taylor series; FFT-based acceleration \cite{taylorser}} & \small \textbf{O}($M(L) L^{1/3}$ $log^{2}(L)$)& {\small Compare in Pairs} & \small \textbf{O}($P$)\\
 {\small Arithmetic-geometric mean iteration \cite{aritgeo}} & \small \textbf{O}($M(L) log(L)$)& {\small Tournament Method} & \textbf{O}($P$)\\
{\small Taylor series; binary splitting + bit burst \cite{taybin}} & \small \textbf{O}($M(L) log^{2}(L)$)& {\small Linear search} & \small\textbf{O}($P$)\\ 

\bottomrule 
 \hline

\end{tabular}

\end{tabular}

\end{table*}
\begin{table}[t] \label{mult}
\centering

\caption{The algorithm complexity of some multiplication techniques. $log^{*}$ stands for Iterated Logarithm.}
\begin{tabular}{c c c}    \toprule
\emph{Techniques} & \emph{Multiplication}   \\\midrule
    &\multicolumn{1}{c|}{Order} & N    \\ \midrule
Furer's algorithm \cite{furer}& \multicolumn{2}{c}{\textbf{O}($L log(L)~ 2^{\textbf{O}(log^{*}(L))}$) }\\    

Schonhage Strassen algorithm \cite{schonhage} & \multicolumn{2}{c}{\textbf{O}($L log(L) log(log(N))$)}\\
    
Mixed level Toom-Cook \cite{knuth} & \multicolumn{2}{c}{\textbf{O}(N $2^{\sqrt{2 log(L)}} log(L)$)}  \\ 
\bottomrule 
 \hline

\end{tabular}

\end{table}
\indent In addition to the core arithmetic block (Hilbert Vs. IpDFT) there are further auxiliary processes that need to take place to acquire the spectral sweep of the laser.\\
\indent For the Hilbert transformation based technique to work, there exists the need to perform an inverse tangent operation as Equation  \ref{k[n]} suggests. This operation usually is done by one of the algorithms listed in Table 3.3, and Table 3.4. $M(L)$ relates to one of the multiplication algorithm used to perform the tangent inverse. \\
\indent For the IpDFT based technique to work, the only auxiliary process that needs to take place is detecting the maximum bin which can be performed by one of the algorithms suggested in Table 3.3. Overall, it can be observed that the auxiliary process for the IpDFT based technique is much less computationally complex compared to that of the Hilbert transformation. \\
\indent It is observed that both core arithmetic and the auxiliary arithmetic is much more computationally intensive in the Hilbert transformation based technique giving the IpDFT based technique a clear advantage over the Hilbert based technique.
\subsection{Results}
\begin{figure}[!t]
	\centering
	\includegraphics[scale=.65]{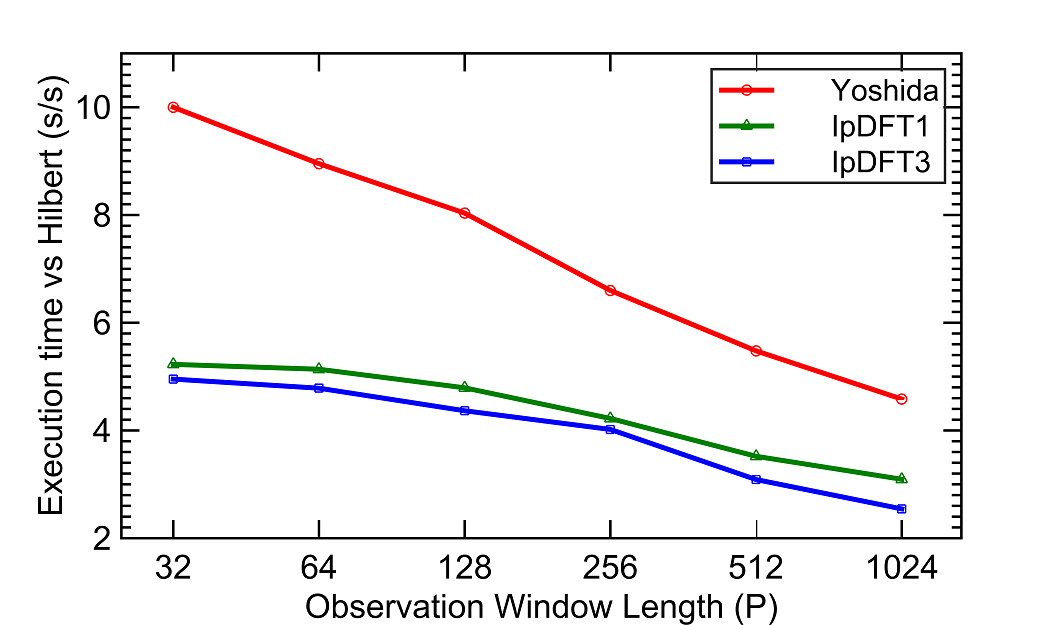}
	\caption{Execution time comparison between IpDFT based demodulation and Hilbert based demodulation. IpDFT1, IpDFT3 and Yoshida are 3 different ways that the IpDFT algorithm is applied. IpDFT1 relates to the RVCI order 1 algorithm and IpDFT3 relates to the RVCI order 3 algorithm.}
	\label{HilbertvIpDFTexec}
\end{figure}

\begin{figure}[t]
\centering
\subfloat{
    \includegraphics[width=.47\textwidth]{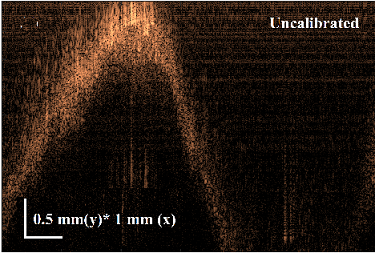}}
\subfloat{
    \includegraphics[width=.47\textwidth]{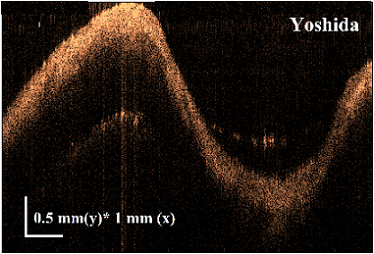}}\\
\subfloat{    
     \includegraphics[width=.47\textwidth]{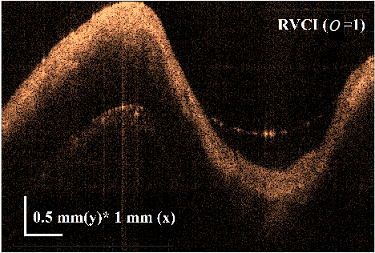}}
\subfloat{
    \includegraphics[width=.47\textwidth]{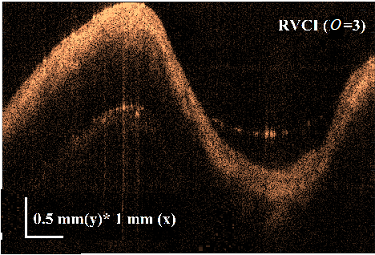}}
    \caption{The performance evaluation of the IpDFT algorithm on the quality of tomographic images acquired. (Top left) Without calibration. (Top Right) The tomographic image obtained by applying the IpdFT technique based on bertoco-yoshida. (Bottom left) The tomographic image obtained by applying the IpdFT technique based on RVCI order 1. (Bottom right) The tomographic image obtained by applying the IpdFT technique based on RVCI order 3}
\label{tomoIpDFT}  
\end{figure}

To evaluate the performance of the IpDFT based demodulation technique, 500 A-scans were captured optically as Figure \ref{configuration} suggests. The MZI and the interferometric signals are then used as Figure \ref{demodulation} suggest to obtain a calibrating clock to sample the interferometric signal at correct time instances. \\
\indent First, the performance of the proposed algorithm is computed in terms of the execution time. The time to execute the IpDFT based technique is measured and compared against the standard Hilbert based technique. As explained before, IpDFT based technique is superior compared to the Hilbert based technique as suggested by Table 3.2. The logarithmic nature of the complexity of the IpdFT technique (As suggested by Table. 3.2) is shown in Figure \ref{HilbertvIpDFTexec}
\indent Further, the axial resolution as the result of performing the demodulation technique is evaluated. The axial resolution first is measured for different axial depths of the sample under test. This experiment was conducted for the three different IpDFT algorithms (Bertocco-Yoshida, RVCI order 1/IpDFT1, RVCI order 3/IpDFT3) and are compared to those obtained from doing the demodulation based on Hilbert transformation. As seen, although the IpDFT cuts the computational complexity by up to 10 times, the axial resolution, a measure to evaluate the quality of the demodulation technique, remains on par with the Hilbert based demodulation technique.\\
\indent Further, the axial resolution is measured when the length of the window $P$ in the IpDFT technique sweeps between different values. As seen, the technique shows the best result for shorter window length for which the system has higher computational advantage compared to the Hilbert transformation technique. \\
\begin{figure}[t]
\centering
\subfloat{
    \includegraphics[width=.47\textwidth]{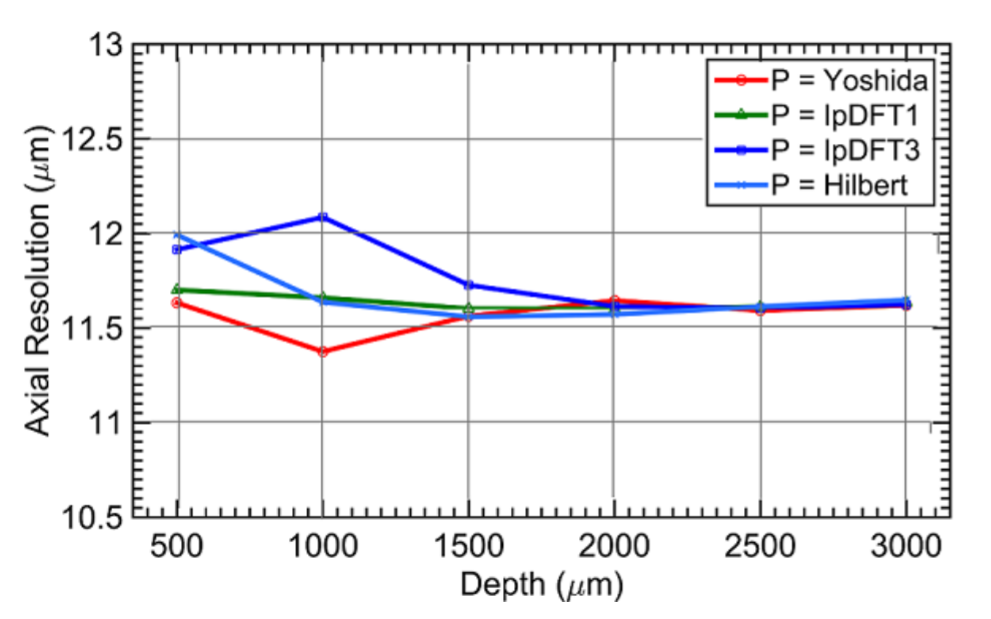}}
\subfloat{
    \includegraphics[width=.47\textwidth]{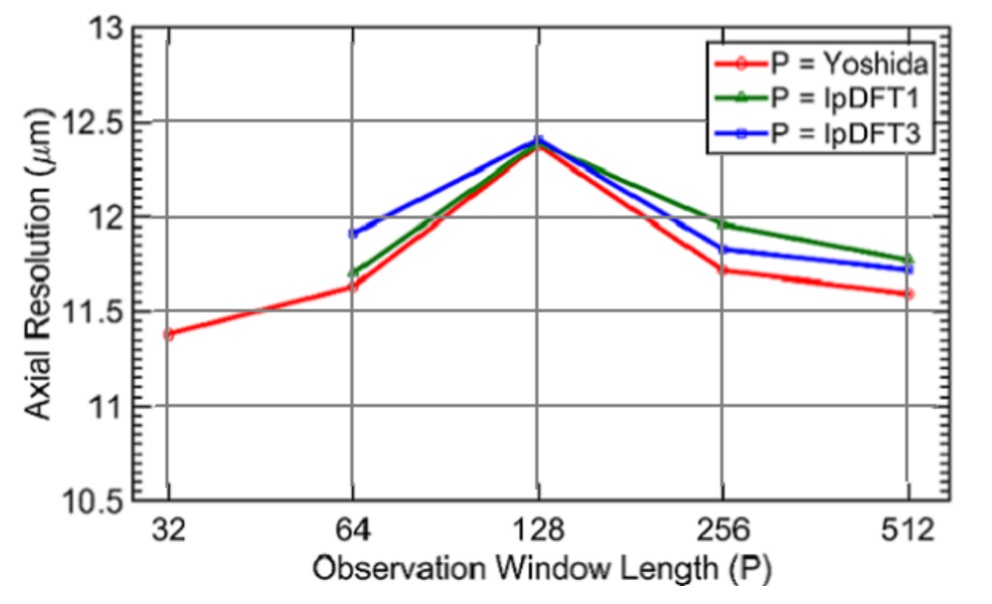}}
    \caption{(Left) Axial resolution for different Axial depths of the subject under test for different demodulation techniques (Right) Axial resolution evaluation for different IpDFT window lengths.}
\label{kalmantomoeff}  
\end{figure}
Furthermore, the effect of the calibration on the quality of tomographic images acquired is evaluated as seen by Figure \ref{tomoIpDFT}, different IpDFT techniques are on par regarding the result they produce and are necessary to obtain distinguishable results.
\subsection{Conclusion}
In this section yet another way of demodulating the MZI signal was presented. The superiority of the technique compared to the Hilbert based demodulation was shown and proven in terms of computational complexity. The method for acquiring a calibrating clock is re introduced and the calibrating clock is used to sample the interferometric signal at correct time instances. The performance of the system was checked in terms of axial resolution and the quality of the tomographic image they result in.
%
%
%
%

\chapter[LEVEL CROSSING SAMPLER]{LEVEL CROSSING SAMPLER \footnote{Reprinted with permission from "Compressed Level Crossing Sampling for Ultra-Low Power IoT Devices" by Jun Zhou, Amir Tofighi Zavareh, Robin Gupta, 
Liang Liu, Zhongfeng Wang, Brian M. Sadler, 
Jose Silva-Martinez, and Sebastian Hoyos, 2017.  IEEE Transactions on Circuits and Systems I: Regular Papers, Volume: 64 , Issue: 9, pp 2495 - 2507, Copyright [2017] by Institute of Electrical and Electronics Engineers (IEEE).} }

\begin{figure}[t]
	\centering
	\includegraphics[scale=1.3]{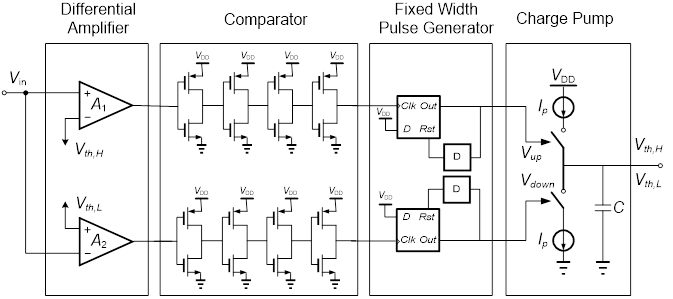}
	\caption{The block diagram representation of the proposed LCS.}
	\label{Chapter4LCS}
\end{figure}

\indent The LCS introduced in \cite{tcas1} provides a solid tool to sample a signal on asynchronous basis. As explained in Section. \ref{3.1.1.2}, the full dynamic range of the input is divided into $2^Q$ equidistant sections the top and the bottom value of each is denoted by $V_{th,L}$ and $V_{th,H}$. The circuit level LCS introduced in this chapter is capable of firing a narrow pulse whenever the input signal hits one of those values. Figure. \ref{Chapter4LCS} provides a block diagram representation of the proposed LCS and is related to the one shown by Figure. \ref{LCS}.\\
\indent Figure \ref{Chapter4LCS} is constituted of two differential amplifiers at the input where the positive pin of both is connected to the input signal. The negative pins of the two differential amplifiers are connected to $V_{th,H}$ (top amplifier) and $V_{th,L}$ (bottom amplifier), respectively. These differential amplifiers are followed by comparators the sensitivity of which is increased by the existence of the differential amplifier. The comparators are followed by a fixed width pulse generator depicted by the D- flip flops and the delay lines that are put in a loop as depicted by Figure. \ref{Chapter4LCS}. Finally, the pulses are driving a charge pump circuit that sets the value of threshold voltages $V_{th,H}$ and $V_{th,L}$.\\
\indent While the differential amplifier is only put there to provide additional gain and could be implemented via very simple circuits, the comparators are built by cascading four inverters where the threshold voltage of the first can be set by changing the size of the transistors that affect the voltage transfer curve of the transistor as is formulated by Equation. \ref{VTC} 
\begin{equation} \label{VTC}
    V_{th}=\dfrac{V_{DD}-|V_{tp}|+V_{tn}\sqrt{\dfrac{K_n}{K_p}}}{1+\sqrt{\dfrac{K_n}{K_p}}},
\end{equation}
where $V_{th}$ is the comparator threshold, $V_{DD}$ is the power supply rail voltage, $V_{tn}$ and $V_{tp}$ are NMOS and PMOS device threshold, respectively, $K_n$ and $K_p$ are the product of device capacitance mobility and dimensions for NMOS and PMOS, respectively. Three additional inverters with increasing sizing are used to increase the sensitivity and the gain of the overall comparators stage. The increasing size of cascaded inverters is making it capable of driving larger loads at its very output. The key is to match the two comparators in the top and the bottom path to have the same threshold voltages. That way the entire system remains more balanced on both rising and falling edges of the input signal making it more suitable fore higher frequency input signals.\\
\indent The benefit of the inverter based comparator relates to its ultrafast nature compared to those based on differential amplifiers that have some sort of regenerative positive feedback within their structure. The drawbacks of the inverter based comparator are two fold. The first shortcoming relates to the sensitivity of the threshold to temperature related quantities such as $K_n$, $K_p$, $V_{tn}$, and $V_{tp}$ that need to be calibrated for at the power up. The second shortcoming is that the value of the input comparator triggering threshold relates to the supply voltage. Any noise on the supply voltage can cause the comparator triggering voltage to fluctuate. Large bypassing capacitor could be used in parallel with the supply voltage to bypass the aforementioned noises. Picking up the minimum width for the first inverter, the overall delay for the comparator stage was simulated to be 45 ps.\\
\indent The fixed width pulse generator is built by putting a delay line in feedback with the reset pin of a D-flip flop as Figure. \ref{Chapter4LCS} suggests. The delay line is built by cascading a group  of inverters. The fixed width pulse is then used to drive a charge pump and set the two value of $V_{th,H}$, and $V_{th,L}$. The delay line as well as the charge pump could be very accurately calibrated by DLL\textbackslash PLL based mechanisms either at the start up or on the background.\\
\begin{figure}[!b]
	\centering
	\includegraphics[scale=1.15]{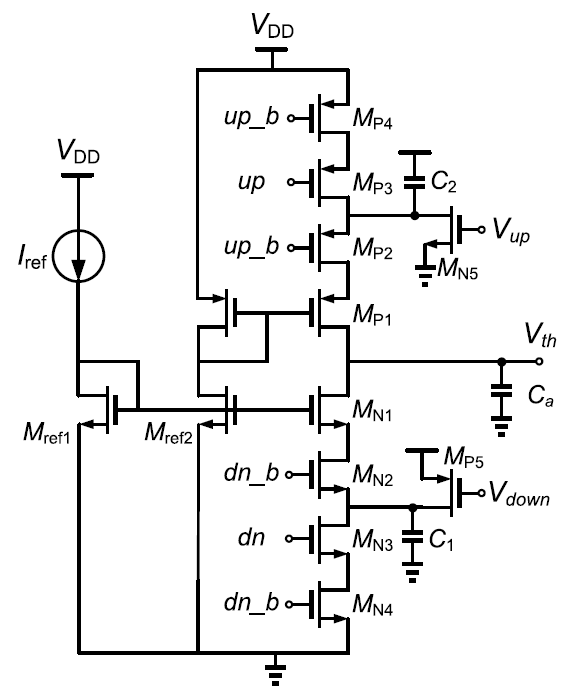}
	\caption{Circuit schematic representation of the charge pump.}
	\label{chargepump}
\end{figure}
\indent The charge pump shown by Figure. \ref{Chapter4LCS} can be realized by the circuit shown in Figure. \ref{chargepump}. Transistors $M_{P3}$ and $M_{N3}$ are used as the main switches that conduct the current to the output capacitor $C_a$ and increase\textbackslash decrease the value of $V_{th}$, respectively. Obviously, two of these charge pumps are used to set the values of $V_{th,H}$ and $V_{th,L}$. Dummy transistors $M_{N2}$, $M_{N4}$, $M_{P2}$, and $M_{P4}$ are added to minimize the effects of charge injection and clock feed-through. Transistors $M_{N5}$, and $M_{P5}$ are used to make turning off $M_{N3}$ and $M_{P3}$ faster.  \\
\indent The higher the value of $I_{ref}$, the faster the output capacitor gets charged and the faster the charge-pump works. The lower the value of the output capacitor, the faster it gets charged. However, decreasing the output capacitor can increase the overall $KT/C$ noise associated to this capacitor. Having all those constraints in mind, a capacitor value of 50 fF and a fixed pulse width that equals 50 ps are picked. The total delay line associated to the loop depicted by Figure. \ref{Chapter4LCS} is around 130 ps which translates to an asynchronous sampling rate that equals 7.7 GHz.\\
\indent The lower the value of the step size in the charge pump, the higher the resolution of the entire sampler becomes. The charge pump step size can be set by the fixed width pulse and the value of the $I_{ref}$. Increasing the resolution, however, limits the slew rate of the input signal. If $U$ denotes the entire dynamic range of the input signal, $2^Q$ is the number of amplitude levels that the sampling happen upon. The maximum slew rate (SR) of the sampler would be set at,
\begin{equation} \label{SR}
    SR=\dfrac{U}{2^Q}f_s=LSB f_s
\end{equation}
\indent To compare this proposed LCS with other similar works reported in the literature, a figure of merit is defined where it captures the amount of power burnt per each conversion of the sampler. 
\begin{equation} \label{FOM}
    FOM=\dfrac{Power}{2^{ENOB}\times2\times BW}
\end{equation}
where ENOB stands for the effective number of bits and should theoretically equal $Q$. The worst power performance configuration for this sampler was picked where the sampler was fed with a ramp with a SR of $1V/\mu s$. Table 4.1 summarizes the performance of this asynchronous sampler compared to other similar works reported in the literature.\\
\begin{table}[H]
	\centering
		\caption{Comparing the proposed LCS to similar works in the literature.}
	\label{Band}
    \begin{tabular}{*8c}   \toprule
    \cite{tcas11} &\cite{tcas12}&\cite{tcas13}&\cite{tcas14}&\cite{tcas15}&\cite{tcas16}& This Work \\
    Simulation &Chip&Chip&Chip&Chip&Chip& Chip \\\midrule
    Architecture & LCS & LCS &LCS&LCS&LCS&LCS&SAR\\ 
    CMOS Technology &0.35 $\mu$m &0.5 $\mu$m&0.18 $\mu$m&90 nm&90 nm&28 nm&65 nm\\
    Power Supply & 1V & 3.3 V &0.8 V&1 V&0.7 V&0.9 V&1 V\\
    Power Consumption (uW) & 30  & 106 &0.582&350 &9.25 &350 &402\\
    Input Bandwidth (MHz) &2e-3& 5e-3  & 3.3e-3 &1e-3&2 &50 &5 \\
    Resolution &6& 5&5  & 8 &8.81&12 &10 \\
    FOM &11.7& 1568.8&1.4  & 683.6 &0.0052&0.0026 &0.039 \\\bottomrule
    \hline
    \end{tabular}

\end{table}
\indent The ultra fast level crossing sampler that is proposed in this dissertation can be used in conjunctions with demodulation techniques proposed in the previous chapter to provide an online all on a chip calibration method that can be used to enhance the SS-OCT device in multitude of applications. These techniques open up the possibility of using SS-OCT systems in ultra fast applications without sacrificing the axial resolution or the imaging range that the SS-OCT device can provide.

\chapter{PROPOSED CALIBRATION MECHANISM}
\begin{sidewaysfigure}
\begin{center}
\includegraphics[width=9in]{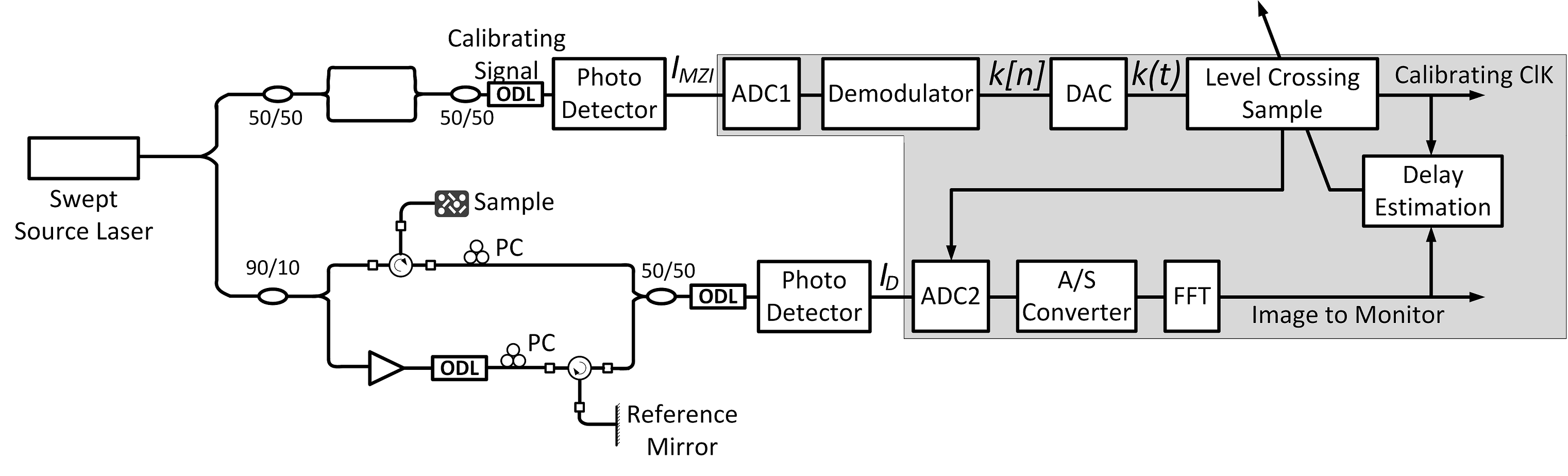}
\end{center}
\caption{Overall block diagram of the spectral calibration method.}
\label{ProposedMechanism}
\end{sidewaysfigure}

Figure \ref{ProposedMechanism} shows the overall system block diagram of how the demodulation techniques, as well as the level crossing sampler, is going to be used to build our proposed calibration engine. As seen in Fig. \ref{ProposedMechanism}, the figure is comprised of two different paths, one produces the calibrating signal by the means of using a MZI and the other produces the interferometric signal by the mean of a Michelson Interferometer.
The calibrating CLK is produced by passing the calibrating signal through the demodulation techniques and then sampled by the level crossing sampler to generate the interferometric sampling clock at the correct sampling times. As expressed before, the interferometric signal produced by the Michelson interferometer can be mathematically expressed as,

\begin{equation} \label{Equ.formulaOCTchap5}
\begin{split}
I_D(k(t))& = \frac{\rho}{4}S(k(t))[R_R+R_{S1}+R_{S_2}+...+R_{Sn}] \\
&+ \frac{\rho}{2}S(k(t))\bigg[\sum_{n\ne m=1}^{\infty}\sqrt{R_{Sm}R_{Sn}}cos\big(k(t)(z_{Sm}-z_{Sn})\big)\bigg]. \\
&+ \frac{\rho}{2}S(k(t))\bigg[\sum_{n=1}^{\infty}\sqrt{R_{R}R_{Sn}}cos\big(k(t)(z_R-z_{Sn})\big)\bigg].
\end{split}
\end{equation}

where $\rho$, $S(K)$, $R_R$, $R_Sn$, $z_R$, and $z_Sn$ are values that depend on the optical configuration of the system depectid by Fig. \ref{ProposedMechanism}. The n-number of values $R_{Sn}$ are the power reflectivity of different particles within the object to be imaged and are the values to be obtained. Once the signal $k(t)$ is produced, the level crossing sampler provides time instances at which the value of $k$ is one of the predefined equidistant values shown by Figure.\ref{LCSEKF}. This dissertation now delves into the building blocks that comprise the proposed system and expresses the benefits it demonstrates compared to the widely utilized conventional system based on interpolation and resampling. 

\section{Optical Setup}

An in house built SS-OCT system (optical side of Fig. \ref{ProposedMechanism}) was used to acquire the data used in this experiment. It includes a Swept Source (ESS, Exalos) with center wavelength at 1310 nm. The MZI optical path is consisted of two 50/50 couplers. The Michelson Interferometer path is consisted of polarization controllers, optical delay lines, optical attenuator, and optical delay lines. 
 
\section{Proposed Data Acquisition and Processing}
This subsection presents the real-time realization of a functional decomposition to linearize the parameter $k$ directly when sampling the interferometric signal. This system utilizes an analog to digital converter (ADC1) that samples the calibrating signal $I_{MZI}(t)$ into the digital domain (Fig. \ref{ProposedMechanism}). The sampled signal $I_{MZI}[n]$ then passes through an $k$-estimation block that uses an extended Kalman filter to extract the parameter $k[n]$. Once the parameter $k$ is extracted, it will be passed through a digital to analog converter. The signal $k(t)$ then will be passed through a level crossings sampler \cite{} to determine the time instances at which $k$ has undergone identical changes. These time instances are denoted by $t_c$ and can be formulated as,
\begin{equation}
    t_{c_n}=k^{-1}(k_{C_n}),
\end{equation}
where $\{k_{C_n}\}$ is a group of equidistant wavenumbers. Those time instances $t_{c_n}$ are then used to sample the interferometric signal $I_D$ at the correct time instances using ADC2. Previously designed level crossing samplers \cite{tcas1} are reported to be able of providing calibrating clock pulses as fast as 5 GHz as the signal $k(t)$ passes the equidistant set of data $\{k_{C_n}\}$. The rate that the level crossing sampler can run at is denoted as $f_{sample}$. The sampled interferometric signal is passed through an asynchronous to synchronous (A/S) converter. This block is a FIFO buffer in which data is inserted via the asynchronous calibrating clock and is fetched via a synchronized clock to be fed to the denoising (CKF) block (Fig. \ref{ProposedMechanism}). The fetched data is the calibrated interferometric signal,
formulated as,

\begin{equation}\label{Equ.formulaOCTchap5cal}
\begin{split}
    &I_{D,Calibrated}[n]=I_D(t_{c_n})= \frac{\rho}{4}S(k_c)[R_R+R_{S1}+R_{S_2}+...+R_{Sn}] \\
&+ \frac{\rho}{2}S(k_c)\bigg[\sum_{n\ne m=1}^{\infty}\sqrt{R_{Sm}R_{Sn}}cos\big(k_c(z_{Sm}-z_{Sn})\big)\bigg]. \\
&+ \frac{\rho}{2}S(k_c)\bigg[\sum_{n=1}^{\infty}\sqrt{R_{R}R_{Sn}}cos\big(k_c(z_R-z_{Sn})\big)\bigg].
\end{split}
\end{equation}
 The calibrated signal then passes through a denoising algorithm that is implemented with a Conventional Kalman filter (CKF). An FFT is then applied on the denoised signal to produce the signal $i_D(z)$.
\section{Skew Estimation and Calibration}

\begin{figure}[h]
	\centering
	\includegraphics[width=0.5\textwidth]{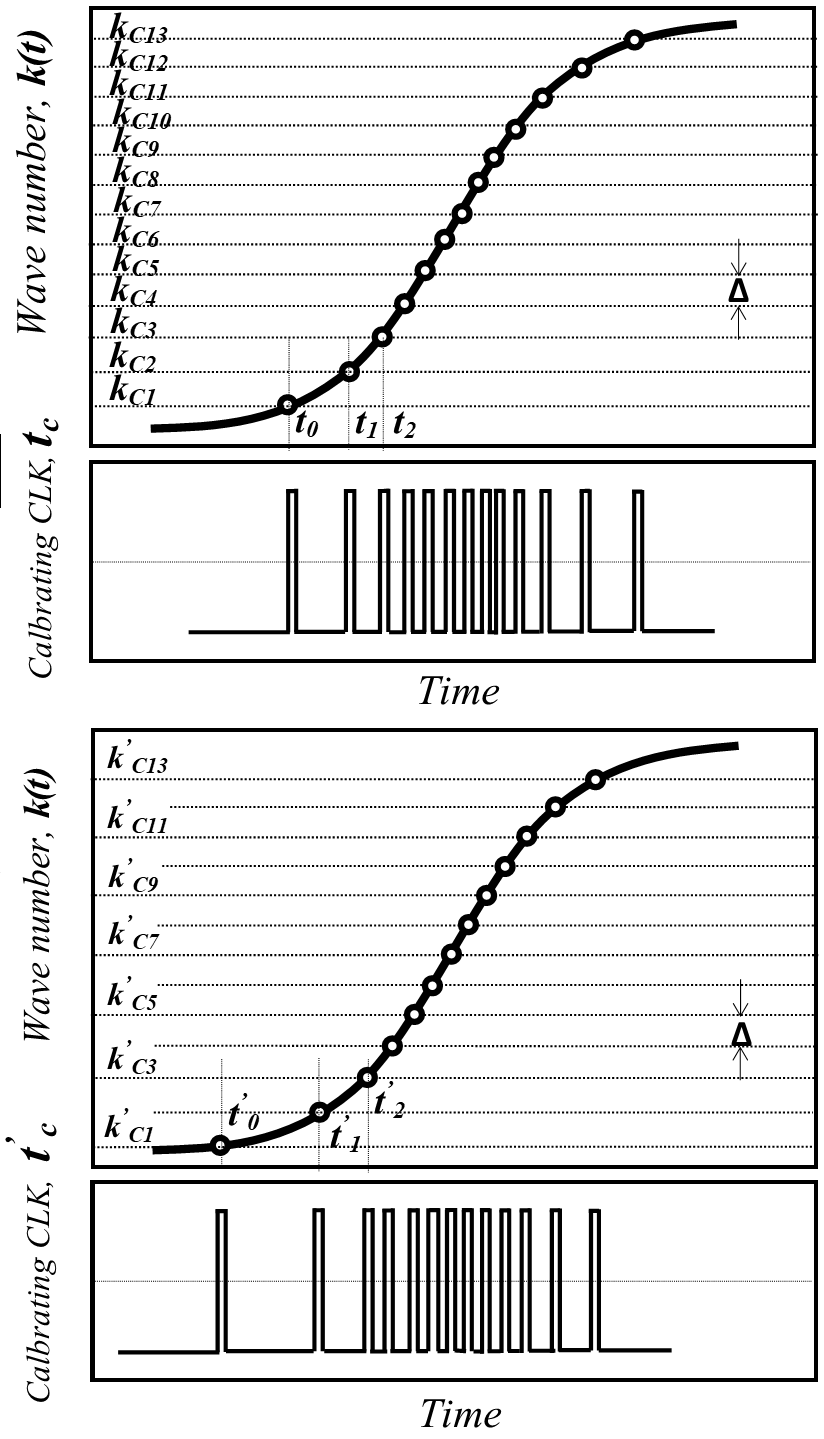}
	\caption{The process of adjusting the equidistant wavenumber values on temporal location of the calibrating clock.}
	\label{skew}
	\end{figure}

To calibrate out the possible time skews that occur between the calibrating and interferometric paths, two mechanism are put in place. These skews are due to different propagation delays produced by layout and device mismatches. First, two optical delay lines (ODL) are used (Fig. \ref{ProposedMechanism}) in the calibrating and interferometric paths to align the optical signals coming out of the paths. These optical delay lines are commercially available and can produce delays as accurate of 0.67 fs with a range of adaptation up to 4 ns. The second mechanism is an adaptive electrical solution that estimates the skew in the electrical domain and adapts the level crossing sampler to compensate it. Let us assume that the optical delay difference between the two paths are ideally equalized by the two ODLs that are used right before the two signals $I_D$ and $I_{MZI}$, as in Fig. \ref{ProposedMechanism}. A trigger signal, available in commercial swept source lasers, is used as a time reference to measure the time it takes for the calibrating signal and the signal $I_D$ to be produced. The trigger signal indicates the reference time at which the wavenumber sweep starts to occur in the swept source laser. Knowing the reference time, the total propagation delay for the signal $I_D$ so that it is sensed by the delay estimation block is $T_{Delay,D}$, and the total propagation delay for the calibrating signal to be produced and sensed by the delay estimation block is denoted by $T_{Delay,Calibrating}$. The skew estimated by the skew estimation block is calculated as $\Delta T_{skew}=T_{Delay,Calibrating}-T_{Delay,D}$. Once this parameter is estimated, it can be used to adapt the levels $ \{k_{C_n}\} $ in the level crossing sampler. Let us assume that a set of the equidistant wavenumber levels $\{k^{'}_{C_n}\} $, are found with the adaptive algorithm (Fig. \ref{skew}), 

\begin{equation}
    t^{'}_{c_n}-{t}_{c_n}=\Delta T_{skew_n},
\end{equation}
where,
\begin{equation}
    t^{'}_{c_n}=k^{-1}(k^{'}_{C_n}).
\end{equation}
The skew estimation unit adaptively estimates the skew between the two paths via the least-mean-square (LMS) algorithm which adjusts the values of $\{k^{'}_{C_n}\}$ until $\Delta T_{skew}$ is minimized.
\section{Proposed Real-time Method vs. Interpolation and Resampling Method}



While modern OCT systems use traditional digital-domain interpolation and resampling methods (Fig. \ref{hilbertconfig}), that rely on a look-up table consisted of a group of equidistant wavenumber values used for the purpose of comparison against the values of $k[n]$, the proposed method uses the correct time instances generated in continuous-time by a level crossing sampler whose time resolution is only limited by electronic noise. The proposed real-time technique directly samples the input analog interferometric signal at the precise instances at which the signal is linear with respect to $k(t)$. This real-time processing completely eliminates the interpolation error, reducing the calibrated signal MSE. 
\subsection{Errors in the Parameter $k[n]$}
In order to evaluate the effectiveness of calibration, a mirror is used as the sample under test and a tomographic image is reconstructed. Such mirror test simplifies Eqn. (\ref{Equ.formulaOCTchap5}) to the following expression, 
\begin{equation} 
\begin{aligned}
  I_D(t) = \sqrt{R_RR_S}(cos(k(t)(z_R-z_S))),
\end{aligned}
\end{equation}
where the sample under test (the mirror) is placed at the distance $z_S$ from the coupler. Let us assume that the data points picked from the interferometric signal correspond to wavenumber values that are off from the perfect equidistant wavenumber values, denoted by $\{k_{cn}\}$, by a variable $\epsilon[n]$. This error models the lack of accuracy when the values of $k[n]$ are compared against $\{k_{cn}\}$. In this scenario, 
\begin{equation} 
\begin{aligned}
  I_{D,Calibrated}[n] = \sqrt{R_RR_S}cos\big((k_{cn}+\epsilon_{k}[n])(z_R-z_S)\big).
\end{aligned}
\label{eq12}
\end{equation}
Assuming that the values of $\epsilon_k[n]$ are sufficiently small, (\ref{eq12}) can be simplified to,
\begin{equation} 
\begin{aligned}
  I_{D,Calibrated}[n] =& \sqrt{R_RR_s}cos\big(k_{cn}(z_R-z_S)\big)-\\&\epsilon_k[n]\sqrt{R_RR_S}sin\big(k_{cn}(z_R-z_S)\big).
\end{aligned}
\end{equation}
The noise signal is modulated around the same distance location of the desired signal ($z_R-z_S$). Once the signal $I_{D,Calibrated}[n]$ is taken a FFT upon to produce the tomographic image, an increased noise variance of $\epsilon_k[n]$ distorts the image produced. In particular, the error $\epsilon_{k}[n]$ is comprised of three different noise sources, the estimation noise $\hat\epsilon_{k}[n]$, the quantization noise $\epsilon_Q[n]$, and the time-resolution noise $\epsilon_{TR}[n]$ introduced by the DSP clock frequency. In the mathematical terms, 
\begin{equation} 
\begin{aligned}
  \epsilon_{k}[n] = \hat\epsilon_{k}[n]+\epsilon_Q[n]+\epsilon_{TR}[n].
\end{aligned}
\end{equation}
While the quantization noise and the time-resolution errors are set by the floating point operations and multiplexing capabilities of the processing units, the estimation noise depends on the demodulation technique used and requires further examination. While the proposed system uses an extended Kalman filter to $k$-estimate the MZI signal, the interpolation and lookup table method often utilizes a Hilbert transformation to do the same.

\begin{figure}[h]
	\centering
	\includegraphics[width=0.7\textwidth]{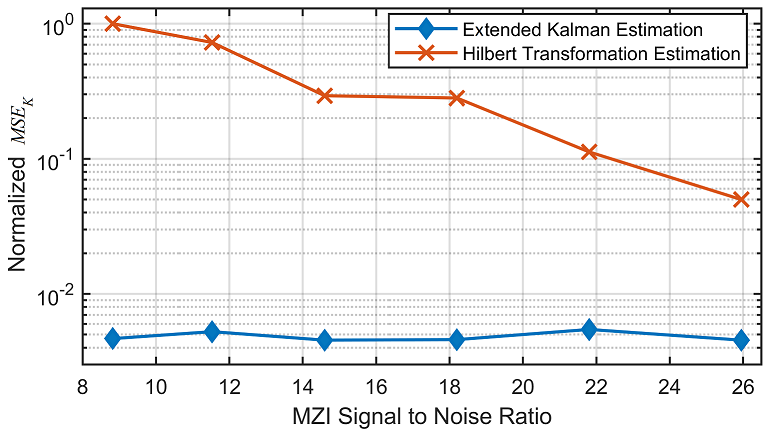}
	\caption{Comparison between the estimation accuracy of the MZI argument $k[n]$ when the Kalman filter and the Hilbert transformation are applied. The errors are measured in the MSE sense.}    
	\label{HilbertVsKalmanError}	
\end{figure}

Fig. \ref{HilbertVsKalmanError} demonstrates the effectiveness of the extended Kalman filter based estimation in extracting the parameter  $k[n]$ from the argument of the MZI signal. The amplitude of the MZI signal is first normalized. Additive White Gaussian noise (AWGN) with variance $\sigma^2_{N_{MZI}}$ is then added to the MZI signal ($I_{MZI}(t)+W(t)$). The argument of the MZI signal $k_{W,Kalman}[n]$ is extracted from the noisy MZI signal with the extended Kalman filter and the result is compared with the case when Hilbert transformation extracts the parameter $k_{W,Hilbert}[n]$. The MSE of the two $k$-estimation techniques, compared to the noise-free MZI argument $k[n]$, are calculated and normalized. As Fig. \ref{HilbertVsKalmanError} illustrates, the extended Kalman filter yields a $MSE_k$ that is at least an order of magnitude superior to the resulting error using  Hilbert transformation. $MSE_k$ can be formulated as,
\begin{equation}
     MSE_k = \mathbf{E}\big[\big| k_W[n]-k[n] \big|^2\big].
\end{equation}

\begin{figure}[h]
	\centering
	\includegraphics[width=0.7\textwidth]{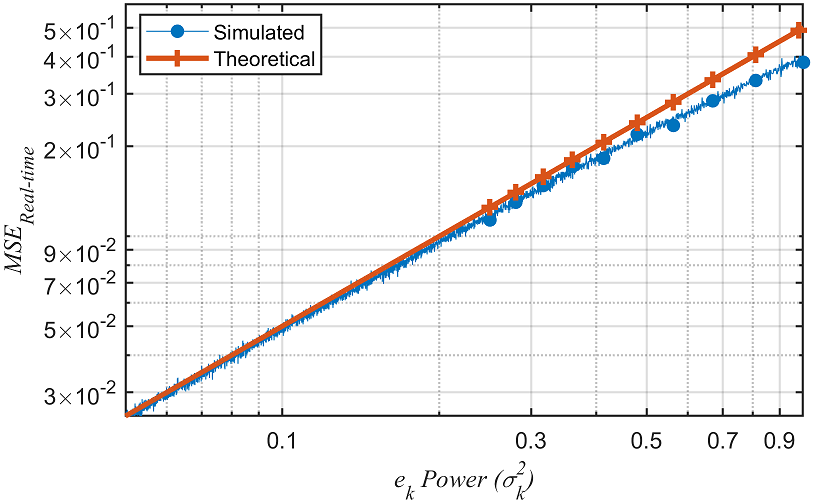}
	\caption{The relationship between the value of MSE and $\sigma_k^2$, assuming $\sqrt{R_RR_S} = 1.$}    
	\label{MSEvsEnoise}	
\end{figure}

In the MSE sense, the calibrated interferometric signal is different from the an ideally $k$-calibrated interferometric signal by,
\begin{equation} 
\begin{aligned}
   MSE_{Real-time}  =& \mathbf{E}\big[\big|I_{D,Calibrated}-I_{D,ideal}\big|^2\big]\\
   =& \mathbf{E}\big[\big|\epsilon_k[n]\sqrt{R_RR_S}sin(k_{cn}(z_R-z_S))\big|^2\big].
\end{aligned}
\label{MSEhilbert}
\end{equation}
Knowing that $\epsilon_k[n]$ and $k_{cn}$ are two independent variables,
\begin{equation} 
\begin{aligned}
   MSE_{Real-time}  =\frac{1}{2}r_Rr_S \mathbf{E}\big[\big|\epsilon_k[n]\big|^2\big],
\end{aligned}
\label{MSEhilbert1}
\end{equation}
with $\mathbf{E}\big[\big|\epsilon_k[n]\big|^2\big]=\sigma_k^2$. The superior estimation error when an extended Kalman filter is used significantly reduces the MSE value as derived in (\ref{MSEhilbert1}) and illustrated in Fig. \ref{MSEvsEnoise}, showing good agreement between theory and simulation.

\subsection{Interpolation Error for Calibration Mirror Test}
To understand the advantages of the real-time system, the effect of the interpolation rate during the calibration mirror test is investigated. The hypothesis in this paper is that the error produced by interpolation and look-up table deteriorates the performance of the SS-OCT spectral calibration. The remaining of this section presents analysis and simulation results to fully validate this hypothesis. The interpolated signal $I_{D_I}$ can be formulated as, 
\begin{equation} 
\begin{aligned}
   I_{D_I}(t) = \sum_{n} I_{D}[n]\Phi\big(t-nT_s\big),
\end{aligned}
\label{interpolationformula}
\end{equation}
where $\Phi(t)$ can be any interpolation function. The resampled interferometric signal then can be written as,
\begin{equation} 
\begin{aligned}
   I_{D_{Resampled}}[n] =  I_{D_{Interpolated}}(t_I) = I_{D_I}\big(k^{-1}(k_c)\big),
\end{aligned}
\label{resmpledinterferometricformula}
\end{equation}

\begin{figure}[H]
	\centering
	\includegraphics[width=0.6\textwidth]{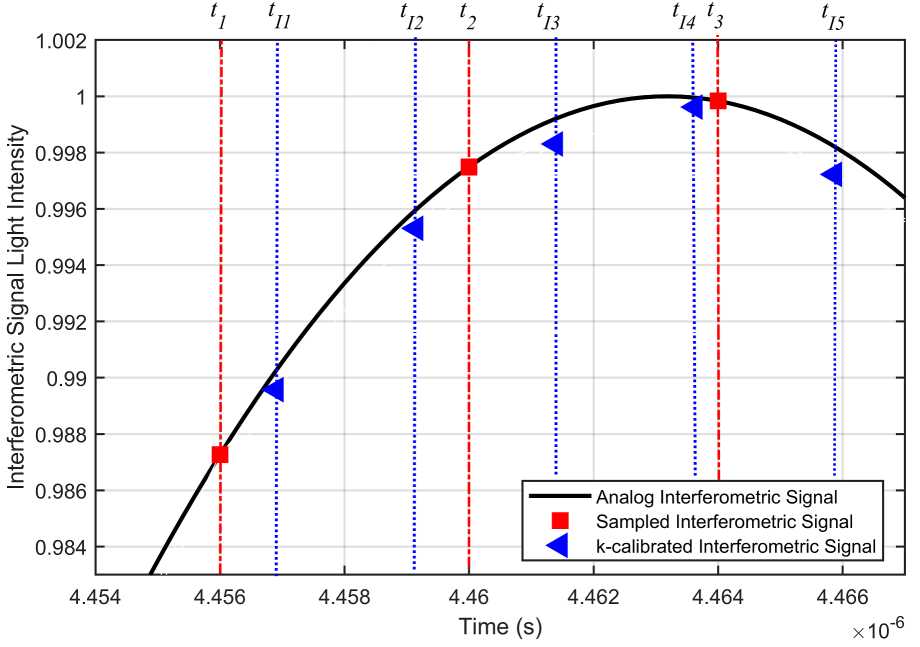}
	\caption{A demonstration of the analog interferometric signal (black), the sampled interferometric signal after ADC2 in Fig. \ref{hilbertconfig} (red), and the $k$-calibrated samples (blue). Linear interpolation is used and the interpolation rate is two. }
	\label{interpolator error}
\end{figure}

where $t_I=k^{-1}(k_c)$ denotes the time instances that corresponds to the equidistant wavenumber values $\{k_c\}$. The resampled data is off from the signal $I_{D_{Ideal}}$ as the interpolated signal $I_{D_{Interpolated}}$ differs from the original analog signal $I_D$ sampled at the correct time instances. Fig. \ref{interpolator error} shows how the interpolation causes the $k$-calibrated samples to differ from the samples from the ideal analog interferometric signal. This error can be formulated in the MSE sense,
\begin{equation} 
\begin{aligned}
  MSE_{Resampling} = \mathbf{E}\big[\big|I_{D_{Resampled}}[n]-I_{D_{Ideal}}[n]\big|^2\big],
\end{aligned}
\label{resmpledinterferometricformula1}
\end{equation}
where $I_{D_{Resampled}}[n]$ denotes the samples that are acquired after interpolation and resampling. The accuracy of the interpolation function $\Phi$, directly affects the value of $MSE_{Resampling}$. The more advanced the interpolation function, the less the value of $MSE_{Resampling}$ becomes. For simple interpolation functions such as \textit{Next} and \textit{Previous}, no further arithmetic are needed to be carried out to perform the interpolation operation, however, interpolations such as \textit{Spline} are rather computationally intensive and increase the design overhead in the interpolation and resampling based spectral calibration. 
\begin{figure}[h]
	\centering
	\includegraphics[width=0.6\textwidth]{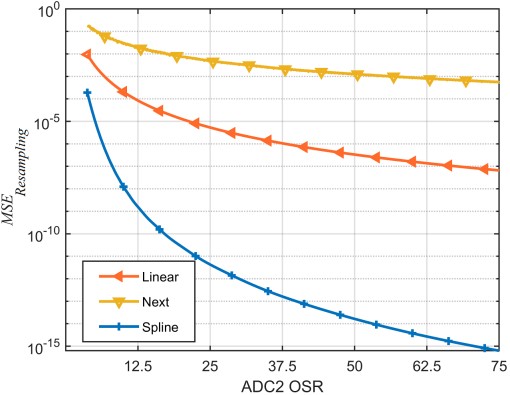}
	\caption{The value of $MSE_{Resampling}$ Vs. ADC2 oversampling rate (OSR) in Fig. \ref{hilbertconfig} for different interpolation functions. The Nyquist rate is twice the bandwidth of the interferometric signal.  }
	\label{MSEFs}
\end{figure}
This MSE error is expected to be larger if an A-scan on an arbitrary object is taken because the remaining nonlinearities will introduce distortion not accounted by interpolation, leading to artifacts after resampling. This fundamental limitation will be evaluated in the next subsection.


\subsection{Sampling Rate Effect in the Interpolation and Look-up Table Based Method}

Unlike the proposed system, the accuracy of the $k$-calibrated inteferometric signal in the interpolation and look-up table based technique is a function of the sampling rate of ADC2 in Fig. \ref{hilbertconfig}. As Fig. \ref{MSEFs} shows, the MSE deteriorates if the ADC2 sampling rate is not high enough. Note that this simulation is for the mirror calibration test and the performance shown does not predict the performance in an A-scan with an arbitrary object. We will show that such performance is object dependent for the resampling approach because there is residual nonlinearities that are not present with the real-time approach which $k$-linearizes the signal during sampling.  
\begin{figure}[h]
	\centering
	\includegraphics[width=0.6\textwidth]{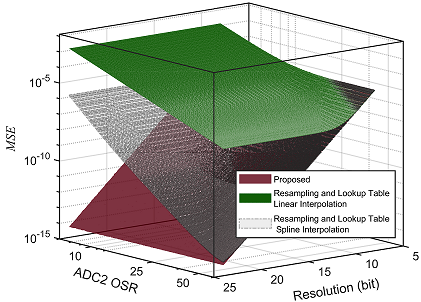}
	\caption{Comparison between the MSE of the proposed real-time calibration and the MSE of conventional resampling method.  }
	\label{3Dimage}
\end{figure}

\begin{figure}[h]
\centering
\includegraphics[width=.5\textwidth]{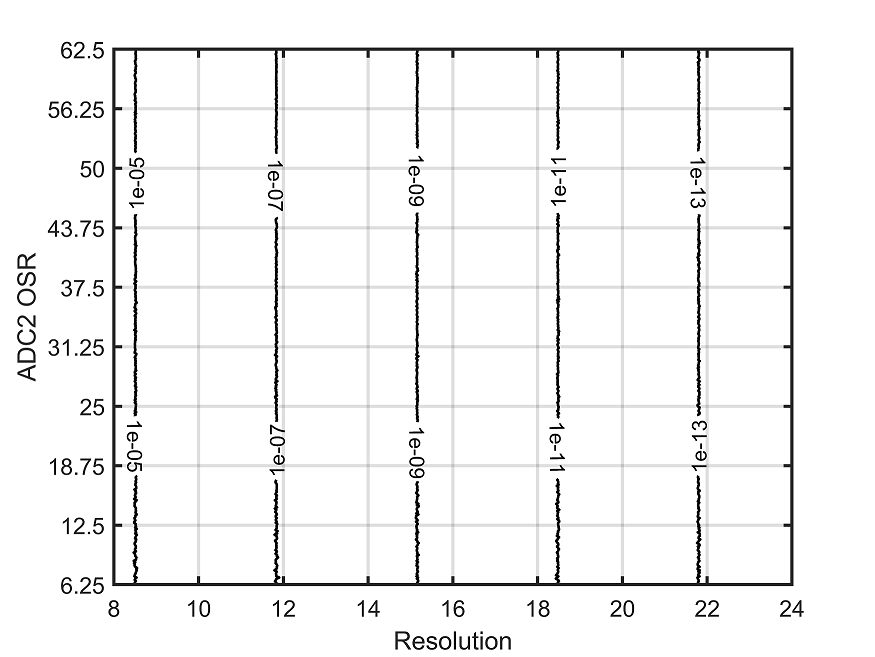}

\break
\includegraphics[width=.5\textwidth]{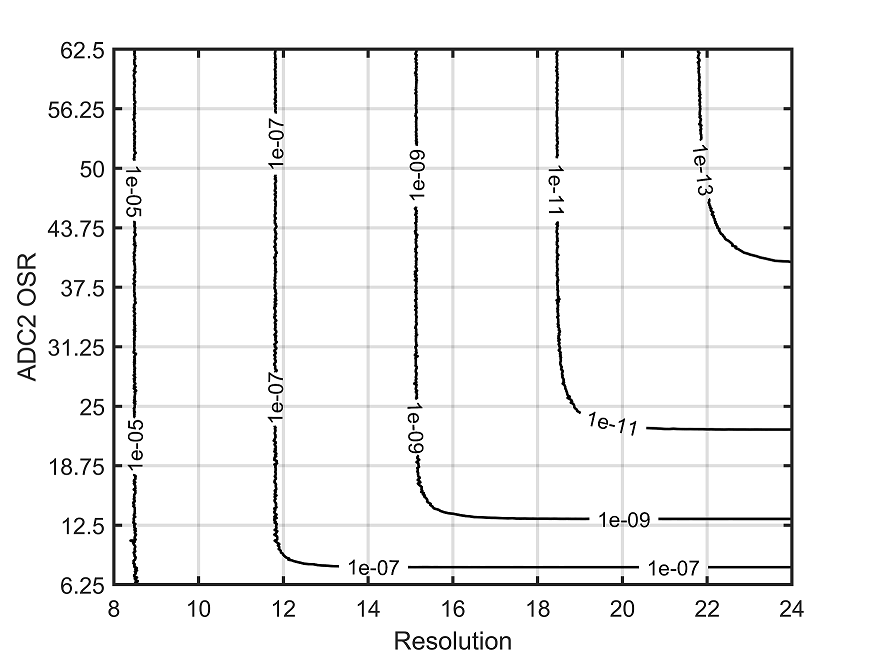}
\caption{Contours to achieve specific values of MSE. (Top) The MSE values for the proposed technique. (Bottom) The MSE values for the Spline interpolation technique}\label{contour}
\end{figure}

\subsection{Digital Resolution Effect in the Interpolation and Look-up Table Based Method}

\begin{figure}[H]
\centering
\includegraphics[width=.55\textwidth]{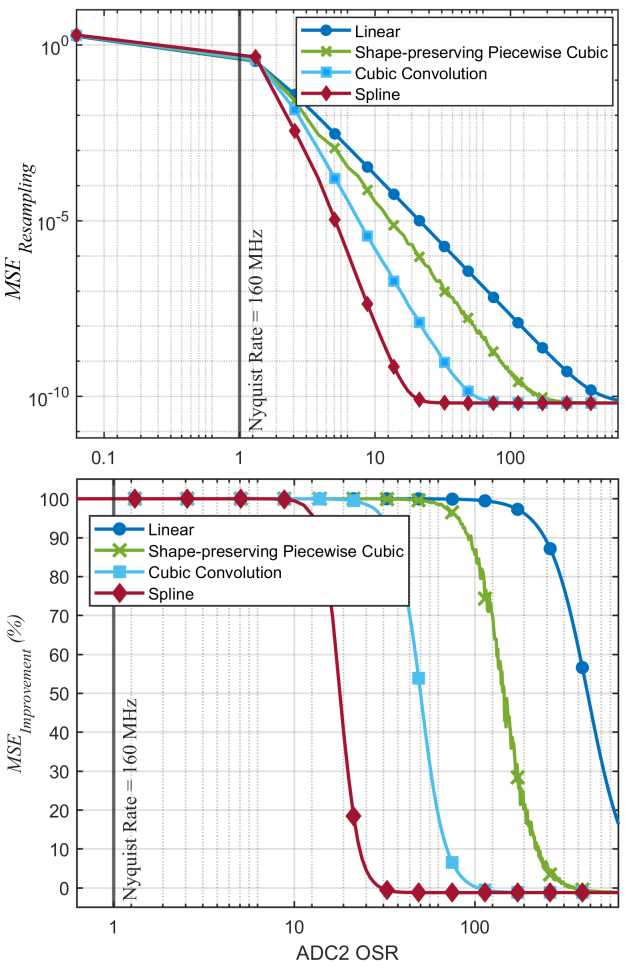}
\caption{(Top) The $MSE_{Resampling}$ performance of spectral calibration technique based on interpolation and resampling as a function of ADC2 oversampling rate. (Bottom) The MSE improvement using the proposed technique compared with the spectral calibration technique based on interpolation and resampling as a function of ADC2 oversampling rate. A-scan repetition rate is 1 MHz.}\label{comparison}
\end{figure}

Another drawback that the system based on interpolation and resampling exhibits is the sensitivity to the digital bit resolution of ADC2 in Fig. \ref{hilbertconfig} and its effect on the accuracy of the interpolation unit. As the resolution of ADC2 decreases, the interpolation unit results in a higher amount of error in the calculated $k$-calibrated interferometric signal. As Fig. \ref{3Dimage} shows, the value of  $MSE$ significantly worsens as the bit resolution of ADC2 decreases. In the proposed system, however, the MSE error is only set by the quantization noise and the interferometric $k$-calibrated time-instances are accurately obtained using the level crossing sampler. As Fig. \ref{3Dimage} shows, an increased sampling and resolution rates are required in the interpolation and resampling based technique to yield the same results that the proposed system achieves with an average sampling set at Nyquist rate. Moreover, when simpler interpolation techniques such as \textit{Linear} are used, the value of $MSE_{Resampling}$ can not reach that of the proposed system regardless of the values of the resolution and sampling rate of ADC2. Further, the high oversampling rate required for the spline interpolation technique to yield the same MSE as the proposed technique makes it impossible to be used in future versions of OCT devices where the A-scan repetition rate will be much higher. Our simulations indicates that for a MHz repetition rate, an ADC2 sampling rate of more than 10 GHz is required. Fig. \ref{contour} depicts the contours to yield specific MSE values. While the MSE value is only a function of the ADC resolution in the proposed system, the method based on interpolation and resampling require an increased sampling rate to achieve MSE values on par with the proposed technique. Figure \ref{comparison} shows that for the case of a 1 MHz repetition rate, the MSE improvement that can be yielded from using the proposed technique is significant for feasible sampling rates of ADC2.

\begin{equation} 
\begin{aligned}
MSE_{Improvement}=\dfrac{MSE_{Resampling}-MSE_{Real-time}}{MSE_{Resampling}}100\%
\end{aligned}
\label{IDideal}
\end{equation}
\subsection{Practical Limitation of SS-OCT Spectral Calibration on an A-scan of an Arbitrary Object}

\begin{figure}[h]
\centering
\includegraphics[width=.52\textwidth]{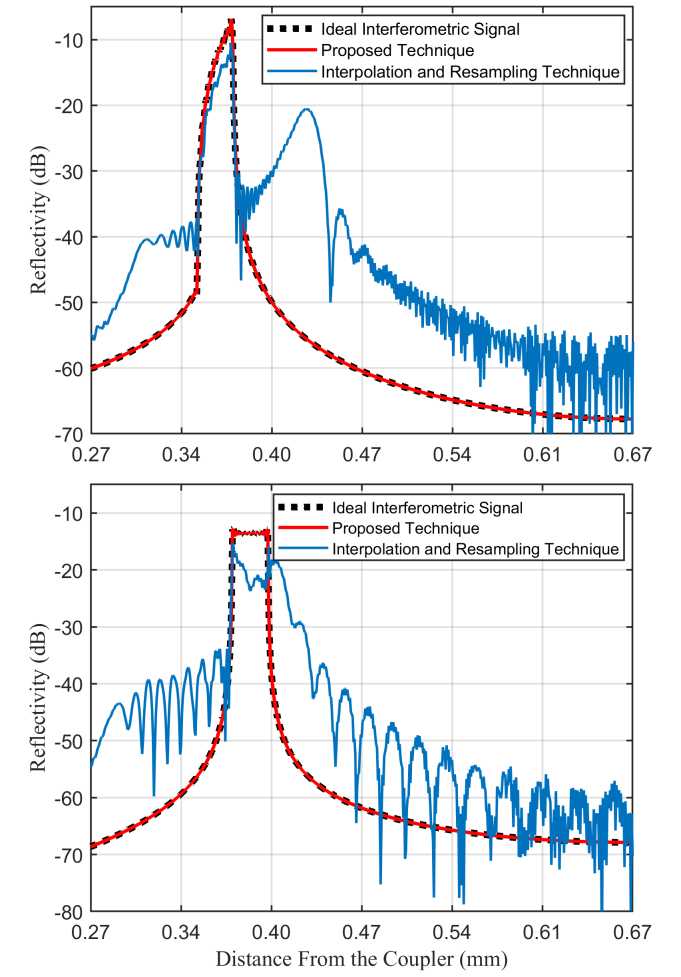}
\caption{The performance of the proposed spectral calibration technique compared with the method based on Spline interpolation and resampling on object under test with (top) ramp shape reflctivity profile $R_{S_i}=iR_{S,ramp}(Z-Z_{S_i})$, OSR is 1.81 and the ADC2 sampling rate is 4 GHz, (bottom) squared shape reflectivity profile $R_{S_i}=R_{S,square}~\text{for all}~Z_{S_i}$, $R_{S,square}$ and $R_{S,ramp}$ are constant values.  $R_{S_i}$ is defined in Eqn. (\ref{Equ.formulaOCTchap5}).  A-scan repetition rate is 1 MHz.}\label{object}
\end{figure}

The previous MSE results are useful for the evaluation of the calibration system using the mirror test. However, such results do not predict the performance during the A-scan of an arbitrary object because the nonlinearities will become a strong function of the modulation produced by the object into the interferometric signal. Figure \ref{object} shows that while the performance of the proposed technique is insensitive to the reflectivity profile of the object under test, that of the calibration technique based on interpolation and resampling is heavily dependent on the reflectivity profile of the sample under test. 
\begin{figure}[h]
\centering
\includegraphics[width=.6\textwidth]{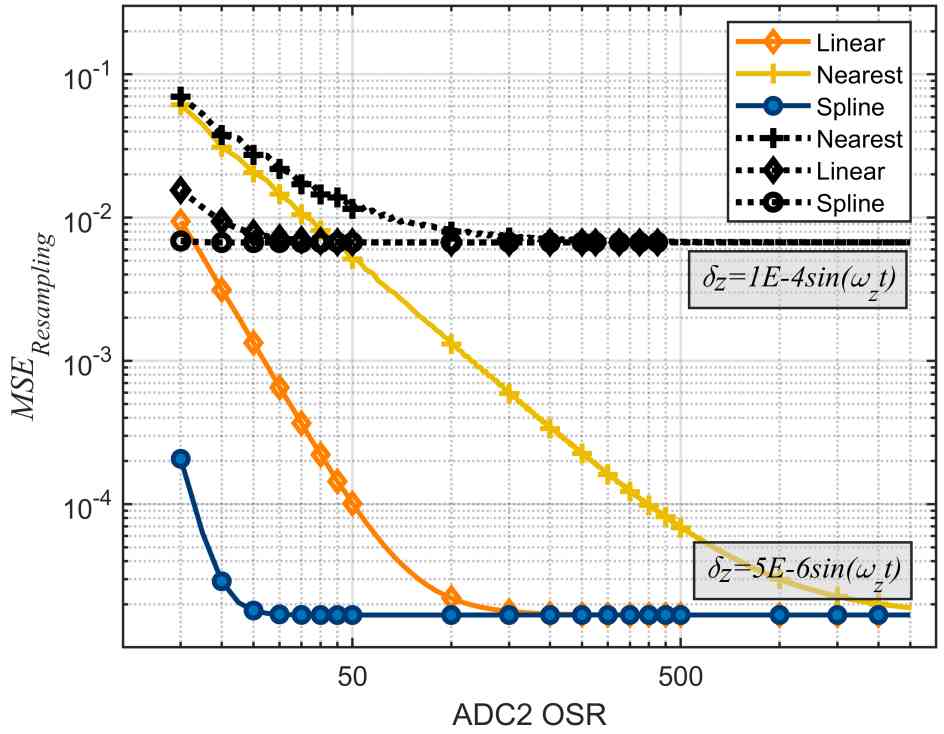}
\caption{The effect of subject under test physical location perturbation on setting a floor on the value of the $MSE_{Resampling}$, assuming $\sqrt{R_RR_S}=1$.  }\label{zperturbation}
\end{figure}
Another practical limitation is the physical movements that the subject under test may exhibit during the time span in which an A-scan is being acquired. As an example, such movements can be attributed to the involuntary movements that a patient's eye exhibits during an eye test. 
Let us assume that the ideal interferometric signal is still denoted by $I_{D_{Ideal}}$ but has the new formulation given by,
\begin{equation} 
\begin{aligned}
I_{D_{Ideal}} = \sqrt{r_Rr_S}\big(cos(k(Z_R-Z_S))\big) = \sqrt{r_Rr_S}\big(cos(k\Delta Z)\big),
\end{aligned}
\label{IDideal}
\end{equation}
and,
\begin{equation} 
\begin{aligned}
I_{D_{Resampled}} =  \sqrt{r_Rr_S}\big(cos\big(k(\Delta Z+\delta_ Z(t_{Resampled}))\big)\big),
\end{aligned}
\label{IDidealresampled}
\end{equation}
where $\delta_{Z}(t)=A_{\delta_z}sin(\omega_zt)$ models the physical perturbations of the sample under test.
Fig. \ref{zperturbation} demonstrates how having a perturbation on the physical movement of the subject under test sets a floor on the MSE regardless of the type of interpolation or ADC2 oversampling.

\subsection{Fundamental Trade-off in SS-OCT that Highlights Real-time Approach Speed-up Factor}
\begin{figure}[h]
	\centering
	\includegraphics[width=0.7\textwidth]{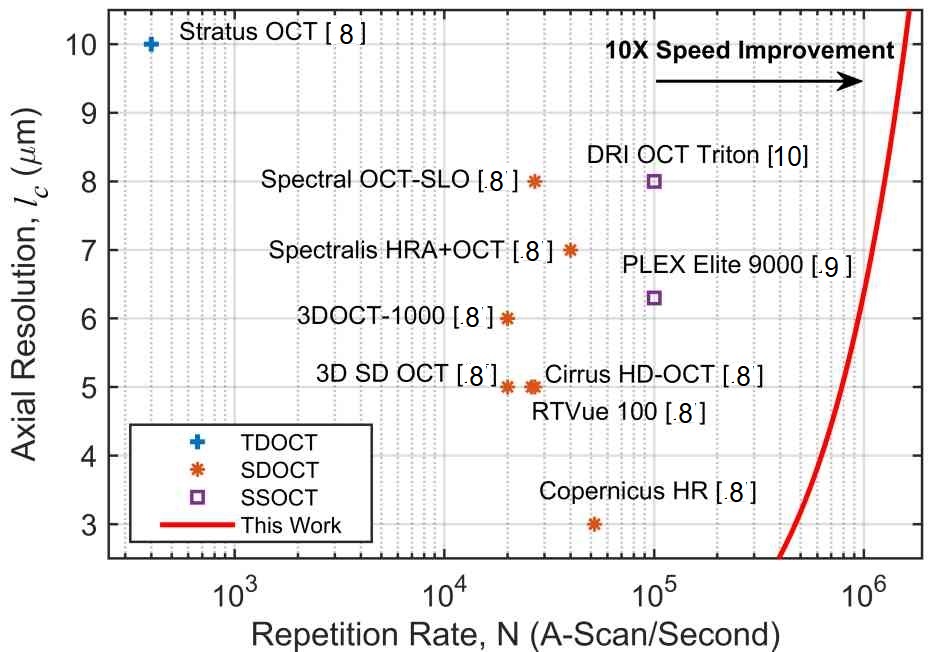}
	\caption{Comparison between commercially available products \cite{commercial1,commercial2,commercial3} and the proposed real-time system. Red curve illustrates trade-off
between resolution and A-scan speed.  }
	\label{commercialfig}
\end{figure}

The proposed OCT system has a critical trade-off between axial resolution and A-scan rate that is
characterized by \cite{OCTOriginal}:
\begin{equation} 
\begin{aligned}
   l_c = 4\sqrt{Ln(2)} \frac{N}{\pi}\frac{Z_{max}}{f_{sample}},
\end{aligned}
\label{commercial}
\end{equation}
Fig. \ref{commercialfig} shows a comparative study \cite{commercial1,commercial2,commercial3} with the imaging rate (Repetition
Rate $N$) and the axial resolution $l_c$ of several commercially available products and
the proposed system. $Z_{max}$ denotes the imaging range of the proposed OCT system and $f_{sample}$ indicates the maximum calibrating clock frequency that the level crossing sampler in Fig. \ref{ProposedMechanism} can provide. Substituting the extraordinary high value of the $f_{sample}$ that the level crossing sampler proposed in \cite{tcas1} in (\ref{commercial}), this work achieves an order of magnitude higher imaging speed at comparable axial resolution specifications.
The poor imaging range resulted from the zero crossing approach as well as an extensively high processing overhead and lack of efficiency in the arithmetic processes done by a Hilbert transformation has been explained as the reason why other demodulation techniques such as envelope detection, IpDFT, and KF are proposed. One other drawback the Hilbert based technique has is the requirement to sample both the calibrating and interferometric signals into the digital domain doubling the amount of data that is processed. The proposed calibrating engine cuts the amount of data by only sampling and processing the calibrating signal. Furthermore, our proposed technique provides an ultra-fast level crossing sampler that enables a custom-built SS-OCT calibrating engine that can provide the imaging rates corresponding to such high sweep rate as high as what new swept source lasers can provide. Our research shows \cite{tcas1} that the level crossing sampler proposed is capable of providing asynchronous sampling rates as high 5 GHz. This is a significantly high data rate compared to other reported samplers in the literature. It should be noted that the proposed LCS can provide such high asynchronous sampling rates while simultaneously inverts the effects of the non-linear sweep of the wavenumber as Equation. \ref{Equ.formulaOCTchap5cal} suggests. This revolutionary level crossing sampler combined with the efficient demodulation techniques, KF and IpDFT, enable us to have the first ever SS-OCT spectral calibrating mechanism that can provide a calibrating clock on the fly via an all integrated on a single chip circuit as the interferometric and the calibrating signals are being produced. To make it more clear, an interferometric signal being produced by the wavenumber sweep of the laser needs to be sampled at around 1000 points per sweep. Being confident that our demodulation techniques can provide the signal $k(t)$ available to the level crossing sampler, we can propose an SS-OCT calibrating mechanism that can provide sampling clocks for laser sweep rates in the MHz realm. Such high sweeping rate can be achieved with only one laser beam pointing at the object under test. By interleaving multiple laser beams that are pointing to the same object at the same time, the imaging rate can be increased even more \cite{multimegahertz}. \\
\section{The Benefit of Having a High $f_{sample}$}
\indent The fundamental goal of the proposed calibration system is to increase the imaging rate while keeping the resolution at the highest setting possible. Let us assume that the number of interferometric signals per second is denoted as $N$, and the number of points sampled per interferometric signal is denoted by $M$. The sampling rate that the LCS can provide then can be formulated as $f_{sample}  = M\times N$. Furthermore, the wavenumber range that the laser sweeps the wavenumber across is denoted by $\Delta K$. Then, the wavenumber distance between two consequent sampled points within every scan is expressed as $\delta_s k$, where $\Delta K=M\delta_s k$. Having these expressions in mind, the axial resolution $l_c$, would be equal to $\dfrac{2\sqrt{ln(2)}}{\Delta k}$, and the imaging range\textbackslash depth $(z_max)$  is  $\dfrac{\pi}{2\delta_s k}$. It should be noted that the parameters $\Delta K$ and $N$ are set by the laser that is used in the OCT system and are given a priori. The extra-ordinarily high sampling rate $f_{sample}$ that our LCS can provide in combination with its inherent non-uniform sampling time stamps allows to increase the number $M$ for a given $N$ effectively sampling every interferometric signal at a sufficiently high speed non-uniformly distributed time instances. An increased number of $M$, results in a lower value of $\delta_s k$ and an increased imaging range $z_{max}$.\\ \indent Thus, an increased $f_{sample}$ makes it possible for OCT manufacturers to simultaneously increase the value of $N$ and $\Delta K$ while keeping $M$ and $\delta_s k$ at reasonable values to maintain a high imaging rate, an enhanced resolution, and a better imaging depth. To clarify more, new research \cite{widebandlaser} shows the possibility of having lasers with double the spectral bandwidth compared to that of commonly available commercial lasers. Incorporating such spectral sweep bandwidths within the system doubles the axial resolution. In common commercial products, the value for the imaging depth is designed to be a couple of millimeters, the axial resolution is about 10 $\mu$m, and the imaging rate is at around couple of hundreds of thousands per second. Our proposed calibration mechanism will enable OCT manufacturers to achieve MHz imaging rates while keeping the axial resolution and the imaging depth at values same or better than they currently are. This is a 10X scan speed improvement and is the fundamental specification that will make our product to stand out in comparison with current state-of-art commercially available OCT instruments.

\chapter{CONCLUSION}
As explained in previous chapters of this dissertation, the main premise of this work is to provide an online all on a chip spectral calibration method that enables SS-OCT devices to work on the fly and enables these devices to work in the MHz imaging acquisition speed. This improvement in turn can increase the imaging range and\textbackslash or the axial resolution if the designer chooses to. These enhancements in terms of image acquisition speed, axial resolution and imaging depth can be useful in different major OCT applications as will be discussed below.
\section{Improvements in Terms of Imaging Depth}
\subsection{Improvements Possible for Anatomical OCT Devices, A-OCT}
"A-OCT is an endoscopic optical modality designed to provide quantitative cross-sectional images of large internal hollow organ anatomy over extended observational periods. It has previously been used in assessment of sleep disorders in the upper airwa, and preliminary work has been presented on its application in the lower airways. It acquires anatomical cross-sectional images of lumens over a range of several centimeters (fast scanning rate can increase the range). \cite{A-OCTrelationtoSSOCT}"\\
\indent As explained in the previous chapter the imaging depth\textbackslash range of a SS-OCT system can be increased using the ideas presented in this dissertation making it possible to make more vivid images from hollow organs of the body that require ranges more than few centimeters.
\section{Improvements in Terms of Resolution}
\subsection{Improvements Possible for Microscopic OCT Devices, $\mu$OCT}
\indent The main application of $\mu$OCT devices relates to scenarios when the required lateral resolution are too fine where extra care needs to be taken into account. The improvement of axial resolutions is obviously beneficial and possible via the means explained in previous chapters of of this dissertation. \\
\indent Often times, the value of the lateral resolution is restricted by the speed the laser can sweep its spectral frequency and therefore the speed it can sweep the laser beam across the subject under test. The possibility of having devices with higher acquisition speed capabilities enables designers to use light sources with smaller beam size which in turn improves the lateral resolution of the system. Having a higher A-scan acquisition speed makes it possible for the designer to sweep the same length on the subject under test at the same time that a wider beam size would have required, effectively giving the designer another degree of freedom to further improve the lateral resolution an therefore improve the specifications of a $mu$ OCT device
\section{Improvements in Terms of Image Acquisition Speed}
\subsection{Improvements Possible for OCT Angiography Devices}
OCT angiography basically works by acquiring different images at consequent times at measuring the difference between them to determine the movement of particles including the blood particles in vessels. Essentially, the movement of blood is captured and the location of the vessel is observed. \\
It is obvious that the faster the acquisition of tomographic images can happen consequently, the angiograph can be be obtained with a higher quality

\subsection{Improvements Possible for OCT Elastography Devices}

Elastogrpahy is the practice of acquiring a tomographic image, applying pressure to the subject under test, acquiring asecond tomographic image and measuring the difference. By doing so, the stiffness or softness of the tissue under test can be dtermined.\\
It is obvious that increasing the image acquisition capacity can improve the quality of information obtained from an OCT elastography device.


\let\oldbibitem\bibitem
\renewcommand{\bibitem}{\setlength{\itemsep}{0pt}\oldbibitem}
\bibliographystyle{ieeetr}

\phantomsection
\addcontentsline{toc}{chapter}{REFERENCES}

\renewcommand{\bibname}{{\normalsize\rm REFERENCES}}

\bibliography{data/myReference}


\end{document}